\documentclass[showpacs,preprint,prb,eqsecnum]{revtex4}
\usepackage{amsmath}
\usepackage{epsfig,color}
\usepackage{graphicx}
\usepackage{dcolumn}
\usepackage{bm}
\newcommand{\be}{\begin{equation}}
\newcommand{\ee}{\end{equation}}
\newcommand{\beq}{\begin{equation}}
\newcommand{\eeq}{\end{equation}}
\newcommand{\bea}{\begin{eqnarray}}
\newcommand{\eea}{\end{eqnarray}}
\newcommand{\bwt}{\begin{widetext}}
\newcommand{\ewt}{\end{widetext}}

\begin{document}

\title{Quasiparticle interaction function in a 2D Fermi liquid near an
antiferromagnetic critical point}
\author{Andrey V. Chubukov}
\affiliation{Department of Physics, University of Wisconsin-Madison, Madison, Wisconsin
53706, USA}
\author{Peter W\"olfle}
\affiliation{Institute for Condensed Matter Theory and Institute for Nanotechnology,
Karlsruhe Institute of Technology, 76021 Karlsruhe, Germany}
\date{\today}

\begin{abstract}
We present the expression for the quasiparticle vertex function $\Gamma
^{\omega }(K_{F},P_{F})$ (proportional to the Landau
interaction function) in a 2D Fermi liquid (FL) near an instability towards
antiferromagnetism. This function is relevant in many ways in the context of
metallic quantum criticality. Previous studies have found that near a
quantum critical point, the system enters into a regime in which the fermionic self-energy
 is large near hot spots on the Fermi surface (points on the Fermi surface connected by the
antiferromagnetic ordering vector $q_\pi=(\pi,\pi)$)
 and has much stronger dependence on frequency than on momentum.
 We show that in this regime, which we termed a critical FL,
the conventional random-phase-approximation- (RPA) type approach
  breaks down, and to properly
calculate the vertex function one has to sum up an infinite series of terms
which were explicitly excluded in the conventional treatment.
 Besides, we show that, to properly describe the spin component of $\Gamma^{\omega }(K_{F},P_{F})$ even in an ordinary FL,
one has to add Aslamazov-Larkin (AL) terms to the RPA vertex.
We show that
the total $\Gamma^{\omega }(K_{F},P_{F})$  is larger in a critical FL than in an ordinary FL,  roughly by  an extra power of
magnetic correlation length
  $\xi$, which diverges at the quantum critical point. However, the enhancement of
$\Gamma ^{\omega }(K_{F},P_{F})$ is highly non-uniform: It holds only when,
 for one of the two momentum variables,  the distance from a hot spot along the Fermi surface
 is much larger than for the other one. This fact renders our case different from
quantum criticality at small momentum, where the enhancement of $\Gamma ^{\omega }(K_{F},P_{F})$
was found to be homogeneous.
  We show that
  the charge and spin components of
  the total vertex function
   satisfy the universal relations
   following from the Ward identities related to the conservation of the particle number and the total spin.
   We show that in a critical FL,  the Ward identity involves $\Gamma^{\omega }(K_{F},P_{F})$ taken between particles on the FS.
  We find that the charge and spin components of $\Gamma^{\omega }(K_{F},P_{F})$ are identical to leading order in the magnetic correlation length. We
   use our results for $\Gamma ^{\omega }(K_{F},P_{F})$ and
 for the quasiparticle residue
  to derive the Landau parameters $F_{c}^{l=0} = F_s^{l=0}$, the density of states,
  and the uniform ($q=0$)
 charge and spin susceptibilities $\chi ^{l=0}_{c} = \chi^{l=0}_s$. We show that the density of states $N_F$
 diverges as $\log \xi $, however $F_{c,s}^{l=0}$ also diverge as $\log \xi$, such that the total $\chi _{c,s}^{(l=0)} \propto N_F
 /(1 + F_c^{l=0})$
remain finite at $\xi = \infty$. We show that at weak coupling
these susceptibilities are
 parametrically smaller than for free fermions.
\end{abstract}

\maketitle

\section{Introduction}

Fermi liquid (FL) theory is arguably the most successful low-energy theory of
interacting fermions. It states that, as long as the system can be
adiabatically transformed from free to interacting fermions, its low-energy
properties are determined by fermionic quasiparticles,
which are qualitatively similar to
 free fermions ~\cite{landau,agd,baym,rg}.
Fundamental characteristics of fermionic quasiparticles, such as the Fermi
velocity $v_{F}^{\ast }$, the effective mass $m^{\ast }=p_{F}/v_{F}^{\ast }$,
the residue $1/Z$, and the velocities of collective two-particle
excitations, like zero-sound waves or spin waves, are expressed via the
fully renormalized, antisymmetrized four-fermion interaction
 $\Gamma_{\alpha\beta,\gamma\delta}
(K,P;K+Q, P-Q)$ ,
taken in the
limit of small momentum and frequency transfer $Q=(\mathbf{q},\omega )$ (we
use the short-hand notation $K=(\mathbf{k},\omega _{k})$,
the spin indices follow the order of $K$)

In Galilean-invariant systems, the effective mass of quasiparticles and the
thermodynamic properties,
like specific heat and magnetic susceptibility, are determined by the
interaction between fermions
$K$ and $P$ right on the Fermi surface (FS) [$K=K_{F}=(\mathbf{k}_{F},0)$ and $P =P_{F}=(%
\mathbf{p}_{F},0)$]  and are
expressed
in terms of $\Gamma^{\omega }_{\alpha\beta,\gamma\delta} (K_{F},P_{F};K_{F},P_{F})\equiv \Gamma
^{\omega }_{\alpha\beta,\gamma\delta} (K_{F},P_{F})$,
 which is the limit $\mathbf{q}=0$ and $\omega_q \rightarrow 0$
of $\Gamma_{\alpha\beta,\gamma\delta}(K,P;K+Q, P-Q)$.
  This function is proportional to the quasiparticle
interaction function introduced phenomenologically by Landau ~\cite{landau}.
 Its
counter-part $\Gamma ^{q}_{\alpha\beta,\gamma\delta} (K_{F},P)$, defined by setting $\omega_q=0$ first and then taking $q \rightarrow 0$,
 determines the
quasiparticle scattering properties.
 The quasiparticle residue $1/Z_k$ is not determined by the properties right on the FS, but
  nevertheless is expressed via an integral involving $\Gamma^{\omega}_{\alpha\beta,\alpha\beta} (K_{F},P)$, in which  one of the two momenta is on
the FS and the other is generally away from the FS (Ref. \cite{agd}).

 In
lattice systems, the thermodynamic properties of fermions are not determined
by $\Gamma ^{\omega }$ taken right on the FS, but Ward identities,
associated with the conservation of the total number of fermions~\cite{agd,pitaevskii}
and total spin~\cite{kondratenko},
 still
allow one to express the fermionic residue $1/Z$,
  the effective mass $%
m^{\ast }$,
 and the effective magnetic $g$ factor
via integrals involving
 spin and charge components of
$\Gamma ^{\omega }_{\alpha, \beta,\gamma\delta} (K_{F},P)$ (Ref.[\onlinecite{nozieres}]).

\begin{figure}[htbp]
\includegraphics[width=0.6\columnwidth]{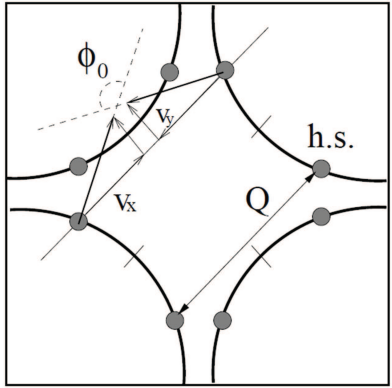}
\caption{The Fermi surface with hot spots (labeled as h.s.). Conjugate pairs of hot spots are connected by $Q= (\pi,\pi)$,
 and $\theta$ is the angle between Fermi velocities at $k_{F,hs}$ and $k_{F,hs} +Q$.
 From Ref.\cite{acs}.}
\label{fig1}
\end{figure}

The subject of this paper is the analysis of the form of the
 fully renormalized $\Gamma ^{\omega }(K_{F},P)$ in a FL near a
quantum-critical point (QCP). We specifically consider a $\mathbf{q}_\pi=(\pi ,\pi )$
commensurate spin-density-wave (SDW) QCP in a 2D metal with a FS like that
of the high $T_{c}$ cuprates (Fig. \ref{fig1}).
Previous works\cite{acs,ms,efetov,wang} have demonstrated that, unless certain
certain vertex renormalizations are strong ~\cite{peter}, the system
near a  SDW  QCP
enters into a
 regime in which the fermionic self-energy
 $\Sigma (k,\omega )$ develops a much stronger dependence on frequency than on
$\mathbf{k-k_{F}}$ and $Z_{k}=1 +
\partial \Sigma (\mathbf{k}_{F},\omega )/\partial
\omega $ gets large
in "hot regions",  where
shifting $\mathbf{k}_{F}$ by $\mathbf{q}_\pi$ does not take a fermion far away
from the FS.~\cite{acs,ms,efetov}
 We will be calling this regime a critical FL (CFL), following the notation in Ref. \cite{cm_nematic}.
 Because the self-energy predominantly depends on frequency, the effective mass approximately
scales as $Z$, and
  we will see that
the renormalization of $Z$
 comes from fermions in the
vicinity of the FS
and involves $\Gamma ^{\omega }_{\alpha\beta,\gamma\delta} (K_{F},P_{F})$.
We emphasize that our calculation of the $Z$-factor refers to the renormalization induced
by antiferromagnetic spin fluctuations. These renormalization is usually on top of renormalizations
generated at higher energies. The latter are thought to be included in the Hamiltonian we use as our starting point.

The
"common sense" approach to construct a microscopic theory near a SDW type
QCP is to replace the original four-fermion interaction by the effective
 interaction mediated by soft collective bosonic fluctuations in the spin
channel~\cite{scalapino} (the spin-fermion model). This replacement
 can only be justified in the random phase approximation (RPA)
 (see Sec. \ref{conv} below), but the outcome is
physically plausible and we adopt the
spin-fermion model as the microscopic low-energy model for a CFL. The model
describes fermions with the FS as shown in Fig. \ref{fig1} and with spin-spin
interaction
\beq
{\cal H}_{\text{spin-spin}} = \sum_{\mathbf{q}} V_{\alpha\beta,\gamma\delta} (\mathbf{q}) \sum_{\mathbf{k}, \mathbf{k'}} c_{{\mathbf{k},\alpha}}^{\dagger }c_{{\mathbf{k'},\beta}}^{\dagger } c_{\mathbf{k'}-\mathbf{q},\gamma}c_{\mathbf{k}+\mathbf{q},\delta}
\label{fri_1}
\eeq
 where
 \beq
V_{\alpha\beta,\gamma\delta} (\mathbf{q}) = V(\mathbf{q}) \vec{\sigma }_{\alpha \delta} \vec{\sigma }_{\beta\gamma}
\label{fri_2}
\eeq
 and the summation over spin indices is assumed.
 The interaction potential $V(q)$ is peaked at $\mathbf{q}=\mathbf{q}_\pi$ and
near the peak can be approximated by
\begin{equation}
V({\bf q})=\overline{g}/[(\mathbf{q}-\mathbf{q}_\pi)^{2}+\xi ^{-2}],  \label{mo_3}
\end{equation}%
where $\xi $ is the spin correlation length and $\overline{g}$ is the
effective spin-fermion coupling.
The low-energy model is valid when interactions do not take the fermions far
away from the FS. This requires $\overline{g}$ to be smaller than the
fermionic bandwidth, which for the FS in Fig. \ref{fig1} is comparable to the Fermi
energy $E_{F}$.
We assume that the relation $\overline{g}<E_{F}$ holds.
 We will keep $\xi $ large but finite and consider the system's behavior at
energies below
$\omega _{sf}\sim (v_{F}\xi ^{-1})^{2}/\overline{g}$, where the system is
still in the FL regime, At higher frequencies, which we do not consider
here, the system crosses over into a quantum-critical regime, where it
displays a non-FL behavior with $\Sigma (k,\omega )\propto \omega ^{a}$ with $%
a=1/2$ at the tree level (Refs. \cite{acs,ms} and \cite{max_last}).
 It is worth repeating that the "bare" quantities $v_{F}E_{F},\overline{g}$ are those of
quasiparticles
already
 renormalized by
  processes (e.g., the Kondo effect) occurring
at higher energies and not considered here.

For an ordinary FL with short-range interaction $U(q)$, the condition that the interaction
 is weaker than $E_{F}$
implies that the weak coupling approximation is valid.
 In that situation, the leading term
in the vertex function $\Gamma ^{\omega }_{\alpha\beta,\gamma\delta} (K_{F},P_{F})$ is just the
 antisymmetrized interaction
$U(0) \delta_{\alpha \gamma} \delta_{\beta \delta} - U({\bf k}_F-{\bf p}_F) \delta_{\alpha \delta} \delta_{\beta \gamma}$.
 It seems natural, at first glance, to apply
the same strategy near a QCP, i.e., identify $\Gamma^{\omega }_{\alpha\beta,\gamma\delta} (K_{F},P_{F})$
in a CFL with the effective interaction $V_{\alpha\beta,\gamma\delta} (K_F-P_{F})$,
 i..e., identify
 \beq
 \Gamma^{\omega }_{\alpha\beta,\gamma\delta} (K_{F},P_{F}) =
V(\mathbf{k}_{F}-\mathbf{p}_{F}) \vec{\sigma }_{\alpha \delta} \vec{\sigma }_{\beta\gamma} = V(K_F-P_F)
\left(\frac{3}{2}  \delta_{\alpha \gamma} \delta_{\beta \delta} - \frac{1}{2}  {\vec \sigma}_{\alpha \gamma} {\vec \sigma}_{\beta \delta}\right)
\label{fri_4}
\eeq
where $V(K_F-P_F) \equiv V(\mathbf{k}_{F}-\mathbf{p}_{F})$ is given by (\ref{mo_3}).
Additional antisymmetrization of $V_{\alpha\beta,\gamma\delta} (K_F-P_F)$ is not
required here because the effective spin-mediated interaction is already
obtained from an antisymmetrized original four-fermion interaction (see Sec. %
\ref{conv}).

The argument for the identification of $\Gamma ^{\omega }_{\alpha\beta,\gamma\delta}(K_{F},P_{F})$ with
$V_{\alpha\beta,\gamma\delta}(K_F-P_F)$ in a CFL
is seemingly
 quite general
(and applicable
 also
 to the case $\overline{g}\sim E_{F}$)
 because within the
conventional FL strategy $\Gamma ^{\omega }_{\alpha\beta,\gamma\delta}(K_{F},P_{F})$ includes all
renormalizations from fermions away from the FS, and $V_{\alpha\beta,\gamma\delta}
(K_F-P_F)$ is also assumed to include all relevant renormalizations
from outside of the FS. However, if we identify
 $\Gamma ^{\omega
}_{\alpha\beta,\gamma\delta}(K_F,P_F)$ with $V_{\alpha\beta,\gamma\delta}(K_F-P_F)$
and use the  Ward identity between the charge component of
$\Gamma ^{\omega }_{\alpha, \beta,\alpha,\beta}(K_{F},P_F)$ and the quasiparticle $Z_k$
factor (Eq. (\ref{ch_2_1}) below), we find a huge discrepancy between $Z_{k}$
along the FS, computed this way, and $Z_{k}$ computed using the direct
perturbative expansion in powers of $\overline{g}$. Namely, for a fermion
at a hot spot,
 $Z_{k}$ computed in the perturbative expansion~%
\cite{acs,ms} scales as $\overline{g}/(v_{F}\xi ^{-1})$ (modulo
logarithmical corrections~\cite{acs,ms,ms1,senthil}), while $Z_{k}$ computed
using
the FL formula with $V(K_F-P_F)$ scales as $(%
\overline{g}/(v_{F}\xi ^{-1}))^{1/2}$.

There is an even stronger discrepancy with the spin Ward identity~\cite{kondratenko}, which relates the  quasiparticle $Z_k$ with the spin component of
$\Gamma ^{\omega }_{\alpha, \beta,\alpha,\beta}(K_{F},P_F)$. Namely, if we use
$-(1/2) V (K_F-P_F) {\vec \sigma}_{\alpha \gamma} {\vec \sigma}_{\beta \delta}$ for the spin component of $\Gamma ^{\omega }_{\alpha, \beta,\alpha,\beta}(K_{F},P)$, as in Eq. (\ref{fri_4}), we find that the quasiparticle $Z_k$ becomes smaller than 1, which is obviously incorrect (we recall that we define $Z_k$ as the prefactor of the $\omega$ term in the quasiparticle Green's function).

As will be shown here, the resolution of the above problem requires two amendments to the standard microscopic derivation of Fermi liquid theory  ~\cite{agd}. The first originates from the fact that the effective interaction Eq. (\ref{mo_3}) develops a dynamical character when the QCP is approached. Physically the dominating dynamical properties are generated by the Landau damping of spin fluctuations (see Eq. (\ref{fr_11})).
  Once the interaction becomes dynamic,
  the standard argument, showing that quasiparticle contributions do not enter the vertex function $\Gamma ^{\omega }_{\alpha \beta, \gamma\delta} (K_{F},P_F)$, does not hold any more. Instead, it is necessary to sum up a ladder series of "forbidden" diagrams involving an irreducible dynamical vertex $V^{eff}$  and the quasiparticle-quasihole propagator. The quasiparticle-induced contributions to  $\Gamma ^{\omega }$ lead to an enhancement of $\Gamma ^{\omega }$ over
  $V^{eff}$, and the enhancement becomes singular at the SDW QCP.

The second amendment is the judicial choice of the irreducible vertex $V^{eff}$. A plausible guideline is the concept of "conserving  approximation" proposed by Baym and Kadanoff~\cite{baym_1}. It amounts to deriving the irreducible vertex by functionally differentiating the self energy with respect to the single particle Green's function.  If we take  the self-energy in one loop approximation (see Fig.\ref{fig6c_new}) and differentiate it, we then obtain, observing that the spin fluctuation propagator (the wavy line) is composed of an RPA bubble series, that the irreducible vertex $V^{eff}$ is given by the sum of two different contributions: a single
 spin-fluctuation propagator and a certain combination of two spin fluctuation propagators, traditionally called Azlamazov-Larkin (AL) terms (see Fig. \ref{fig_hw_1}b). We will show below that the AL terms play a decisive role for the spin
 part of $\Gamma ^{\omega}$, in fact flipping the sign and renormalizing the magnitude, such that the critical contributions to $\Gamma ^{\omega }$ diverge equally strong in the charge and spin channels.

 Both amendments have been earlier included in the analysis of a FL near a
 nematic instability~\cite{cm_nematic,cm_fm}, where one encounters a CFL driven by nematic fluctuations concentrated near $\mathbf{q}=0$. In that case the RPA result for $\Gamma ^{\omega }$ was found to be (i) corrected by AL terms~\cite{cm_fm} and (ii) renormalized~\cite{cm_nematic} by the divergent factor of $Z_{k}$.
 The renormalizations due to AL terms also play substantial role near a metal-insulator transition in 2D disordered systems~\cite{sasha}.

In this paper, we
 analyze the role of both amendments for  $\Gamma ^{\omega }_{\alpha\beta,\gamma\delta} (K,P)$ near a SDW
transition with $\mathbf{q}_\pi=(\pi ,\pi )$. We show that $\Gamma ^{\omega
}(K,P)$ again differs substantially from the spin-fermion
interaction $V(K-P)$.
We first demonstrate that AL corrections to the spin part of
$\Gamma^\omega_{\alpha\beta,\gamma\delta}$ in Eq. (\ref{fri_4})  have to be included even in the weak coupling regime, when the quasiparticle $Z_k$ is close to 1.
We show that the  AL contribution to  $\Gamma^\omega_{\alpha\beta,\gamma\delta}(K,P)$ for $K$ and $P$  near the Fermi surface is precisely
$2 V(K-P) {\vec \sigma}_{\alpha \gamma} {\vec \sigma}_{\beta \delta}$. Adding the AL contributions to the $V_{\alpha\beta,\gamma\delta}$ in  (\ref{fri_4}) we obtain the irreducible vertex $V^{eff}$, which
 represents
 the new "bare"
 vertex function
\beq
 V^{eff} (K,P) = \frac{3}{2} V(K-P)
\left(\delta_{\alpha \gamma} \delta_{\beta \delta} + {\vec \sigma}_{\alpha \gamma} {\vec \sigma}_{\beta \delta}\right)
\label{fri_4_1}
\eeq
where $V(K-P)$ is given by
 Eq. (\ref{mo_3}) for $K$ and $P$ on the FS (i.e., $K=K_F = (\mathbf{k}_{F},0)$ and  $P=P_F = (\mathbf{p}_{F},0)$) and by Eq.
 (\ref{fr_11})
  for $K$ and $P$ slightly away from the FS.
 We show that $V^{eff} (K,P)$  satisfies the charge and spin Ward identities in the weak coupling regime, as it should.

We then identify the series of additional contributions to  $\Gamma^\omega (K_F-P_F)$, which are small in an ordinary FL at weak coupling but become $O(1)$ in a CFL.
 Employing the spin
structure of the vertex $\Gamma^{\omega}_{\alpha\beta,\gamma\delta } (K,P) =\Gamma _{c} (K,P)\delta _{\alpha \gamma }\delta
_{\beta ,\delta }+\Gamma _{s} (K,P)\vec{\sigma }_{\alpha \gamma }\vec{\sigma
}_{\beta ,\delta }$, we obtain and solve two
integral equations for $\Gamma _{a}(K,P)$, $a=c,s$, in  the charge
and spin sectors.
 These equations are simplified for $K \approx K_F$ and $P \approx P_F$ and may be solved
by a suitable ansatz once we introduce  $f_{k,p}^{a}=\Gamma _{a}(K_{F},P_{F})/V^{eff}_{a}(K_F-P_F)$,
where scalar variables $k$ and $p$ are deviations along the FS from the corresponding hot spots
and $V^{eff}_{a} (K_F-P_F) = (3/2) V({\mathbf k}_F - {\mathbf p}_F)$ are the
 spin and charge
 components of $V^{eff}_{\alpha\beta,\gamma\delta } (K_F-P_F)$ in (\ref{fri_4_1}), defined
analogously to $\Gamma _{a}$.
 We show that  $f_{k,p}^{c} = f_{k,p}^{s} = f_{k,p}$, i.e., the $\delta_{\alpha \gamma} \delta_{\beta \delta} + {\vec \sigma}_{\alpha \gamma} {\vec \sigma}_{\beta \delta}$ structure of the vertex survives. We compare  the quasiparticle residue $%
1/Z_{k}$ obtained in (i) the direct diagrammatic calculation and (ii) using spin and charge  Ward identities with the  total $\Gamma^{\omega}_{\alpha\beta,\gamma\delta } (K,P)$, and show
  that they
 agree with each other, as they should.
 In other words,   the  total vertex function $\Gamma^{\omega}_{\alpha\beta,\gamma\delta } (K,P)$,
 which we obtain,
  satisfies spin and charge  Ward identities.
We again emphasize that the contributions to $\Gamma ^{\omega }$ discussed here and in the following come in addition to the usual contribution originating from the bare interaction renormalized by the incoherent part of the Green's function. The former contributions are generated by the critical fluctuations and are found to dominate near the QCP.

 Our results are similar, but not identical, to the ones in the nematic case~\cite{cm_nematic}.
  The key difference is that in our SDW case the function $%
f_{k,p}^{a}$ is
not a constant,
 like it was near a nematic transition, but rather depends strongly
 on the positions of the FS momenta $\mathbf{k}$ and $\mathbf{p}$ relative to
the corresponding hot spots.  Specifically,
 if $k$ and $p$  are comparable, $f_{k,p}=O(1)$, i.e.,
$\Gamma^\omega (K_{F},P_{F})$
is roughly the same as the spin-fermion
interaction $V^{eff}(K_F-P_F)$. If, however, one deviation
is parametrically larger than the other, $f_{k,p} $ becomes large and,
roughly, acquires an extra factor of $Z$, like in the case of a nematic QCP.
The physical consequence of this momentum-sensitive enhancement of $%
f_{k,p}$ is the
ultimate connection between the
FL description of fermions in the hot and cold regions on the FS. Namely,
$Z_{k}$ for a fermion in a hot region, where the FL becomes
 critical
near the QCP, is determined by
$\Gamma ^{\omega} (K_{F},P_{F})$
in which the
characteristic momenta $\mathbf{p}_{F}$ are located
 at the boundary to a cold region, where $Z_{p}$ remains $O(1)$ even at the
QCP.
This connection between hot and cold fermions is not easily seen in
perturbation theory where $Z_{k}$ for a hot fermion is
 determined
solely by fermions in hot regions, at least at one-loop order.

We also analyze the interplay between the contributions to $f_{k,p}$
from processes with even and odd numbers of spin-fermion scattering events.
For the processes with an odd number of scatterings, $\mathbf{k}_{F}$ and $%
\mathbf{p}_{F}$ differ by approximately $\mathbf{q}_\pi$, and for an even number of
scattering events, $\mathbf{k}_{F}$ and $\mathbf{p}_{F}$ are close to each
other. A similar separation into vertices with small and large
$\mathbf{k}_{F}-\mathbf{p}_{F}$ has been performed in Ref.\cite{max_last} in the
context of the calculation of the conductivity near a QCP. In our case, the
corresponding contributions to $f_{k,p}$ are $f_{k,p}^{\pi }$ and $%
f_{k,p}^{0}$. The full $f_{k,p}$  is the sum of the
two:
$f_{k,p}=f_{k,p}^{+} = f_{k,p}^{\pi }+f_{k,p}^{0}$. We compute $f_{k,p}^{\pi }$
and $f_{k,p}^{0}$ separately and find that their difference $%
f_{k,p}^{-}=f_{k,p}^{\pi}-f_{k,p}^{0}$ is also a highly non-trivial
function of  $k$ and $p$. Our results for $f_{k,p}^{+} = f_{k,p}$ and $%
f_{k,p}^{-}$ are summarized in Figs. \ref{fig10} and \ref{fig11}.

We use the result
 for $f_{k,p}$
to determine the density of states (DOS), the Landau function, and the
uniform spin and charge susceptibilities. We show that the DOS $N_F$
diverges as $\log \xi $ upon the approach to the SDW QCP.
We introduce the Landau function $F (K_F,P_F)$ by straightforward extension of the formula relating $F (K_F,P_F)$ and $\Gamma^\omega (K_F,P_F)$ in the isotropic case.
Because the quasiparticle $Z_k$ depends on the position on the FS, the Landau function
$F(K_F,P_F)$ depends separately on $\mathbf{k}$ and $\mathbf{p}$ rather than on their difference. In this situation,
one cannot use FL formulas relating partial components of $F(K_F,P_F)$ to charge and spin susceptibilities  and has to obtain the  susceptibilities by explicitly summing up bubble diagrams with self-energy
and vertex corrections. We demonstrate how to do this and pay special attention to the difference between contributions coming from the infinitesimal vicinity of the FS and from states at
small but still finite distances from the FS. We show that
the charge and spin susceptibilities ($\chi_c$ and $\chi_s$, respectively) are identical and for both  higher-loop terms form a geometrical series,
like in an isotropic FL. We argue that, in this situation, one can effectively describe $\chi_{c,s}$ by using a FL-like formula in which
$\langle F_c (K_F,P_F) \rangle$, averaged over both momenta, plays the role of the $l=0$ Landau interaction component.

The paper is organized as follows. In the next section we briefly review
basic facts about FL theory, viewed from a microscopic perspective,
introduce the fully renormalized, anti-symmetrized vertex function, and
present the relation between $\Gamma ^{\omega }$ and $Z_{k}$ . In Sec.\ref{conv}
we
derive the vertex function between low-energy fermions near a SDW QCP, but before the system enters the CFL regime.
We first present, in Sec. \ref{sec_new_1},  a conventional,
 RPA-type,
  "common-sense" derivation of $\Gamma ^{\omega }$.
 We show that the "common-sense" $\Gamma ^{\omega
}(K,P)$  coincides with the
 effective dynamical four-fermion interaction $V(K-P)$  mediated by soft collective excitations in
the spin channel.
In Sec. \ref{sec_new_2} we argue that the RPA analysis is incomplete near a magnetic transition, even in the ordinary FL regime.
We show that AL terms are as important as RPA-type terms and the $\Gamma^\omega$, which satisfies both spin and charge Ward identities in the
 ordinary FL regime, is not given by the direct spin-mediated interaction $V(K-P)$, but by the modified, effective
 interaction between near-critical fermions, $V^{eff}_{\alpha\beta, \gamma\delta }(K-P)$,  which is the sum of the direct spin-fluctuation exchange and AL terms.
In Sec.\ref{sec_3} we argue that Landau damping can
be neglected only if one is interested in the behavior of interacting electrons
above a certain frequency $\omega _{L}$, but must be kept if one is
interested in properties of fermions at the smallest frequencies. We discuss
the Landau damping induced crossover between the ordinary\ (non-critical) FL
and the CFL, evaluate $Z_{k}$ in a direct loop expansion
 and show
 that
there is a discrepancy between $Z_{k}$ obtained
 this way and
 $Z_{k}$  obtained using the FL formula,
  if the
   effective
   interaction $V^{eff}_{\alpha\beta ;\gamma\delta }(K_F-P)$ is used for $\Gamma ^{\omega }_{\alpha\beta ;\gamma\delta } (K_F,P)$. Sections \ref{sec_4}-\ref{sec_7} are the central
sections of the paper.
In Sec. \ref{sec_4} we argue that $\Gamma^{\omega}_{\alpha\beta, \gamma\delta }(K,P)$ in a CFL must differ
from $V^{eff}_{\alpha\beta, \gamma\delta }(K-P)$ and identify the diagrammatic
series for the fully renormalized $\Gamma ^{\omega }_{\alpha\beta, \gamma\delta }(K,P)$  for $K \approx K_F$ and $P \approx P_F$,  in which
$V^{eff}_{\alpha\beta, \gamma\delta }(K-P)$ is the first term. We show
that the expression for $Z_{k}$, obtained using the
 Ward identities with this fully renormalized $\Gamma ^{\omega }$, coincides with $%
Z_{k}$ obtained in a direct loop expansion.
 In Sec.\ref{sec_5} we present our solution of
the integral equation for $f_{k,p} = \Gamma ^{\omega} (K_F,P_F)/V^{eff}(K_F-P_F)$
and discuss the
momentum-selective enhancement of $f_{k,p}$ and the inter-connection
between hot and cold regions on the FS.
 We also discuss in this section the
interplay between $f_{k,p}^{\pi }$ and $f_{k,p}^{0}$. In Sec. \ref{sec_7} we present the full expression for the vertex function
  and discuss the quasiparticle interaction function (the Landau function), the density of states, and the uniform spin and
charge susceptibilities.
 Sec. \ref{sec_8} presents our
conclusions.

\section{Fermi liquid, basic formulas}
\label{sec_FL}

Fermi liquid  theory was first developed as a phenomenological theory for isotropic
 fermionic systems, obeying Galilean invariance and the conservation laws
for total charge and spin,
 on the postulate that interactions do
not change the relation between the fermionic density and the volume of the
FS. Subsequently, the FL theory was applied to conduction electrons in a metal~\cite{abrikosov},
 on which we naturally focus here,
 and
  was
  also re-formulated in the microscopic
(diagrammatic) language, using the notion of the fully renormalized,
anti-symmetrized vertex function $\Gamma _{\alpha\beta ;\gamma\delta } (K,P,Q)$, taken in the limit of small momentum transfer $\mathbf{q}$ and
small frequency transfer $\omega _{q}$ (we will continue to use 3D notation %
 $Q=(\mathbf{q},\omega _{q})$ , etc.).
 The universal relation between
fermionic density and the volume of the FS has been shown diagrammatically,
in order-by-order perturbative calculations, and is commonly known as
 the Luttinger theorem [\onlinecite{agd}].

We will work with Matsubara fermionic Green's functions in the limit
  when the
temperature $T\rightarrow 0$.
We split the fermionic Green's function into quasiparticle and incoherent parts as
\begin{equation}
G(k,\omega )=G_{qp}+G_{inc}.
\end{equation}%
The quasiparticle part of the Green's function has the form
\begin{equation}
G_{qp}=\frac{1}{Z_{k}}[{i\omega -v_{F,k}^{\ast }(k-k_{F})]}^{-1},  \label{s_1}
\end{equation}%
 where
$Z_{k}$ is the inverse quasiparticle residue, and $v_{F,k}^{\ast
}=k_{F}/m^{\ast}_{k}$, with $m^{\ast}_{k}/m=Z_{k}$, is the renormalized Fermi velocity.
We omitted the quasiparticle damping, considering that $G_{qp}$ will only be used sufficiently close to the Fermi surface.
In isotropic systems, $%
Z $ and $v_{F}^{\ast }$ are constants, in lattice systems both generally
depend on the location of $\mathbf{k}_{F}$ along the FS. $G_{inc}$ accounts
for the incoherent part of the fermionic propagator. In the following we will make
use of the fact that near a QCP the relevant fermionic states are located close
to the Fermi surface, and are described by the quasiparticle part of the Green's function.
The effects of the incoherent part are assumed to be included in the effective
interaction to be described later.

\begin{figure}[htbp]
\includegraphics[width=\columnwidth]{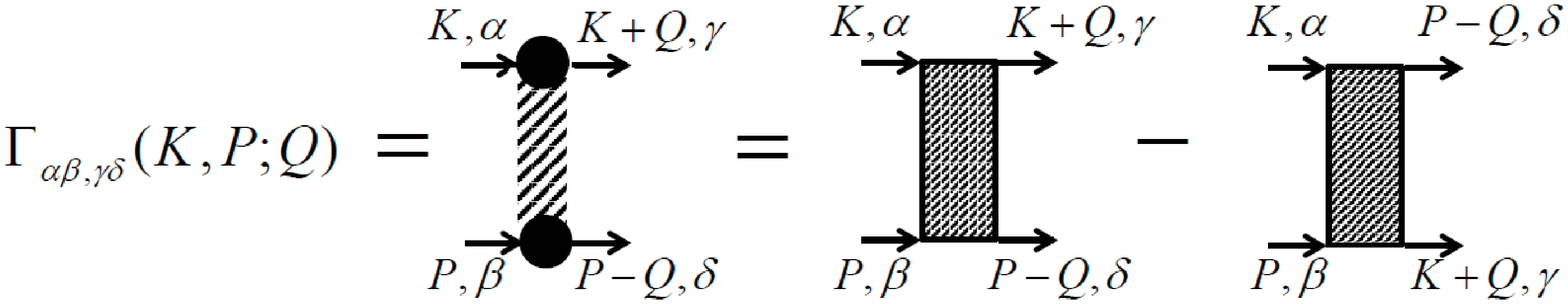}
\caption{Diagrammatic representation of the vertex function $\Gamma_{\protect\alpha,\protect\beta,\protect\gamma,\protect\delta}(K,P,Q)$. The
shaded vertices are fully renormalized 4-fermion interactions}
\label{fig2}
\end{figure}

The  vertex
 function
 $\Gamma_{{\alpha\beta, \gamma \delta }}(K,P)$ is the
 fully renormalized and anti-symmetrized interaction between quasiparticles.
   Graphically, $\Gamma_{{\alpha\beta, \gamma \delta }}(K,P,Q)$
   is the combination of the two terms  shown in Fig. \ref{fig2}.
   The second one is obtained from the first one
   by interchanging the two
outgoing fermions and changing the overall sign as required by the Pauli
principle.

Of particular interest for the FL theory is the limit of $Q= (\mathbf{q}, \omega_q)$ when $\mathbf{q}$ is
strictly zero and $\omega _{q}$ tends to zero, known as the "$\omega $%
-limit":
\begin{equation}
\Gamma^{\omega }_{\alpha\beta,\gamma\delta }(K,P,q)=\lim {}_{\omega
_{q}\rightarrow 0}\Gamma_{\alpha\beta,\gamma\delta }(k,0;p,0|k,\omega _{q};p,-\omega _{q}) =
\Gamma^{\omega }_{\alpha\beta,\gamma\delta }(K,P)
 \notag
\end{equation}%
According to
FL theory,
 the function $\Gamma^{\omega }_{\alpha\beta,\gamma\delta }(K_F,P_F)$,  taken between particles at the FS ($K_F = ({\bf k}_F,0)$), is
 proportional to the quasiparticle
interaction function (the Landau function) $F_{\alpha\beta,\gamma\delta
}(K_{F},P_{F})$.
 For isotropic, Galilean-invariant systems $Z_{k}=Z$ and $m^*_{k} = m^*$ are
 independent of the location of $\mathbf{k}_{F}$ on the FS, and the relation between $\Gamma^{\omega }_{\alpha\beta,\gamma\delta }(K_F,P_F)$ and
 $F_{\alpha\beta,\gamma\delta}(K_{F},P_{F})$ is~\cite{agd}
\begin{equation}
F_{\alpha\beta,\gamma\delta }(K_{F},P_{F})=2N_{F}Z^{-2}\Gamma _{\alpha\beta,\gamma\delta }^{\omega }(K_{F},P_{F})  \label{y_1}
\end{equation}%
 where
 $N_{F}\propto m^{\ast }$
 is the fully renormalized DOS per spin at the Fermi level.
 Although Eq. (\ref{y_1}) looks like a
simple proportionality, it is actually a highly non-linear integral relation
as $m^{\ast }$ is expressed via a particular partial component of $F_{\alpha\beta,\gamma\delta }$ and $Z$ is expressed via an integral over $\Gamma
_{\alpha\beta,\gamma\delta }^{\omega } (K_F,P)$
(see Eq. (\ref{ch_2_1}) below).
Still, as long as $Z$ and $m^{\ast }/m$ are some constants, the functional
form of $F_{\alpha\beta,\gamma\delta }(K_{F},P_{F})$ is the same as
that of $\Gamma _{\alpha\beta,\gamma\delta }^{\omega }(K_{F},P_{F})$.
The other limit (termed the "q limit"), $\omega _{q}=0$ and $q\rightarrow 0$
defines $\Gamma _{\alpha\beta,\gamma\delta }^{q}(K,P)$. The latter is
related to the quantum mechanical quasiparticle scattering amplitude .
Partial components of $\Gamma ^{q}$ and $\Gamma ^{\omega }$ are simply
related ~\cite{agd}.
 For non-isotropic systems, the definition of the Landau function is more nuanced
 as in general there is no straightforward relation between its partial amplitudes and observables.
  For practical purposes, it is convenient to introduce
$F_{\alpha\beta,\gamma\delta }(K_{F},P_{F})$ via  a relation similar to (\ref{y_1}) but with
  $N_F Z^{-2}$  replaced by $N_F/(Z_k Z_p)$, where now $N_F \propto <m^*_k>$
  is the total DOS ($m^*_k$ is the momentum dependent effective mass, and $<...>$ is the  average over the FS).
  We will return to this issue in Sec. \ref{sec_7}.

That the Landau function is expressed via $\Gamma ^{\omega }$ (as opposed to, e.g., $\Gamma
^{q}$) has a clear physical meaning. We will see in the next Section that in
ordinary FL theory $\Gamma ^{\omega }(K_{F},P_{F})$ includes all possible
renormalizations of the interactions between fermions at the FS, which come
from virtual processes with intermediate fermions away from the FS. The
processes in which all intermediate fermions
 in the immediate vicinity of the FS are explicitly
excluded from $\Gamma ^{\omega }(K_{F},P_{F})$ (but these are present in $%
\Gamma (K_{F},P_{F},Q)$ at an arbitrary ratio of $\omega _{q}$ and $v_{F}q$,
and, in particular, in $\Gamma ^{q}(K_{F},P_{F})$). In other words, $\Gamma
^{\omega }(K_{F},P_{F})$ represents the fully irreducible interaction
between fermions at the FS. Similarly, in Landau FL theory, the Landau
function $F$ has the meaning of the effective interaction between
quasiparticles on the FS, which absorbs all contributions from virtual
fermion excitations outside of the FS. Obviously, $F$ and $\Gamma ^{\omega
}(K_{F},P_{F})$ have to be expressed via each other. The remaining
renormalizations, coming from fermions
 in the immediate vicinity of the FS,
are all captured within FL theory which relates the observables with the
partial components of $F$.

For $SU(2)$ spin-invariant systems, the vertex $\Gamma^\omega_{\alpha,\beta,%
\gamma\delta} (K_F,P_F)$ and the Landau function $F_{\alpha,\beta,%
\gamma\delta} (K_F,P_F)$
can be decoupled into spin and charge components
as
\begin{eqnarray}
\Gamma^\omega_{\alpha,\beta,\gamma\delta} (K_F,P_F)&=& \Gamma_c
(K_F,P_F)\delta_{\alpha \gamma} \delta_{\beta,\delta} + \Gamma_s (K_F,P_F)%
\mathbf{\sigma}_{\alpha \gamma} \mathbf{\sigma}_{\beta,\delta} \nonumber \\
F _{\alpha\beta,\gamma\delta} (K_F,P_F)&=& F_c (K_F,P_F)\delta_{\alpha \gamma} \delta_{\beta,\delta} + F_s (K_F,P_F)%
\mathbf{\sigma}_{\alpha \gamma} \mathbf{\sigma}_{\beta,\delta}   \label{thr1}
\end{eqnarray}
In isotropic systems $\Gamma_c (K_F,P_F)$ ($F_c (K_F,P_F)$)  and $\Gamma_s (K_F,P_F)$ ($F_s (K_F,P_F)$) depend on the angle between ${\bf K}_F$ and ${\bf P}_F$.   Partial components of $F_c$ determine, e.g.,
 the effective mass $m^{\ast }$, the specific heat, and the velocities of zero-sound collective modes, and partial components of $F_s$
 determine the spin susceptibility and the properties of spin-wave excitations. In non-isotropic systems, the relations are a bit more involved, and to obtain $m^*$ one generally needs to extend the Landau function to the case when one of the momenta is away from the FS [Refs. \onlinecite{agd,maslov_last}].

In this paper we will be especially interested in the relation between $%
\Gamma _{\alpha\beta,\gamma\delta }^{\omega } (K,P)$ and the inverse quasiparticle
residue $Z_k = \partial G^{-1} (k,\omega) /\partial (i\omega_m)|_{\omega=0}$ for ${\bf k}$ on the FS.
 The quasiparticle $Z_k$ is not determined within the Landau FL (even in the isotropic case),
  as
  it generally
  does not come from fermions in the immediate vicinity of the FS.
 Still, there exist two exact relations between $Z_k$ and charge and spin components of $\Gamma^\omega (K_F,P)$.  These relations follow from the Ward identities associated with
the conservation laws for the total number of particles (in other words, the
charge) and the total spin. These relations do not require Galilean invariance and hold even
when
$Z=Z_{k}$ depends on the position of $\mathbf{k}_{F}$ along the FS.

 The Ward identity associated with the particle number conservation identifies $-\partial G^{-1}_{\alpha\gamma} (k,\omega_k)/\partial \delta \mu = \delta_{\alpha\gamma} \partial G^{-1} (k,\omega_k) /\partial (i\omega_k)$, where $\delta \mu$ is a small time-dependent and spatially homogeneous variation
   of the chemical potential, with the triple charge vertex $\Lambda^c_{\alpha\gamma} (\Omega,K) = \Lambda_c (\Omega,K) \delta_{\alpha\gamma}$, where $K = ({\bf k}, \omega_k)$ and $\Omega $ is set to be infinitesimally small~\cite{agd,pitaevskii}:
\begin{equation}
\frac{\partial G^{-1} (k,\omega)}{\partial (i\omega)} = \Lambda_c (\Omega,K)_{|\Omega \to 0}
\label{ex_1}
\end{equation}
\begin{figure}[htbp]
\includegraphics[width=\columnwidth]{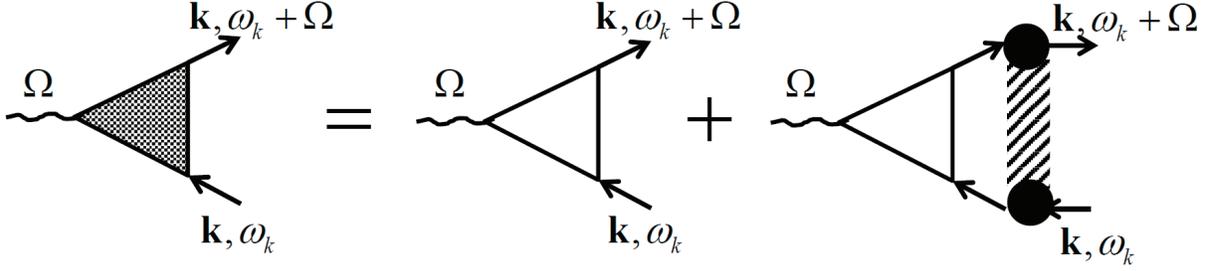}
\caption{The diagrammatic representation of the triple vertex $\Lambda_{c,s} (\Omega, K)$ ($K = ({\bf k}, \omega_k)$),
 involved in the Ward identities. For the charge Ward identity (associated with the conservation of the total number of fermions, i.e., the total charge),
  the bare vertex
  (the unshaded triangle)
  is $\delta_{\alpha,\gamma}$, where $\alpha$ and $\gamma$ are spin components of incoming and outgoing fermion.
 For the spin Ward identity (associated with the conservation of the total spin), the bare vertex is  $\sigma^z_{\alpha,\gamma}$.}
\label{fig7}
\end{figure}

The triple vertex $\Lambda_c (\Omega,K)$ is in turn expressed in terms of the fully renormalized
 charge component of $\Gamma^\omega$ as shown diagrammatically in Fig. \ref{fig7}.
 In analytical form
 \begin{eqnarray}
 \Lambda_c (\Omega,K) \delta_{\alpha\gamma}  &=&
 \delta_{\alpha\gamma} + \sum_{\beta,\delta}\int \delta_{\beta\delta} \Gamma _{\alpha \beta ,\gamma \delta }^{\omega}(K,P)\left\{ G^{2}(P)\right\} _{\omega }\frac{d^{3}P}{(2\pi )^{3}} \nonumber \\
&& =  \delta_{\alpha\gamma} \left(1 + 2 \int \Gamma^c (K,P)\left\{ G^{2}(P)\right\} _{\omega }\frac{d^{3}P}{(2\pi )^{3}}\right)
\label{ex_2}
\end{eqnarray}
 where $d^{3}P=d^{2}\mathbf{p}d\omega _{p}$.
For $K =K_F$, $\partial G^{-1} (k,\omega) /\partial (i\omega)|_{\omega=0} = Z_k$, and we have
\begin{equation}
Z_k = \Lambda_c(\Omega,K_F) = 1 + 2 \int \Gamma_c (K_F,P)\left\{ G^{2}(P)\right\} _{\omega }\frac{d^{3}P}{(2\pi )^{3}}
\label{ch_2_1}
\end{equation}
Similarly, the Ward identity associated with the conservation of the total spin identifies $-\partial G^{-1}_{\alpha\gamma} (k,\omega)/\partial \delta (\mu_ H^z)|_{\omega=0} = \sigma^z_{\alpha\gamma} \partial G^{-1} (k,\omega) /\partial (i\omega)|_{\omega=0}$ with a triple charge vertex $\Lambda^{s}_{\alpha\gamma} (\Omega,K) = \Lambda_s(\Omega,K) \sigma^z_{\alpha\gamma}$, where  now $\delta (\mu_H^z)=\delta (\mu_B H^z)$ is the Zeeman energy shift induced by a small infinitely slowly time-dependent and spatially homogeneous variation of a magnetic field $H^z$,
 which for definiteness we direct along the $z-$axis. Then the same partial derivative $\partial G^{-1} (,\omega)/\partial (i\omega)$ is~\cite{kondratenko}
\begin{equation}
\frac{\partial G^{-1} (k,\omega)}{\partial (i\omega)}|_{\omega=0} = \Lambda_s (\Omega,K)|_{\Omega \to 0}
\label{ex_3}
\end{equation}
The triple vertex $\Lambda_s(\Omega,K)$ is expressed in terms of the fully renormalized
 spin component of $\Gamma^\omega$
 \begin{eqnarray}
 \Lambda_s (\Omega,K) \sigma^z_{\alpha\gamma}  &=&
 \sigma^z_{\alpha\gamma} + \sum_{\beta \delta}\int \sigma^z_{\beta \delta}\Gamma _{\alpha \beta ,\gamma \delta}^{\omega}(K,P) \left\{ G^{2}(P)\right\} _{\omega }\frac{d^{3}P}{(2\pi )^{3}} \nonumber \\
 && =  \sigma^z_{\alpha\gamma} \left(1 + 2 \int \Gamma_s (K,P)\left\{ G^{2}(P)\right\} _{\omega }\frac{d^{3}P}{(2\pi )^{3}}\right)
\label{ex_4}
\end{eqnarray}
For $K =K_F$,  this reduces to
\begin{equation}
Z_k =1 + 2 \int \Gamma_s (K_F,P)\left\{ G^{2}(P)\right\} _{\omega }\frac{d^{3}P}{(2\pi )^{3}}
\label{ch_2_1_1}
\end{equation}
The fact that the left hand side of (\ref{ch_2_1}) and (\ref{ch_2_1_1}) are identical implies
 that the spin and charge components of the fully renormalized $\Gamma^\omega_{\alpha\beta,\gamma\delta} = \Gamma_c \delta_{\alpha\gamma}\delta_{\beta\delta} +
 \Gamma_s {\bf \sigma}_{\alpha\gamma}{\bf \sigma}_{\beta\delta}$  are related by
\begin{equation}
\int \Gamma_c (K_{F},P)\left\{ G^{2}(P)\right\} _{\omega }\frac{d^{3}P}{(2\pi )^{3}} =
\int \Gamma_s (K_{F},P)\left\{ G^{2}(P)\right\} _{\omega }\frac{d^{3}P}{(2\pi )^{3}}
\label{ch_2_1_n}
\end{equation}
To the best of our knowledge, this relation has not been explicitly presented in the literature,
 although Eqs. (\ref{ch_2_1}) and (\ref{ch_2_1_1})
 have been presented in Refs. [\onlinecite{agd,pitaevskii}] and [\onlinecite{kondratenko}], respectively

In a generic FL the integration over $P$
 in (\ref{ch_2_1_n}) is not
confined to the
  FS, and this is the reason why the
 fermionic $Z$ factor is considered as an input
  for Landau FL theory rather than an integral part of
  it
   (This is presumably also the reason why the relation (\ref{ch_2_1_n}) has not been discussed in the past.)
   Like we said in the
Introduction, we will see that typical $P-P_F$
in Eq. (\ref{ch_2_1}) get
progressively smaller as the system approaches a QCP, and
 in a CFL regime
 near a
QCP the integrals for $Z_{k}$ in
(\ref{ch_2_1}) and (\ref{ch_2_1_1}) are predominantly determined by
  $P \approx P_F$, for which ${G^2(P)}_\omega$ can be approximated by the quasiparticle part ${G^2_{qp} (P)}_\omega$.
     In this limit, $Z_k$
      becomes an integral part of FL theory, and
        Eqn. (\ref{ch_2_1_n}) establishes the fundamental relation between charge and spin components of the vertex function $\Gamma^\omega$ between the particles on the FS.

The quasiparticle residue and the vertex function $\Gamma^\omega$ can both be computed
 in direct perturbation theory, and we will perform such calculations near a SDW QCP.
  Equations (\ref{ch_2_1}), (\ref{ch_2_1_1}),  and (\ref{ch_2_1_n})  must be satisfied at any order of perturbation theory
   and can be viewed as  "consistency checks" for  perturbative calculations.

\section{Quasiparticle vertex function in an ordinary Fermi liquid}
\label{conv}

In this Section we present the derivation of the quasiparticle vertex
function $\Gamma _{\alpha,\beta,\gamma\delta}^{\omega}(K_{F},P_F)$ using the
rules applicable to an ordinary FL. This "conventional" vertex function will
serve as a bare quasiparticle function in our subsequent analysis  of the vertex function in a CFL.

\begin{figure}[htbp]
\includegraphics[width=\columnwidth]{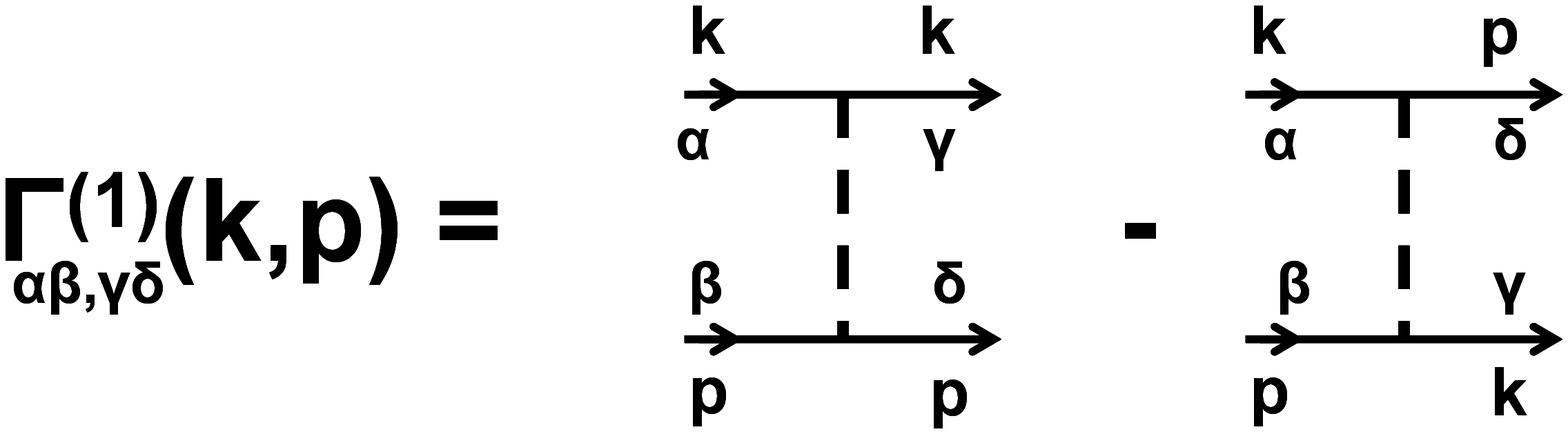}
\caption{The vertex function $\Gamma^\protect\omega$ to first order in the
interaction.
The dashed line is the interaction potential $U(q)$ ($q=0$ in the first term and $q = |{\bf k} -{\bf p}|$ in the second.}
\label{fig3a}
\end{figure}

Consider a system of fermions with some short-range (screened Coulomb)
interaction $U(q)$. To first order in $U(q)$, the vertex function is simply
the anti-symmetrized interaction (see Fig. \ref{fig3a}).
\begin{equation}
\Gamma _{\alpha \beta,\gamma \delta }^{\omega }(K,P)=U(0)\delta _{\alpha
\gamma }\delta _{\beta \delta }-U(|\mathbf{k}-\mathbf{p}|)\delta _{\alpha
\delta }\delta _{\beta \gamma }  \label{fr_2}
\end{equation}%
We use the same overall sign convention as in Ref.\cite{cm_nematic}.

The first-order result can be re-expressed in terms of spin and charge
components as
\begin{equation}
\Gamma^\omega_{\alpha\beta,\gamma\delta} (K,P) = \left(U(0) - \frac{U(|\mathbf{k}-\mathbf{p}|)}{2}\right)
\delta_{\alpha\gamma} \delta_{\beta\delta}  - \frac{U(|\mathbf{k}-\mathbf{p}|)}{2} {\vec \sigma}_{\alpha\gamma} {%
\vec \sigma}_{\beta\delta}  \label{fr_3}
\end{equation}

\begin{figure}[htbp]
\includegraphics[width=\columnwidth]{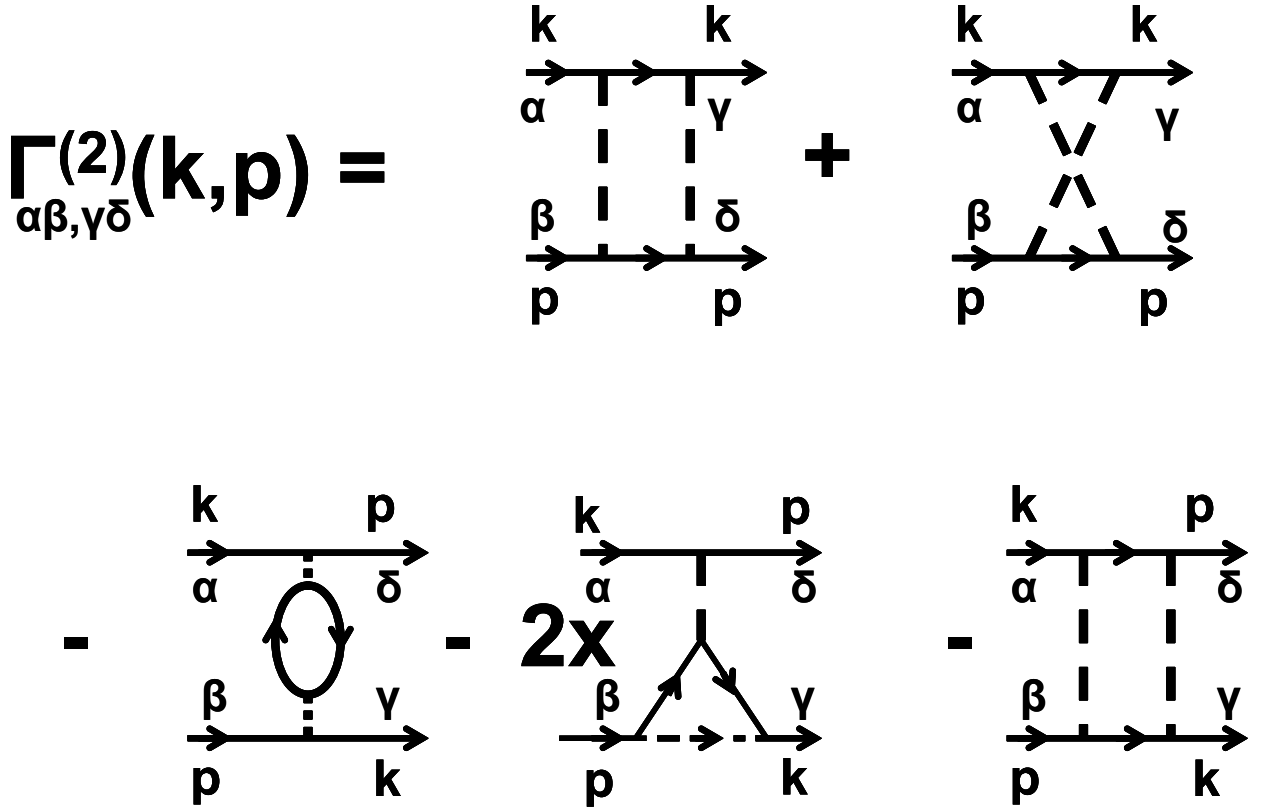}
\caption{The diagrams that contribute to $\Gamma^\protect\omega$ to second
order in the interaction}
\label{fig3}
\end{figure}

\begin{figure}[htbp]
\includegraphics[width=0.7\columnwidth]{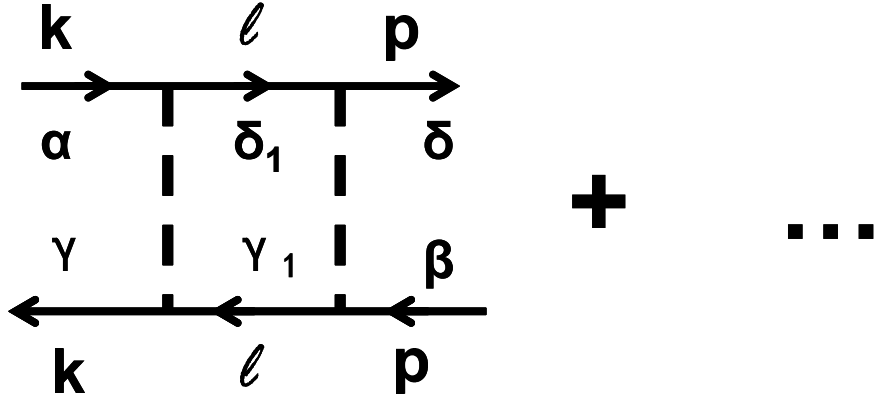}
\caption{An example of a "forbidden" diagram for $\Gamma^\protect\omega$.
This diagram has an internal particle-hole bubble with zero momentum transfer, and vanishes
for a static interaction (see the text).}
\label{fig6}
\end{figure}

The
 generic rule how to compute $\Gamma _{\alpha \beta,\gamma \delta }^{\omega }(K,P)$ beyond first order is that one has to sum
up all diagrams except the ones which contain a particle-hole bubble with
zero momentum transfer and vanishingly small but finite frequency transfer.
The diagrams that contribute to $\Gamma^\protect\omega$ to second
order in the interaction are shown in Fig. \ref{fig3}, and an
  example of a "forbidden"  second-order diagram  is shown in Fig. \ref{fig6}.
 The argument why the "forbidden" diagrams should not be included is that the corresponding bubble contains
\begin{equation}
\int d\mathbf{\mathbf{l}}d\omega _{l}G_{qp}(\mathbf{l},\omega _{l})G_{qp}(\mathbf{l},\omega _{l}+\omega _{q}).
\label{fr_5}
\end{equation}%
The integral over Matsubara frequency
$\omega _{l}$
has to be done first because that integration extends over infinite
range. Integrating, we find that it vanishes because the poles in the two propagators are located in the
same half-plane of complex frequency.

\begin{figure}[htbp]
\includegraphics[width=\columnwidth]{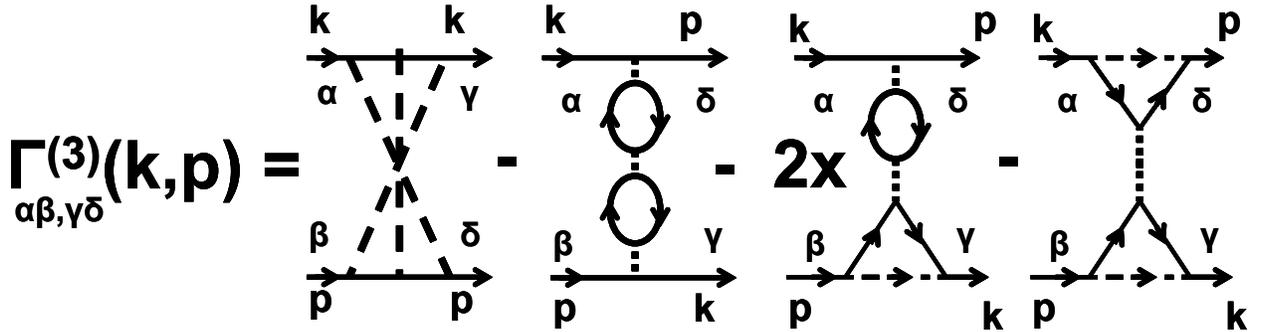}
\caption{
RPA-type
 diagrams for $\Gamma^\protect\omega$ to third order in the
interaction}
\label{fig4}
\end{figure}

\subsection{vertex function in RPA}
\label{sec_new_1}

Returning to terms of second order in $U$ (Fig. \ref{fig3}) we see that they  contain separately particle-hole and
particle-particle polarization bubbles. At next (third) order the two get
mixed, and there is no controllable way to proceed.
 A commonly used
approximation
 for a system near a SDW instability
 is to entirely neglect the renormalization in the
particle-particle channel, because by itself such a renormalization does not
lead to magnetic order, approximate the interaction by a constant $U$, and
sum  the
RPA-type series of diagrams which contain particle-hole polarization
bubbles
 $\Pi (|\mathbf{q}|)=\int d%
\mathbf{l}d\omega _{l}G_{qp}(\mathbf{l},\omega _{l})G_{qp}(\mathbf{l+q},\omega _{l})$
at $\mathbf{q}=\mathbf{k}_{F}-\mathbf{p}_{F}$. The
corresponding diagrams, shown in Fig. \ref{fig4}  to third order in $U$, form a
ladder series and can be summed explicitly. The result is the familiar RPA-type expression~\cite{scalapino}
\begin{widetext}
  \bea
 && \Gamma^{\omega, RPA}_{\alpha\beta,\gamma\delta} (K_F,P_F) = \frac{U}{1-U \Pi (\mathbf{k}_{F}-\mathbf{p}_{F})}\delta_{\alpha\gamma} \delta_{\beta\delta} -
  \frac{U}{1-U^2 \Pi^2 (\mathbf{k}_{F}-\mathbf{p}_{F})}\delta_{\alpha\delta} \delta_{\beta\gamma} \nonumber\\
 &&  = \frac{U}{2(1-U \Pi (\mathbf{k}_{F}-\mathbf{p}_{F}))}{\vec \sigma}_{\alpha\delta} {\vec \sigma}_{\beta\gamma} - \frac{U}{2(1+U \Pi (\mathbf{k}_{F}-\mathbf{p}_{F}))}
\delta_{\alpha\delta} \delta_{\beta\gamma}
\label{fr_6}
\eea
\end{widetext}
Using the last line, one may split
 $\Gamma _{\alpha\beta,\gamma\delta }^{\omega, RPA }(K_{F},P_{F})$ into terms containing $\sigma$ matrices and $\delta-$functions
\begin{eqnarray}
&&\left( \Gamma _{\alpha\beta,\gamma\delta }^{\omega }\right) _{spin}=%
\frac{U}{2(1-U\Pi (\mathbf{k}_{F}-\mathbf{p}_{F}))}{\vec{\sigma}}_{\alpha
\delta }{\vec{\sigma}}_{\beta \gamma }  \notag \\
&&\left( \Gamma _{\alpha\beta,\gamma\delta }^{\omega }\right)
_{charge}=-\frac{U}{2(1+U\Pi (\mathbf{k}_{F}-\mathbf{p}_{F}))}\delta
_{\alpha \delta }\delta _{\beta \gamma }  \notag
\end{eqnarray}%
Note, however, that this splitting is not the same as in Eq. (\ref{thr1})
because the combinations of spin indices in the $\delta \delta $ and ${\bf \sigma} {\bf \sigma} $
 terms
 differ  from those in (\ref{thr1}). We will return to the
same notations as in (\ref{thr1}) later in this section.

A Stoner-type magnetic instability occurs when $U>0$ and $U\Pi (\mathbf{k}-%
\mathbf{p}))=1$ for a particular $\mathbf{k}-\mathbf{p}$ (later assumed to
be $\mathbf{q}_\pi$), which (for constant $U$) is determined by the
structure of the fermionic dispersion. Once $\left( \Gamma _{\alpha\beta,\gamma\delta }^{\omega }\right) _{spin}$ gets enhanced,
 it is natural to
 neglect the charge component of the vertex and
approximate the full anti-symmetrized vertex by its spin component, i.e.,
set
\begin{equation}
\Gamma _{\alpha\beta,\gamma\delta }^{\omega, RPA }(K_{F},P_{F})=\frac{U}{%
2(1-U\Pi (\mathbf{k}_{F}-\mathbf{p}_{F}))}{\vec{\sigma}}_{\alpha\delta }{%
\vec{\sigma}}_{\beta ,\gamma }  \label{fr_8}
\end{equation}%
 This $\Gamma _{\alpha\beta,\gamma\delta }^{\omega, RPA }(K_{F},P_{F})$ can
be considered as an effective interaction between fermions at the FS,
mediated by collective magnetic excitations. To make this more transparent,
one can expand near $\mathbf{q}_\pi=(\pi ,\pi )$ and extend Eq. (\ref{fr_8}) to
fermions not necessarily on the FS (i.e., to non-zero frequencies $\omega
_{k}$ and $\omega _{p}$).
At a finite frequency, the particle-hole polarization operator contains a dynamical
 term which describes Landau damping of a collective boson by interaction with the particle-hole continuum.
  Keeping this term, we find
\begin{equation}
1-U\Pi (K-P)\propto \xi ^{-2}+|\mathbf{k}-\mathbf{p}-\mathbf{q}_\pi|^{2}+\gamma
|\omega _{k}-\omega _{p}|  \label{fr_10}
\end{equation}%
where $\gamma $ is the Landau damping coefficient and $\xi $ is the magnetic
correlation length, which diverges at the SDW transition. Substituting into (%
\ref{fr_8}), we obtain
\begin{equation}
\Gamma _{\alpha\beta,\gamma\delta }^{\omega, RPA }(K,P)=V(K-P){\vec{\sigma}}%
_{\alpha \delta  }{\vec{\sigma}}_{\beta ,\gamma }  \label{thr2}
\end{equation}%
where
\begin{equation}
V(K-P)=\frac{\overline{g}}{\xi ^{-2}+|\mathbf{k}-\mathbf{p}-\mathbf{q}_\pi%
|^{2}+\gamma |\omega _{k}-\omega _{p}|}  \label{fr_11}
\end{equation}%
can be viewed as the effective four-fermion
 dynamical
 interaction mediated by spin
fluctuations.   The effective coupling $\overline{g}$ in (\ref{fr_11}) is of
order $U/a^{2}$, where $a$ is the interatomic spacing. The damping coefficient $%
\gamma $ also scales with $\overline{g}$: $\gamma \sim \overline{g}%
/v_{F}^{2} $, because the damping comes from the $U\Pi $ term.

\begin{figure}[htbp]
\includegraphics[width=0.8\columnwidth]{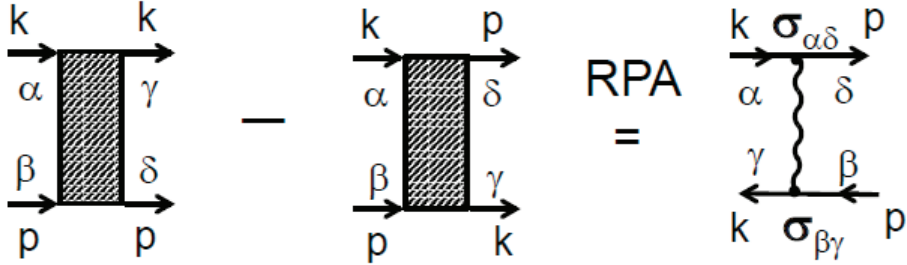}
\caption{
Graphical representation of the quasiparticle vertex function $\Gamma _{\alpha\beta,\gamma\delta }^{\omega, RPA }(K_{F},P_{F})$
 in the RPA scheme. Note that the combination of spin indices in the two $\sigma-$matrices is the same as in the "anti-symmetrized" component of the vertex. }
\label{fig1N}
\end{figure}

We present this $\Gamma _{\alpha\beta,\gamma\delta }^{\omega, RPA }(K,P)$ graphically in Fig. \ref{fig1N}.
Note that the combination of spin indices in the two $\sigma-$matrices  in (\ref{thr2}) is the same as in the
 "anti-symmetrized" component of the vertex.

Re-expressing ${\vec{\sigma}}_{\alpha \delta  }{\vec{\sigma}}_{\beta \gamma
}$ via $\delta -$ and $\sigma -$ matrices involving combinations $(\alpha
\gamma)$ and $(\beta \delta)$, as in (\ref{thr1}), we find
\begin{equation}
{\vec{\sigma}}_{\alpha \delta  }{\vec{\sigma}}_{\beta \gamma  }=-{\delta}_{\alpha \delta  }{\delta }_{\beta \gamma  }+2{\delta }_{\alpha \gamma }{%
\delta }_{\beta \delta }  = \frac{3}{2}{\delta }_{\alpha \gamma }{\delta }_{\beta \delta }-\frac{1}{%
2}{\vec{\sigma}}_{\alpha \gamma }{\vec{\sigma}}_{\beta \delta }
\label{thr3}
\end{equation}%
and hence
\begin{equation}
\Gamma^{\omega,RPA}_{\alpha\beta,\gamma\delta }(K,P)= V(K-P) \left( \frac{3}{2}{\delta }_{\alpha \gamma }{\delta }_{\beta \delta }-\frac{1}{2} {\vec{%
\sigma}}_{\alpha \gamma }{\vec{\sigma}}_{\beta \delta }\right)
\label{thr4}
\end{equation}
Placing $K$ and $P$ on the FS, we obtain ($K=K_{F}=(\mathbf{k}_{F},0)$, $P=P_{F}=(\mathbf{p}_{F},0)$)
\begin{equation}
\Gamma _{\alpha\beta,\gamma\delta }^{\omega,RPA }(K_{F},P_{F})=%
\frac{\overline{g}}{\xi
^{-2}+|\mathbf{k}_{F}-\mathbf{p}_{F}-\mathbf{q}_\pi|^{2}} \left( \frac{3}{2}{\delta }_{\alpha \gamma }{\delta }_{\beta \delta }-\frac{1}{2} {\vec{%
\sigma}}_{\alpha \gamma }{\vec{\sigma}}_{\beta \delta }\right)    \label{fr_12}
\end{equation}%

\subsection{The role of AL diagrams}
\label{sec_new_2}

 At first glance, Eq. (\ref{fr_12}) is a natural choice for the quasiparticle vertex function in a situation when scattering by spin fluctuations is much more relevant than scattering by charge fluctuations. Upon closer expection, however, we see that this interaction does not satisfy the condition set by
  Ward identities.  Indeed, according to (\ref{thr4}),  the spin and charge components of $\Gamma^\omega$ have the same dependence on $K-P$
   through $V(K-P)$, but the overall factors differ in sign and magnitude.  The Ward identity (\ref{ch_2_1_n}), on the other hand, requires that the two must have the same prefactors, if they both scale as $V(K-P)$.
  Clearly then, the expression for $\Gamma^\omega$ in (\ref{fr_12}) is incomplete and one has to include further contributions for $\Gamma^\omega$ which
    do not fit into the RPA scheme.

\begin{figure}[htbp]
\includegraphics[width=0.8\columnwidth]{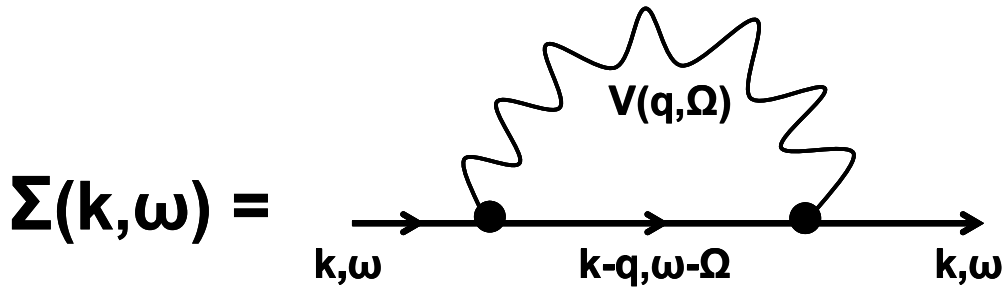}
\caption{
One loop diagram for the fermionic self-energy }
\label{fig6c_new}
\end{figure}

On physical grounds, it is natural to use the RPA-type spin-mediated interaction as a building block for constructing further contributions to
$\Gamma^\omega$, i.e.,  express all non-RPA contributions to $\Gamma^\omega$ in terms of RPA-renormalized, spin-mediated interaction $\Gamma^{\omega, RPA}$ rather than the bare interaction $U$.  In this nomenclature,  the RPA interaction is the "first-order" term (one wavy line) and all other terms contain more than one interaction line.   At first glance, including higher-order terms cannot resolve the issue posed by Ward identities, as higher-order terms contain higher powers of ${\overline g}$, while Ward identities
  are valid for any coupling and hence must hold independently at each order in ${\overline g}$.
However, we show  below that in some higher-order terms, extra powers of ${\overline g}$ come in the combination ${\overline g}/\gamma$, where $\gamma$ is the rate of Landau damping of collective excitations. The latter is by itself of order ${\overline g}$, i.e., the ratio ${\overline g}/\gamma$ is of order one. Because of this, certain higher-order terms are actually of the same order in ${\overline g}$ as $\Gamma^{\omega, RPA}$.

\begin{figure}[htbp]
\includegraphics[width=0.8\columnwidth]{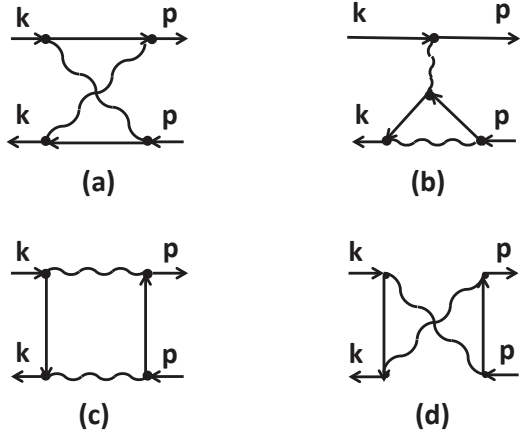}
\caption{The four  "two-loop" diagrams for the vertex function $\Gamma^\omega$ in an ordinary FL.
  The difference with Fig. \protect\ref{fig3} is that now the interaction line (the wavy line) is the RPA potential $V(K-P)$ rather than the bare interaction potential $U$.
 To avoid double counting, we didn't include terms which contain  $\Pi (K-P)$ as these terms are already included into the RPA renormalization from $U$ to $V$.
  Like before, we did not include the diagram which contains a particle-hole bubble with zero momentum and finite frequency transfer.}
\label{fig_hw_4}
\end{figure}

We follow the same strategy as before and do not include terms containing particle-hole bubbles with strictly zero momentum and vanishing frequency. We also do not include terms which contain particle-hole bubbles with transferred 4-momentum $K-P$, as  such terms are already incorporated into the RPA scheme.
The four remaining second-order diagrams are shown in Fig. \ref{fig_hw_4}. The first diagram [Fig.(10a)] is of order ${\overline g}^2$ without $\gamma$ in the denominator, and is not a competitor to the first-order in ${\overline g}$ term.  The second diagram [Fig.(10b)] contains $\Gamma^{\omega,RPA} (K-P)$ multiplied by the product of two Green's functions and one interaction line.  Such a term represents a correction to one of the two triple vertices in the effective interaction. It does contain additional ${\overline g}$ in the combination ${\overline g}/\gamma$, and, moreover, the internal integration yields $\log \xi$ (Refs.\cite{acs,ms}). At the same time, the summation over internal spin indices in this diagram shows that the spin structure of the interaction term given by diagram in Fig. (10b)
 is the same ${\vec{\sigma}}_{\alpha \delta  }{\vec{\sigma}}_{\beta \gamma }$ as in (\ref{thr2}), i.e., this term just renormalizes the overall coupling ${\overline g}$.  The presence of such a term is not unexpected because, as we said, there is no physical small parameter which would distinguish the RPA series.
 Still, the fact that the diagram in Fig. (10b) preserves the spin structure of $\Gamma^{\omega,RPA}$ implies that
  this particular
  vertex renormalization is irrelevant to the issue
   of Ward identities. In a more formal way, this vertex renormalization can be made small  by extending the theory to $N \gg 1$ fermionic flavors~\cite{acs,ms,senthil}.

The last two diagrams [Figs. (10c) and (10d)] serve our purpose in the sense that, on one hand, an additional ${\overline g}$ appears in the combination ${\overline g}/\gamma$ (see Eq.(\ref{esu_5}) below), and, on the other hand, the spin structure of the effective interaction given by any of these diagrams does not match that of
$\Gamma^{\omega, RPA}$.  Specifically, the diagram in Fig. (10c) yields a spin structure in the form
\beq
\sum_{s,t} \left({\vec \sigma}_{\alpha s} {\vec \sigma}_{t\delta}\right)  \left({\vec \sigma}_{s\gamma} {\vec \sigma}_{\beta t}\right) = 4 \delta_{\alpha\delta}\delta_{\beta\gamma} +  \delta_{\alpha\gamma}\delta_{\beta\delta} = 2 {\vec \sigma}_{\alpha\gamma} {\vec \sigma}_{\beta\delta} + 3
\delta_{\alpha\gamma}\delta_{\beta\delta}
\label{esu_1}
\eeq
and the diagram in Fig. (10d) yields a spin structure of the form
\beq
\sum_{s,t} \left({\vec \sigma}_{\alpha s} {\vec \sigma}_{\beta t}\right)  \left({\vec \sigma}_{s \gamma} {\vec \sigma}_{t \delta}\right) = 5 \delta_{\alpha\gamma}\delta_{\beta\delta} -4\delta_{\alpha\delta}\delta_{\beta\gamma}  = -2 {\vec \sigma}_{\alpha\gamma} {\vec \sigma}_{\beta\delta} + 3 \delta_{\alpha\gamma}\delta_{\beta\delta}
\label{esu_2}
\eeq
 The diagrams in Figs. (10c) and (10d)  have the same structure as the diagrams considered by Azlamazov and Larkin (AL) in the theory of the superconducting fluctuation contribution to the conductivity above $T_c$ in layered superconductors~\cite{al,varlamov}, and by analogy we call these two diagrams for $\Gamma^\omega$ the AL terms. As mentioned in the introduction, these AL terms are generated naturally within the conserving approximation scheme of Baym and Kadanoff~\cite{baym_1}, i.e. they are expected to help conserve particle number and spin, and this is indeed what we find.
The internal part of each of the two AL diagrams contains the combination of two interactions
$V^2(Q)$ and two Green functions, which for the diagram in Fig. (10c) are $G(K+Q) G(P+Q)$ and for diagram in Fig. (10d) are $G(K+Q) G(P-Q)$.  For external momenta  on the Fermi surface, $\omega_{p-q} = -\omega_q$. We assume and then verify that typical $q_{\perp}$ transverse to the FS are small and linearize the dispersion $\epsilon_{p-q}$ around ${\bf p}_F$ as $\epsilon_{p-q} \approx - v_{F,p} q_\perp$.
In this approximation, we have  $G(P-Q) = - G(P+Q)$, i.e., the internal parts of the two diagrams in Figs. (10c) and (10d) differ by a minus sign. Multiplying Eq. (\ref{esu_2}) by $(-1)$ and adding to (\ref{esu_1}) we find that the spin structure of the effective interaction from the diagrams (10c) and (10d) is
\beq
 4 {\vec \sigma}_{\alpha\gamma} {\vec \sigma}_{\beta\delta}
\label{esu_3}
\eeq
This is also a spin-spin interaction, but it contains a {\it different} combination of spin indices compared to that in the RPA interaction in Eq. (\ref{fr_2}).

\begin{figure}[htbp]
\includegraphics[width=0.8\columnwidth]{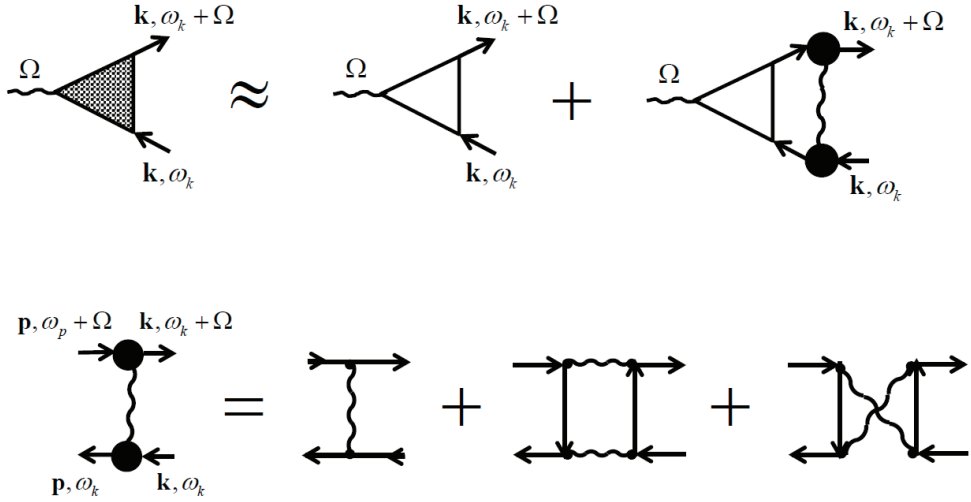}
\caption{a) The contributions from direct spin-fluctuation exchange and from the two AL diagrams to the vertex function in an ordinary FL.
 b) The full irreducible vertex function in an ordinary FL (see Eq. (\protect\ref{esu_9})).}
\label{fig_hw_1}
\end{figure}

Let us now compute the internal part of this diagram. Because we assumed that typical  $q_{\perp}$ are small (i.e., much smaller than the Fermi momentum), we
 approximate the full $G(K) = G(k, \omega_k)$ by its quasiparticle part. We first take ${\bf k}$ and ${\bf p}$ to be right at hot spots separated by $q_\pi = (\pi,\pi)$ (we recall that
  a hot spot $k_h$ is a $\mathbf{k}_{F}$ point on the FS in Fig. \ref{fig1} for which $\mathbf{k}_{F}+\mathbf{q}_\pi$ is
also located on the FS). For ${\bf k}$ and ${\bf p}$ at the hot spots, $V(K-P) = {\overline g} \xi^2$.
 Using~\cite{acs} $v^*_{F,{k+q}} = v^*_x q_x + v^*_y q_y$, $v^*_{F,{k+q +q_\pi}} = v^*_x q_x - v^*_y q_y$, and the fact that, by symmetry, the quasiparticle weight $Z_{k_h}$ must be the same at each of the hot spots, we express $I_c= -I_d = \int G(K+Q)G(P+Q) V^2 (Q) d^3 Q/(2\pi)^3$ with $Q = (q+ q_\pi,\omega_q)$ as
 \bea
 I_c &=& {\overline g}^2\int \frac{dq_x dq_y d \omega_q}{(2\pi)^3 Z^2_{k_h}}
 \frac{1}{i\omega_q - (v^*_x q_x + v^*_y q_y)} \frac{1}{i\omega_q - (v^*_x q_x - v^*_y q_y)} \nonumber \\
 &&\frac{1}{(q^2 + \xi^{-2} + \gamma |\omega_q|)^2}
\label{esu_4}
\eea
Because typical $q_x$ and $q_y$ in the two Green's functions are of order $\omega_q$ and typical $q$ in the bosonic proparator (the interaction $V(Q)$)
 are of order $\sqrt{\xi^{-2} + \gamma |\omega_q|}$, i.e., much larger,  one can integrate over momenta in the two fermionic propagators (this integral is ultra-violet convergent) and set $q=0$ in the bosonic propagator.  The 2D integration over $d q_x d q_y$ yields $- \pi^2/(Z^2_{k_h} v^*_x v^*_y)$ independent of $\omega_q$. Integrating over $\omega_q$ in the bosonic propagator we then obtain
 \beq
 I_c = - {\overline g} \xi^2 \frac{\overline g}{\gamma} \frac{1}{4\pi Z^2_{k_h} v^*_x v^*_y}
 \label{esu_5}
 \eeq
The bosonic damping rate has been calculated before~\cite{acs} and we  cite just the result:
\beq
\gamma = {\overline g} \frac{2}{\pi Z^2_{k_h} v^*_x v^*_y} = {\overline g} \frac{4}{\pi (Z_{k_h} v^*_F)^2 \sin{\theta}}
  \label{esu_6}
 \eeq
 where $\theta$ is the angle between the velocities at the hot spots ($\sin \theta = 2v^*_x v^*_y/(v^*_F)^2$).
 Substituting into (\ref{esu_5}) we obtain
  \beq
 I_c = - \frac{1}{8} {\overline g} \xi^2
 \label{esu_7}
 \eeq
 We see that $I_c$ has the same structure ${\overline g} \xi^2$ as the first-order term.
 To properly compare the overall factors, we now recall that we need to substitute the two terms into the right hand side of the diagrammatic expression for
  the triple vertex in Fig.\ref{fig7}, i.e., compare the diagrams shown in Fig.\ref{fig_hw_1}.
    For the interaction terms given by Figs. (10c) and (10d), the internal fermion loop yields an additional $-1$ and, besides, one needs to sum over all
     hot regions ${\bf k}_F$ which are separated from the external ${\bf p}_F$ by $(\pm \pi, \pm \pi)$.  The constraint is that the velocity at ${\bf k}$ should not be antiparallel to that at ${\bf p}_F$, otherwise the integral in (\ref{esu_4}) over one of momentum components of $q$ would vanish. A simple experimentation shows that there are four allowed hot regions of ${\bf k}_F$, i.e., the additional  factor for the second-order diagrams of Figs. (10c) and (10d) is $-4$.

     Incorporating this extra factor into $I_c$, we find that the nominally second-order contributions to $\Gamma^\omega$ from the AL diagrams [Figs. (10c) and (10d)] add an extra term to the vertex function for $K$ and $P$ at the hot spots of the form
     \beq
     \Gamma^{\omega, AL} = 2 {\overline g} \xi^2 {\vec \sigma}_{\alpha\gamma} {\vec \sigma}_{\beta\delta}
     \label{esu_8}
     \eeq
 Combining this with  $\Gamma^{\omega, RPA} = {\overline g} \xi^2 {\vec \sigma}_{\alpha\delta} {\vec \sigma}_{\beta\gamma}$, we obtain
 \beq
  \Gamma^{\omega, RPA+AL}_{\alpha \beta,\gamma\delta} = \Gamma^{\omega, RPA}_{\alpha \beta,\gamma\delta}  + \Gamma^{\omega, AL}_{\alpha \beta,\gamma\delta} = {\overline g} \xi^2 \left({\vec \sigma}_{\alpha\delta} {\vec \sigma}_{\beta\gamma} + 2{\vec \sigma}_{\alpha\gamma} {\vec \sigma}_{\beta\delta}\right) = \frac{3}{2} {\overline g} \xi^2 \left(\delta_{\alpha\gamma} \delta_{\beta\delta} + {\vec \sigma}_{\alpha\gamma} {\vec \sigma}_{\beta\delta}\right)
  \label{esu_9}
 \eeq
  We see that now spin and charge components are of equal sign and magnitude, a precondition for the Ward identities to be satisfied.

  The analysis of Figs. (10c) and (10d) can be extended to $K$ and $P$ away from hot spots. Performing the integration in the same way as before we
   find, to leading order in $\xi$,  $I_c = -(1/8) V(K-P)$ and $ \Gamma^{\omega, AL} = 2 V(K-P) {\vec \sigma}_{\alpha\gamma} {\vec \sigma}_{\beta\delta}$. The total
   $\Gamma^{\omega,RPA+AL} = \Gamma^{\omega,RPA} + \Gamma^{\omega,AL}$ is given by
  \beq
  \Gamma^{\omega,RPA+AL}_{\alpha \beta,\gamma\delta} (K,P) = V^{eff} (K-P)
  \label{esu_10}
 \eeq
 where
 \beq
 V^{eff} (K-P) =  V(K-P) \left({\vec \sigma}_{\alpha\delta} {\vec \sigma}_{\beta\gamma} + 2{\vec \sigma}_{\alpha\gamma} {\vec \sigma}_{\beta\delta}\right) = \frac{3}{2} V(K-P) \left(\delta_{\alpha\gamma} \delta_{\beta\delta} + {\vec \sigma}_{\alpha\gamma} {\vec \sigma}_{\beta\delta}\right)
  \label{esu_10_1}
 \eeq
 This vertex function has equal spin and charge components and obviously satisfies the relation (\ref{ch_2_1_n}) imposed by the Ward identities.  We emphasize, however, that the equivalence between the components $\Gamma_c (K,P)$ and $\Gamma_s (K,P)$ holds only as long as both have the same dependence on $K-P$ (given, in our case, by $V(K-P)$).  This
   is true
   only when the system is sufficiently close to a magnetic transition, such that the charge component of the RPA interaction
   can be neglected,
    together with non-RPA contributions.
    In a generic FL, $\Gamma_c (K,P)$ and $\Gamma_s (K,P)$ have different functional forms and do not have to be equal, although Eq. (\ref{ch_2_1_n}) must, indeed, hold.

\begin{figure}[htbp]
\includegraphics[width=0.8\columnwidth]{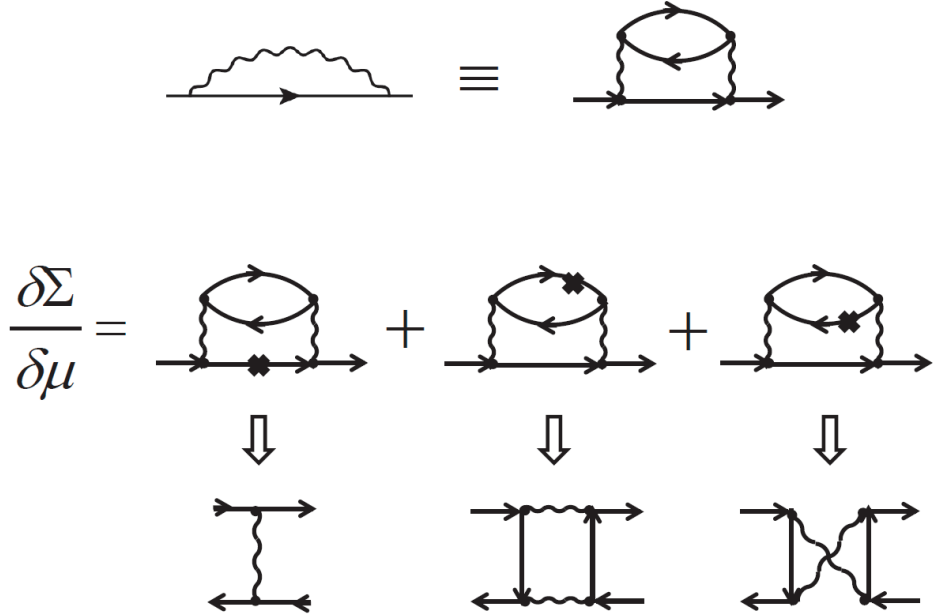}
\caption{The relation between the derivative of the self-energy with respect to a time-dependent correction to the chemical potential $\delta \mu (t)$
 and the direct and AL terms in the particle-hole irreducible vertex function.  The cross shows which Green's function is varied.  The AL terms are generated by varying the bosonic
  propagator $V(q)$ which contains particle-hole bubbles with transferred momentum $q \sim q_\pi$.}
\label{fig_hw_5}
\end{figure}

 The importance of AL terms for the proper description of a FL has been emphasized earlier in various contexts: in the analysis of the effect of  fluctuations near a superconducting~\cite{al,varlamov} or a ferromagnetic transition~\cite{cm_fm}, or also at the disorder driven metal-insulator transition in disordered systems~\cite{sasha}.  That these terms must generally be included into the irreducible vertex function within the conserving approximation scheme
  can also be seen by explicitly differentiating the diagrammatic expression for $G^{-1} ({\bf k}, \omega) = \omega -v_F (k-k_F) + \Sigma ({\bf k}, \omega)$.
   The argument that the spin-mediated interaction is a building block for the diagrammatic expansion
    implies that the fermionic self-energy
    can be expressed via
    $V(K-P)$.
  The one-loop self-energy diagram is shown in Fig. \ref{fig6c_new}.  A variation
     of this self-energy caused by an external perturbation (either a weakly time-dependent component of the chemical potential or a weakly time-dependent magnetic field, both homogeneous in space) generates two parts -- one comes from the variation of the internal fermionic line in Fig. \ref{fig6c_new} and the other is generated by the variation of a Green's function within the RPA interaction line~\cite{varlamov}.
      By varying the internal fermionic line, we reproduce Fig. \ref{fig7} with $\Gamma^\omega = \Gamma^{\omega, RPA}$. To see the effect of the variation of
      the interaction line, we recall that $\Gamma^{\omega,RPA}$  is obtained by summing up particle-hole bubbles; using Eq. (\ref{fr_8}) for $\Gamma^{\omega, RPA}$, and varying $\Pi (K-P)$ we see that contributions involving two fluctuation propagators are thereby generated.  We show this procedure graphically in Fig. \ref{fig_hw_5}. Varying each one of the two Green's functions making up $\Pi (K-P)$, we obtain two additional terms for $\Gamma^{\omega}$, which  are exactly the two AL diagrams.

 \subsubsection{The accuracy of the calculation of $\Gamma^\omega$ in an ordinary FL and the crossover to a critical FL}

The consideration above is essentially that of the conserving approximation scheme~\cite{baym_1,martin} in the sense that the guiding principle which forced us to look at additional terms for $\Gamma^\omega$ beyond $\Gamma^{\omega,RPA}$ is the set of charge and spin Ward identities associated with the conservation of the total number of particles and total spin.  In terms of a perturbative expansion, in the derivation of (\ref{esu_10}),
 we collected terms of order ${\overline g}$ and neglected the term of order ${\overline g}^2$ and also the term of order ${\overline g} \log \xi$, which had the same spin structure as the RPA interaction.

 Consider first the ${\overline g}^2$ term [Fig. (10a)]. A simple analysis shows that in 2D the extra power of ${\overline g}$ comes in the form of a dimensionless ratio  ${\overline g} \xi/v_F$.
 For small enough $\xi \sim a$, this ratio is of order ${\overline g}/E_F$, where $E_F \sim v_F/a$.

  For the rest of this paper we  assume that the spin-mediated coupling $\overline{g}$ is smaller than $E_F$
   This will allow us to separate the low-energy sector (energies
smaller than $\overline{g}$) from the high-energy sector (energies of order $%
E_{F}$) and also to keep using the linearized form of the quasiparticle
dispersion near the FS.  To rigorously  justify the assumption  that $\overline{g}$
is small compared to $E_{F}$ one has to consider a sufficiently long-ranged interaction with a range $a\gg 1/k_F$~\cite{cm_nematic,gorkov}, because
 the Stoner
condition holds at $U/a^{2}\sim 1/ma^{2}$, i.e. $\overline{g}\sim
U/a^{2}\sim E_{F}/(ak_F)^2$.
 We will not explicitly keep $ak_F$ large, but rather treat $\overline{g}$ as a phenomenological parameter, not directly
related to $U$.
 Previous works~\cite{acs,ms,cm_nematic} did not find a
qualitative difference in the system behavior in the regimes $\overline{g}\ll E_{F}$
and $\overline{g}\leq E_{F}$. Hence the assumption $\overline{g}\ll E_{F}$ seems to be
safe to make.

 For ${\overline g}\ll v_F/a$, the dimensionless parameter ${\overline g} \xi/v_F$
  can be kept small  even when $\xi$ is already larger than the inter-atomic spacing.
   The condition $\xi \gg a, \overline{g} \xi/v_{F} <1$ specifies what we call the ordinary FL regime near a SDW QCP.
   The expansion in ${\overline g}$, however, necessarily breaks down at larger $\xi$, when ${\overline g} \xi/v_F$ gets large.
  At $\overline{g}\ll E_{F}$, the boundary between the two regimes is at
$\xi_{cr} \sim a (E_{F}/\overline{g})\gg a$. That $\xi_{cr}\gg a$  means that the ordinary FL regime
  extends deep into the range of large $\xi/a$.
 Still, at sufficiently large $\xi > \xi_{cr}$, the
system crosses over into a new regime which we identify as the critical FL.

In the next section  we extend the conserving approximation to this CFL regime (defined more accurately in the next section)
 and show how $\Gamma^\omega$ gets modified.  We show that in the CFL regime the most relevant corrections to
 $\Gamma^\omega$ are not the diagram in Fig. (10a) (and higher-order terms of this kind), but rather the diagrams which we termed "forbidden", i.e., the ones
  which contain particle-hole bubbles with strictly zero transferred momentum and a finite transferred frequency.
   We show below that our initial argument that these diagrams are irrelevant for $\Gamma^\omega$ has to be reconsidered  once we use the dynamical interaction $\Gamma^{\omega,RPA} (K-P)$ as the building block for the  expansion in powers of ${\overline g}$.

 How to handle the terms which yield ${\overline g} \log \xi$ corrections to $\Gamma^{\omega}$ is a more tricky issue. These corrections are relevant already in an ordinary FL, once $\xi$ gets large.  Like we said, the ${\overline g} \log \xi$ term coming from the diagram with two wavy interaction lines
 acquires a small overall factor $1/N$  once we extend the model to $N \gg 1$ fermionic flavors.
  However, at higher orders of perturbation theory, there are additional logarithms associated with backscattering from  composite
 processes~\cite{ms,senthil,ss_lee}, which do not contain $1/N$.  To put it simply, there is no straightforward way at the moment
   to sum up these logarithmical corrections, even the ones which scale as $1/N$~\cite{ms}.  An approach advocated by one of us~\cite{peter}
is to adopt a plausible phenomenological form of the fully renormalized bosonic
propagator, compute the observables, and judge the validity
of the assumption by comparing the results with the experimental data. In
the rest of this paper we simply neglect all logarithmical corrections
and focus on how to find the $\Gamma^\omega$ satisfying
 the Ward identities in the CFL regime.  We expect that the effects associated with the additional $\log \xi$ terms can be incorporated on top of our analysis in which
 our focus will be how to reproduce the leading power-law dependence on $\xi$ in a situation when
 ${\overline g} \xi/v_F$ becomes a large parameter of the theory.

We will see in the next section that the ordinary FL regime near a SDW QCP  is a weak-coupling regime
 in which $Z_{k}\approx 1$ and $m^{\ast }\approx m$.  In this regime,
 Eqs. (\ref{ch_2_1}) and (\ref{ch_2_1_1}) determine the leading small correction to $Z_k \approx 1$, and
  the relation between $\Gamma ^{\omega }$, given by (\ref{esu_10})  and the Landau function $%
F_{\alpha\beta,\gamma\delta }(K_{F},P_{F})$ takes the simple form
\begin{equation}
F_{\alpha\beta,\gamma\delta }(K_{F},P_{F}) \approx 2N_{F}\Gamma _{\alpha\beta,\gamma\delta }^{\omega }(K_{F},P_{F})  \label{fr_14}
\end{equation}%
where $N_F$ is the density of states of free fermions.
In the CFL regime, however, $Z_{k}$ and $m^{\ast }/m$ become large
  and both, $\Gamma^\omega$ and the relation between $\Gamma^\omega$ and the Landau function, change.

In the next two Sections we extend the conserving approximation scheme to the critical FL regime.  We compute the quasiparticle
  self-energy and quasiparticle $Z_k$ in a direct diagrammatic expansion. In Sec. \ref{sec_4} we analyze the form of $\Gamma^\omega$ for the CFL, guided by the fact that it has to satisfy the Ward identities with $Z_k$ obtained diagrammatically in a certain approximation.

\section{Critical Fermi liquid}
\label{sec_3}

\subsection{The strength of the effective coupling and the role of Landau damping}

The quantity $\overline{g}/v_{F}\xi ^{-1}$ is the ratio of the interaction
energy $\overline{g}$ and the typical internal scale $v_{F}\xi ^{-1}$. It is
natural to expect that the magnitude of this ratio determines the strength
of the renormalization of the effective mass and therefore of $Z_{k}$, and the
calculations confirm this (see below). From this perspective, the ordinary
FL regime in the spin-fermion model at $\overline{g} \ll E_F$  is
 a  weak coupling regime,
where $Z_{k}\approx 1$ and $m^{\ast }\approx m$, while the CFL regime is
a strong-coupling regime, in which one can expect strong renormalizations of
both $Z_{k}$ and $m^{\ast }/m$.
To be more precise, we introduce the same dimensionless parameter as in earlier works~\cite{acs,ms,efetov,wang}
\begin{equation}
\lambda =3\overline{g}/(4\pi v_{F}\xi ^{-1})
\end{equation}%
The ordinary and the critical FL regimes correspond to $\lambda <1$ and $%
\lambda >1$, respectively.

The crossover at $\lambda \sim 1$ can also be identified by looking at the bosonic propagator and analyzing the role of the Landau damping.
 Indeed, at a given $k-k_{F}$, the
frequency of a free fermion is $v_{F}(k-k_{F})$, while a typical bosonic frequency for the same
$k-k_{F}$ is
is $(k-k_{F})^{2}/\gamma \sim (v_{F}(k-k_{F}))^{2}/\overline{g}$. Typical $%
k-k_{F}$ are of order $\xi ^{-1}$ simply because this is the only
long-wavelength scale with the dimension of momentum. Landau damping of collective bosons
is irrelevant as long as
these bosons are fast compared to fermions.
This holds
when $\overline{g}<<v_{F}\xi ^{-1}$ i.e., in the ordinary FL regime.
In the CFL regime $\overline{g} >>v_{F}\xi ^{-1}$,  collective bosons are slow modes compared to fermions, and
the Landau damping plays the central role.

\subsection{Loop expansion for the fermionic self-energy in the critical Fermi liquid}

 The quasiparticle residue $1/Z_{k}$ and the mass renormalization can be
obtained from the fermionic self-energy ,
$\Sigma (\mathbf{k},\omega )$ at the smallest $k-k_F$ and $\omega$.
 At these energies  $G_{inc}$ can be neglected, i.e., $G \approx G_{qp}$.
  Using
  the sign convention in which $\Sigma$ is counted as a positive frequency shift, we have
 $G^{-1}_{qp}(\mathbf{k},\omega )=i\omega
-v_{F,\mathbf{k}}(|\mathbf{k}|-k_{F})+\Sigma (\mathbf{k},\omega ) = Z_k (i\omega - v^*_{F} (|\mathbf{k}|-k_F))$.  The one-loop self-energy diagram with the effective interaction $V(K-P)$
 is shown in Fig.\ref{fig6c_new}.
 Given the analysis in the previous section, one may wonder whether one should also include AL diagrams. The answer is no, because one can easily
    make sure that adding AL terms gives extra contributions to the self-energy which contain $V^2(Q) \Pi(Q)$.  Such terms are already included into
     the RPA series for $V(Q)$ and keeping them will result in double counting.

The evaluation of this diagram has been presented
in some detail before~\cite{acs} and we will be brief.  In analytical form we have
\begin{widetext}
\beq
 \Sigma (\mathbf{k},\omega) = - \frac{3 {\overline g}}{8\pi^3} \int d^2 q d\Omega \frac{1}{i(\omega + \Omega) - v_{F,\mathbf{k}+\mathbf{q}+\mathbf{q}_\pi} (|{\bf k}+{\bf q}+ {\bf q}_\pi| - k_F)} \frac{1}{{\bf q}^2 + \xi^{-2} + \gamma |\Omega|}
\label{mo_1}
\eeq
\end{widetext}
For definiteness, let us focus momentarily on a fermion
located infinitesimally close to a hot spot. It is convenient to subtract from Eq. (\ref{mo_1})
a constant term $\Sigma (k_{F},0)$, whose effect -- the renormalization of
the chemical potential, we already incorporated by writing the bare
dispersion as $v_{F}(|\mathbf{k}|-k_{F})$. Approximating $v_{F,\mathbf{k}+\mathbf{q}+\mathbf{q}_\pi}(|\mathbf{k}+%
\mathbf{q}+\mathbf{q}_\pi|-k_{F})$ as $v_{F}(|\mathbf{k}_{F}+\mathbf{q}+\mathbf{q}_\pi|-k_{F})+v_{F,\mathbf{k}_F+\mathbf{q}_\pi} (|\mathbf{k}|-k_{F})$, we re-write Eq. (\ref{mo_1}) as
\begin{widetext}
 \bea
&&\Sigma (\mathbf{k},\omega) - \Sigma (k_F,0) = \frac{3 {\bar g}}{8\pi^3} \times \nonumber \\
 &&\int d^2 q d\Omega \frac{ i\omega - v_{F,\mathbf{k}_F+\mathbf{q}_\pi} (|\mathbf{k}|-k_{F})} {(i(\omega + \Omega) - v_{F,\mathbf{k}+\mathbf{q}+\mathbf{q}_\pi} (|{\bf k}+{\bf q}+ {\mathbf{q}_\pi}| - k_F))(i\Omega - v_{F,\mathbf{k}_F+\mathbf{q}+\mathbf{q}_\pi} (|{\bf k}_F+{\bf q}+ {\mathbf{q}_\pi}| - k_F))}  \times \nonumber \\
 &&\frac{1}{{\bf q}^2 + \xi^{-2} + \gamma |\Omega|}.
\label{mo_2}
\eea
\end{widetext}
The integral is ultra-violet convergent and can be evaluated
by integrating over $d^{2}q$ and $d\Omega $ in any order. Integrating over
momentum first, we get two contributions -- one comes from the poles in the
fermionic propagators and another comes from the pole in the bosonic propagator. The
first contribution is non-perturbative in the sense that it comes from
internal frequencies $|\Omega |\leq \omega $.  Evaluating this term, we find
 that the term $i\omega
-v_{F,\mathbf{k}_{F}+\mathbf{q}_\pi} (|\mathbf{k}|-k_{F})$ in the numerator of (\ref{mo_2}) is canceled out
by the equivalent term in the denominator, after integrating over the
component of $\mathbf{q}$ normal to the FS at $\mathbf{k}_{F}+\mathbf{q}_\pi$.
As a result, the contribution from the fermionic pole yields a self-energy contribution which only depends on frequency $\omega$.
 This term renormalizes both $Z_k$ and $m^{\ast }_k/m$.

  The second term is perturbative and comes from internal $\Omega \gg \omega$.
 This term is
proportional to $i\omega
-v_{F,\mathbf{k}_{F}+\mathbf{q}_\pi} (|\mathbf{k}|-k_{F})$, i.e., it renormalizes the
residue but not the effective mass. The sum of the two self-energy
contributions is
\begin{equation}
\Sigma (\mathbf{k},\omega )-\Sigma (\mathbf{k}_{F},0)=\lambda \left[ i\omega -\left( i\omega
-v_{F,\mathbf{k}_{F}+\mathbf{q}_\pi} (|\mathbf{k}|-k_{F})\right) f(\lambda )\right]   \label{mo_3_1}
\end{equation}%
where $f(0)=1$ and $f(\lambda \gg 1)\sim (\log \lambda )/\lambda \ll 1$.

We see that at small $\lambda $ (the case of the ordinary FL), the
self-energy predominantly depends on momentum: $\Sigma (\mathbf{k},\omega )\approx
\Sigma (\mathbf{k})$. This result is an expected one as in the ordinary FL regime the
interaction is essentially static (Landau damping is a small perturbation).
In the CFL, however, the first term  $\propto i\omega $ in (\ref{mo_3_1}) is the
largest, and $\Sigma (\mathbf{k},\omega )\approx \Sigma (\omega )$. From this
perspective, the CFL regime near a QCP is a regime of self-generated
locality~\cite{comm_1}. In this regime, $\Sigma (k,\omega )\approx i\lambda \omega $, and $%
Z \approx m^{\ast }/m \approx 1+\lambda $ .

Before we proceed, a comment is in order.  From the presentation above it may look as if
$\Sigma (\mathbf{k}, \omega) \approx i\lambda \omega $ would come from internal fermions located in an infinitesimally small
 range of order $\omega$ around the FS. Indeed, the pole contribution comes from this range.  However, a more careful calculation of the perturbative contribution without setting $\omega$ and $\epsilon_k$ to zero in the denominator of (\ref{mo_2})  shows that it can also be  represented as the sum of two terms -- one comes
  from the nearest vicinity of the FS and  another comes from a distance $\xi^{-2}/\gamma \sim \omega_{sf}$ from the FS.  The sum of these two terms is $f(\lambda)$ in
  (\ref{mo_3_1}), which is small at large $\lambda$. However, if we treat  the two perturbative contributions separately, we find that
    the one from  the nearest vicinity of the FS cancels the contribution
   from the fermionic poles,
     and, as a result, $\Sigma (\mathbf{k},\omega )\approx i\lambda \omega $ actually
      comes from fermions located at  distance of order $\omega_{sf}$  away from the FS.  This
     explains why the quasiparticle residue is not determined within Landau theory (which accounts for the effects coming from an infinitesimally small region near the FS) but rather is
     treated an an input quantity.
       That $Z$ comes from fermions at distance $\omega_{sf}$ from the FS can be also seen explicitly if we integrate in (\ref{mo_2}) first over frequency and then over  $\epsilon_p$.  We will not demonstrate this explicitly here but we will discuss in detail a very similar calculation of the vertex function in Sec. (\ref{sec_4_a_a}).

The quasiparticle residue in the CFL regime can be evaluated for momenta
away from the hot spots. To simplify notations, below we set $k$ in $Z_{k}$ to be the distance from a hot spot at $\mathbf{k}_{F,hs}$, along the FS, i.e., define
 a scalar variable $k = k_{\parallel} \equiv (\mathbf{k}_F-\mathbf{k}_{F,hs})|_{\parallel}$,
  which from
  now on
   will denote the component of the momentum vector along the FS, relative to the nearest hot spot. The component normal to the Fermi surface is denoted by $k_{\perp}$, and will turn out to be confined to small values. The result for $Z_{k}$, obtained in Refs.~\cite{acs} and \cite{ms}, may then be expressed as
\begin{equation}
Z_{k}=1+\frac{\lambda }{\sqrt{1+(k\xi \sin \theta )^{2}}}  \label{ch_5}
\end{equation}%
where, we remind that $\theta $ is the angle between the Fermi velocities at $\mathbf{k}_{F,hs}$
and $\mathbf{k}_{F,hs}+\mathbf{q}_\pi$ (see Fig. \ref{fig1}). For a FS like in the
cuprates, $\theta $ is not far from $\pi /2$, i.e., $\sin \theta \approx 1$.
Right at a hot spot ($k=0$), $Z=1+\lambda$. As $k$\ moves
away from a hot spot by a distance $k>\xi ^{-1}$, $Z_{k}$ begins to decrease
and becomes $O(1)$ at $|k|\geq
\overline{g}/v_{F}$.
The effective mass $m^*_{k}$ follows $Z_k$:  $m^*_k/m = Z_k$.

Below we will also need the residue
 $Z_k (\omega)$ at a nonzero frequency. It is given by
 \beq
Z_{k} (\omega)=1+\frac{2\lambda }{\sqrt{1+(k\xi \sin \theta )^{2} + \gamma \xi^2 |\omega|} + \sqrt{1+(k\xi \sin \theta )^{2}}}  \label{ch_5_1}
\end{equation}%
where the Landau damping coefficient is explicitly expressed via $\overline{g}$ as $\gamma = (4/\pi) \overline{g}/(v^2_F \sin \theta)$.
The frequency-independent form of $Z_k$, Eq. (\ref{ch_5}), is valid for $\gamma |\omega| << (k \sin \theta)^{2}$, or, for typical $k \sim \xi^{-1}$,
 for $\omega < \omega_{sf} \sim (v_F \xi^{-1})^{2}/\overline{g}$. When $\gamma |\omega| >> (k \sin \theta)^{2} >> \xi^{-2}$, $Z_{k} (\omega)\approx 1 + ({\bar \omega}/\omega)^{1/2}$, where ${\bar \omega} = 9 \overline{g} \sin{\theta}/(16\pi)$. The quasiparticle residue becomes $O(1)$ at $\omega \sim {\bar \omega} \sim \overline{g}$.
  A more complex form of the self-energy emerges if one keeps a regular $\Omega^2$ term in the bosonic propagator~\cite{sri_last}.

In the next Section we show that the Ward identities, Eqs. (\ref{ch_2_1}) and (\ref{ch_2_1_1}) are {\it not} reproduced  in the
CFL regime if we use $Z_k$, which we just obtained, and approximate $\Gamma^\omega (K,P)$ by $V^{eff} (K-P)$, as in
  Eq. (\ref{esu_10}).
 We
show that  $\Gamma ^{\omega } (K_F,P_F)$ in the CFL regime is different from Eq. (\ref{esu_10})
because our earlier reasoning to identify $\Gamma ^{\omega }$ and $V^{eff} (K-P)$
neglected Landau damping of collective excitations, which, as we now know,
plays a central role in the CFL regime. We obtain the correct $\Gamma
^{\omega }$ and show that Eqs. (\ref{ch_2_1}) and (\ref{ch_2_1_1}) with the correct $\Gamma
^{\omega }$ does reproduce Eq. (\ref{ch_5}), as they indeed should.

\subsection{The accuracy of the loop expansion for $\Sigma (k, \omega)$}

Before we do this, we briefly discuss the accuracy of the loop expansion in the
CFL regime. In the analysis above, we neglected the dependence of the
self-energy on $k-k_{F}$ as the latter does not contain $\lambda $
 as an overall factor.
  Still, $%
\partial \Sigma /\partial (k-k_{F})$ is not small and in fact scales as $%
\log \lambda $ (see Eq. (\ref{mo_3})). This is similar to what we found earlier in the analysis of the corrections to the vertex function.
The lowest-order logarithmical corrections can again be made small by extending the theory to large
  number of fermionic flavors $N$, however,
  as noted earlier, at higher-loop orders there appear  additional logarithms associated with backscattering from  composite
 processes~\cite{ms,senthil,ss_lee}, and these terms do not contain $1/N$.

We follow the same line of reasoning as we outlined in the previous section, neglect
 logarithmical corrections,
and focus on how to reproduce the Ward identity  in the CFL regime.
 We expect that the effects associated with the additional $\log
\lambda $ terms can be incorporated on top of our analysis.

\section{$\Gamma^\protect\omega$ in a critical Fermi liquid}
\label{sec_4}

\subsection{Failure of approach used for an ordinary FL}
\label{sec_4_a}

We recall that  the vertex function is given by
\beq
\Gamma^{\omega}_{\alpha\beta,\gamma\delta} (K,P) = \Gamma_c (K,P) \delta_{\alpha\gamma} \delta_{\beta\delta} + \Gamma_s (K,P)
 {\vec \sigma}_{\alpha\gamma} {\vec \sigma}_{\beta\delta}
  \label{esu_11}
\eeq
In an ordinary FL,
\beq
 \Gamma_c (K,P) = \Gamma_s (K,P)  = \frac{3}{2} V(K-P)
 = \frac{3}{2}\frac{\overline{g}}{\xi ^{-2}+|\mathbf{k}-\mathbf{p}-\mathbf{q}_\pi%
|^{2}+\gamma |\omega _{k}-\omega _{p}|}
 \label{thr6}
\end{equation}%
We assume and later verify that the equality $\Gamma_c (K,P) = \Gamma_s (K,P)$ holds also in the CFL regime and use only $\Gamma_c (K,P)$ in this Section.

The Ward identities between the true $\Gamma^{\omega }$ and $Z_{k}$, Eqs. (\ref{ch_2_1}), (\ref{ch_2_1_1}),
are re-expressed in terms of $\Gamma_c(K_{F},P)$ as
\beq
Z_{k} = 1+I,
\label{5_3_1}
\eeq
where
\begin{equation}
I=2\int \Gamma_c (K_{F},P)\left\{ G_{qp}^{2}(P)\right\} _{\omega }\frac{%
d^{2}\mathbf{p}d\omega _{p}}{(2\pi )^{3}}  \label{ch_2_1a}
\end{equation}%
and
\begin{equation}
\left\{ G_{qp}^{2}(P)\right\} _{\omega }=\lim_{\omega \rightarrow 0}G_{qp}\left(
\mathbf{p},\omega +\omega _{p}\right) G_{qp}\left( \mathbf{p},\omega _{p}\right)
\label{ch_2a}
\end{equation}%
We recall that this relation does not require Galilean invariance.

Let us assume momentarily that
 the relation $\Gamma_c (K_{F},P) = (3/2) V(K_{F}-P)$ extends into the CFL regime. We use the coherent part
of the fermionic propagator in the form $G_{qp}\left( \mathbf{p},\omega
_{p}\right) =1/(i\omega _{p}Z_{p}-v_{F}(p_{\perp}-k_{F}))$ and neglect the
incoherent part because, as we will see, typical $p$ in the integral in the
 right hand side of (\ref{ch_2_1a}) are close to $k_{F}$. We also assume that $\mathbf{p%
}$ is near a hot spot and approximate the Fermi velocity by its value at this
hot spot, which for brevity we label as just $v_{F}$. Substituting this
propagator into the right hand side of Eq.(\ref{ch_2_1a}) we find that the 3D
integral over $d^{2}p d\omega _{p}$ is convergent and can be
evaluated by doing momentum and frequency integration in any order.
Evaluating the integral over momentum first we again find that the integral
is the sum of two contributions, $I=I_{1}+I_{2}$. The first contribution ($%
I_{1}$) comes from the near-degenerate poles in the fermionic propagators,
the other one, $I_{2}$, comes from the pole in $V(K_{F}-P)$. Both terms can
be readily evaluated. The contribution from the fermionic poles comes from
the tiny range when the poles in the two Green's functions, viewed as
functions of complex $x=v_{F}(p_{\perp}-k_{F})$, are in different half-planes of $x$, i.e.  $-\omega<x<0$.
(we assume $\omega $ to be positive). Accordingly,
 $|p_{\perp}-k_{F}|$ for the pole scales with $\omega $ and is vanishingly
small.
 The interaction  $V(K-P)$ can then be safely approximated by the
static $V(K_{F}-P_{F})=\overline{g}/((\mathbf{k}_{F}-\mathbf{p}_{F})^{2}+\xi ^{-2})$, i.e., the Landau damping term formally
 plays no role.
Integrating over $p_{\perp}-k_{F}$ and $\omega _{p}$, we obtain
\begin{equation}
I_{1}=\frac{3\overline{g}}{4\pi ^{2}v_{F}}\int \frac{dp}{Z_{p}}\frac{1}{%
p^{2}-2pk\cos \theta +\xi ^{-2}+k^{2}}  \label{ch_n1}
\end{equation}%
where, we recall, $k$ and $p$ are momenta along the FS counted from the corresponding
hot spots $\mathbf{k}_{F,hs}$ and $\mathbf{k}_{F,hs}+\mathbf{q}_\pi$,
 and  $\theta $ is the angle between Fermi velocities at hot
spots at $\mathbf{k}_{F,hs}$ and $\mathbf{k}_{F,hs}+\mathbf{q}_\pi$.
  If we were to
neglect $Z_{p}$ in the denominator of (\ref{ch_n1}), the integration over $p$
would give exactly the same $Z_{k}$ as in (\ref{ch_5}). However, this is
only legitimate in the ordinary FL regime, when $\lambda $ is small and $%
Z_{p}\approx 1$. In the CFL regime $Z_{p}$ is large near a hot spot and
cannot be neglected. Then the result of the integration over $p$ in (\ref%
{ch_n1}) is much smaller than without $Z_{p}$. As an example, consider $k=0$
(i.e., take a fermion right at the hot spot).
 Typical $p$ in (\ref{ch_n1}) are between $\xi ^{-1}$ and $\lambda/\xi^{-1}$,
 when $Z_{p} \approx \lambda/(p \xi |\sin \theta|)$.
 Substituting this form into (\ref{ch_n1}) and integrating over $p$ we obtain
\beq
I_{1}=\frac{2 |\sin{\theta}|}{\pi} \log{\lambda }
\eeq

The second term ($I_2$) comes from the pole in the interaction, viewed as a function
of $\mathbf{p}$. In this term, the frequency dependence of the interaction
is relevant as typical $(p-k_{F})^2$ are of order $\xi ^{-2}$ and typical $%
\omega _{p}\sim \omega _{sf} \sim \xi^{-2}/\gamma$, such that
typical values of Landau damping term $\gamma \omega _{p}$ are also of order $\xi^{-2}$.
We then obtain, after proper rescaling
\begin{equation}
I_{2}=B\int \frac{dxdy}{\sqrt{x^{2}+y+1}}\frac{1}{(\sqrt{x^{2}+y+1}%
+y C_{x}(y))^{2}}  \label{mo_6}
\end{equation}%
where the variables are $x=p\xi $, $y=\omega _{p}/\omega _{sf}$, $B=O(1)$, $C_{x}(y)=Z_{\xi ^{-1}x}(y)\xi ^{-1}/(\gamma v_{F})\sim Z_{\xi
^{-1}x}(y)/\lambda $, and $Z_{\xi ^{-1}x}(y)$ is the frequency-dependent $Z$%
-factor
 given by (\ref{ch_5_1}).
 It reduces to $Z_{\xi ^{-1}x}$ in the CFL regime, which in these
notations corresponds to $y\leq 1$, and becomes $Z_{\xi ^{-1}x}/\sqrt{y}$ in
the quantum-critical regime at $y>1$. The upper limit of the integration
over $x$ is $\xi /a>>1$. For large $\lambda $ and $Z_{\xi ^{-1}}\leq \lambda
$, relevant $x$ and $y$ in the integral in (\ref{mo_6}) are large, i.e., the
integral predominantly comes from the quantum-critical region.
 Evaluating the integral we find that $I_{2}\sim \log \lambda $.

If we take these results at face value and substitute into Eq. (\ref{ch_2_1a}%
) for $Z$, we find
\begin{equation}
Z-1=
O(\log {\lambda }),  \label{mo_6_1}
\end{equation}%
 clearly inconsistent with the anticipated
 $Z-1=\lambda $.

To make the next step, we observe that the contributions $I_{1}$ and $I_{2}$
differ qualitatively. The contribution $I_{2}$ has the same form
as the prefactor of the $i\omega -v_{F,k_F+Q}(k-k_{F})$ term in the fermionic
self-energy in Eq. (\ref{mo_2}), and the logarithmical dependence on $%
\lambda $ parallels the logarithmical dependence of the renormalization of
the $k-k_{F}$ term. This analogy can be made even more precise if we extend
the model to a large number of fermionic flavors $N$ and assume that the
spin-fermion vertex conserves flavor. In this situation, the Landau damping
term gets enhanced by $N$, and $I_{2}$ and the prefactor of the $i\omega
-v_{F,k_F+Q}(k-k_{F})$ term in Eq. (\ref{mo_2}) both scale as $1/N$.
On the other hand,
$I_{1}$ does not contain $1/N$. Earlier
we argued that we neglect the logarithmical renormalization of the $k-k_{F}$
term in the fermionic propagator
(even without invoking the large $N$ limit) as the corrections to $Z$ from the
corresponding $\partial \Sigma /\partial (k-k_{F})$ come on top of the $%
O(\lambda )$ renormalization  from $i\omega \lambda $ term in the
self-energy. We apply the same strategy in the FL approach and neglect the $%
I_{2}$ term. We show below that for large $\lambda$ there is a series of $O(1)$ corrections
 to $\Gamma_c$ and, when the full $\Gamma_c(K_{F},P_{F})$
is used instead of $(3/2) V(K_{F}-P_{F})$, the $I_{1}$ term becomes exactly
$\lambda/\sqrt{1 + (k \xi)^2 \sin^2 \theta}$, i.e., $Z_k = 1 + I_1$
becomes consistent with the loop-expansion result, Eq. (\ref{ch_5}).

\subsubsection{The role of "forbidden" diagrams for the vertex}
\label{sec_4_a_a}

The reason why $\Gamma_c (K_{F},P_{F})$ in the CFL is different
from $(3/2) V(K_{F}-P_{F})$ becomes clear once we look back at our reasoning regarding
 which diagrams should be included into the calculation of $\Gamma^\omega$.
We argued earlier that the vertex function  is the sum of all diagrams except the ones
which contain fermionic bubbles with zero momentum transfer and vanishingly
small frequency transfer (see Fig. \ref{fig6}).
 The mathematical argument to neglect these last diagrams was that, if one integrates the bubble in the diagram in Fig. \ref{fig6} first over
frequency and then over momentum, the integral vanishes because both poles
in frequency space are in the same half-plane.

\begin{figure}[htbp]
\includegraphics[width=0.7\columnwidth]{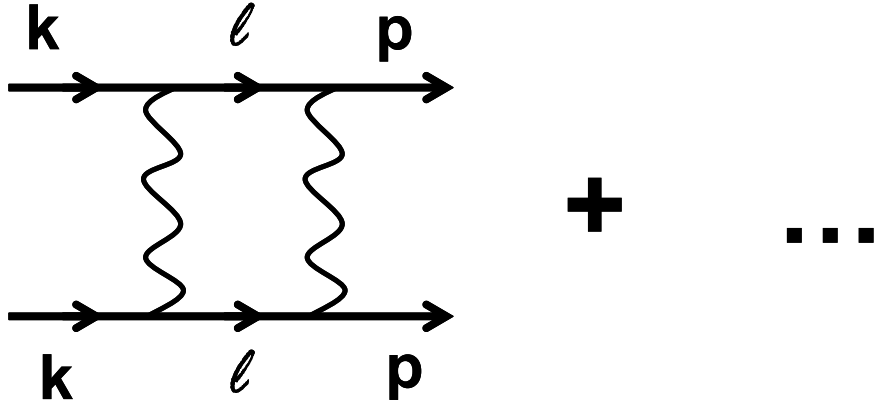}
\caption{An example of a "forbidden" diagram for $\Gamma^\protect\omega$ for diagrammatic series in which the interaction is
 the dynamical $V^{eff} (K-P)$.
This diagram still has an internal particle-hole bubble with zero momentum transfer, but, contrary to the corresponding diagram in Fig.\protect\ref{fig6},
  this one does not vanish because $V^{eff} (K-P)$ and $V^{eff} (L-P)$ have branch cuts in both half-planes of complex Matsubara frequency (see the text).}
\label{fig6_1}
\end{figure}

This is true, however, as long as the interaction can be treated as static.
Once the interaction acquires a Landau damping, the situation changes
because the product of the Green's functions and the interactions (the
combination that we actually have in the diagram in Fig. \ref{fig6_1}) contains, in addition to
poles, also branch cuts associated with the Landau damping, which in Matsubara formalism contains $|\omega| = \sqrt{\omega^2}$.
 Now, when we integrate over frequency, we have both pole and
 branch cut contributions. The contributions from the poles can still be avoided by closing the integration contour in the half-plane of frequency where there are no poles, however the branch cuts are located in both half-planes of frequency and cannot be avoided. As a result, the contributions from "forbidden" diagrams are
 generally nonzero.

 To estimate the strength of "forbidden" contributions, consider as an example the  diagram from Fig. \ref{fig6_1} with wavy lines instead of dashed lines.
 It gives, for external $K = K_F$ and $P = P_F$,
 \beq
\delta \Gamma_c (K_F,P_F)  = 2
\int\frac{d\epsilon_l dl d\omega_l}{(2\pi)^3 v_F} \frac{1}{i(\omega + \omega_l)Z_l - \epsilon_l} \frac{1}{i\omega_l Z_l - \epsilon_l} V(K_F-L)) V(L-P_F)
\label{s1}
\eeq
where, as before, $L = (\mathbf{l},\omega_l)$,  $k$ and $p$ are deviations from a hot spot at $\mathbf{k}_{F,hs}$ along the FS, $l$ is the deviation along the FS from $\mathbf{k}_{F,hs} + \mathbf{q}_\pi$,
$\epsilon_l$ is the linearized dispersion near $\mathbf{k}_{F,hs} + \mathbf{q}_\pi$, the overall factor $3$ comes from the summation over spin indices, and
the dynamical interactions $V(K_F-L)$ and $V(L-P_F)$ are given by Eq. (\ref{fr_11}).  We assume and then verify that the dominant contribution to $\delta \Gamma^\omega$ comes from the interaction between fermions taken right on the FS.  In this case,
\bea
 &&V(K_F-L) V(L-P_F) = V(\mathbf{k}-\mathbf{l}, \omega_l) V(\mathbf{l}-\mathbf{p}, \omega_l) =  \nonumber \\
 &&
 {\overline{g}}^2 \frac{1}{\xi^{-2} + l^2 + k^2 - 2 kl \cos \theta + \gamma|\omega_l|} \frac{1}{ \xi^{-2} + l^2 + p^2 - 2 pl \cos \theta + \gamma|\omega_l|}
\label{s2}
\eea

\begin{figure}[htbp]
\includegraphics[width=0.8\columnwidth]{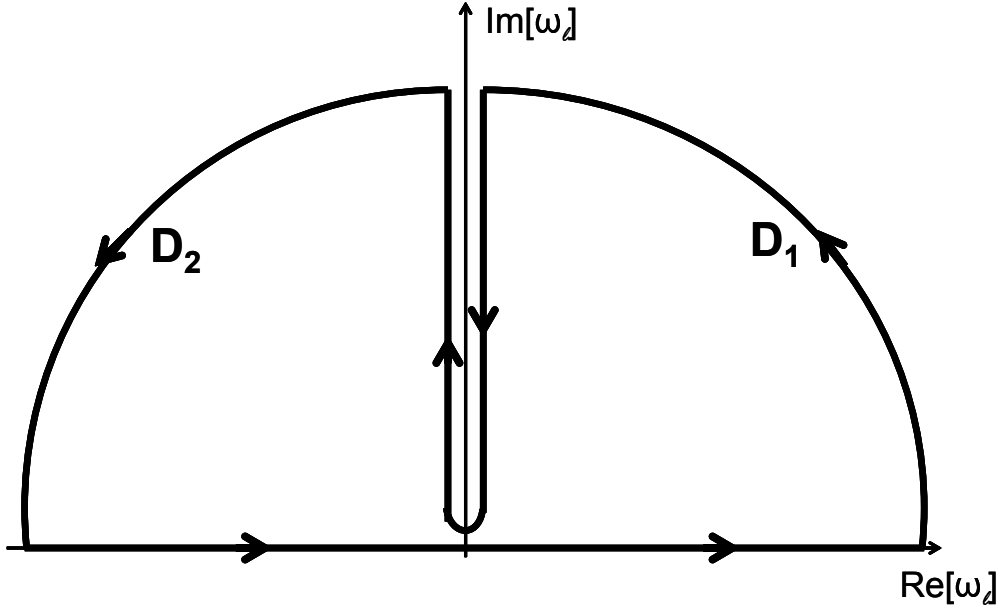}
\caption{The integration contour for the integration over frequency in Eq. (\ref{s4}).}
\label{fig2N}
\end{figure}

Like we did in the case of a static interaction, we integrate first over the fermionic frequency $\omega_l$.
That the  poles in the two Green's functions in (\ref{s1}) are splitted by $\omega$ does not make much difference because both poles are in the same
 half-plane of $\omega_l$. We can then safely set $\omega =0$ and re-write Eq. (\ref{s1}) as
 \bea
&&\delta \Gamma_c (K_F,P_F) = -2 {\overline{g}}^2
\int\frac{d\epsilon_l dl}{(2\pi)^3 v_F Z^2_l}  \int d \omega_l \frac{1}{(\omega_l +i \epsilon_l/Z_l)^2} \nonumber \\
&& \times
  \frac{1}{\gamma |\omega_l| +  \xi^{-2} + l^2 + p^2 - 2 p l \cos \theta} \frac{1}{\gamma |\omega_l| +  \xi^{-2} + l^2 + k^2 - 2 k l \cos \theta} \nonumber \\
 &&
\label{s3}
\eea
The location of the poles in (\ref{s3}) (whether they are in the upper or in the lower half-plane of complex $\omega_l$) is determined by the sign of $\epsilon$.
 For each sign, we choose the half-plane in which there are no poles. The branch cut, coming from the $\gamma |\omega_l|$ term in the interaction, is, however,
 present in each half-plane and cannot be avoided.  One can easily  make sure that contributions to the integral in (\ref{s3}) from positive and negative
  $\epsilon_l$ are the same, i.e., the full result is twice the contribution from $\epsilon_l >0$. Using this, we re-express (\ref{s3}) as
  \bea
&&\delta \Gamma_c (K_F,P_F) = -4 {\overline{g}}^2
\int_0^\infty \frac{d \epsilon_l}{2\pi v_F} \int \frac{dl}{(2\pi)^2 Z^2_l} \int d \omega_l
\int \frac{1}{(\omega_l +i \epsilon_l/Z_l)^2} \times  \nonumber \\
&&
\frac{1}{\gamma |\omega_l| + \xi^{-2} + l^2 + p^2 - 2 p l \cos \theta} \frac{1}{\gamma |\omega_l| +  \xi^{-2} + l^2 + k^2 - 2 k l \cos \theta}
\label{s4}
\eea
 and close the integration contour over $\omega_l$ in the upper half-plane (see Fig.\ref{fig2N}).
   The branch cut renders the $\gamma |\omega_l|$ term ill defined
  along the imaginary frequency axis, where $\omega_l = i \omega$  ($|\omega_l| = \sqrt{\omega^2_l} = \sqrt{(i \omega + \delta)^2} = i \omega {\text{sign}} (\delta)$ for $\omega >0$).
  Choosing the contour to avoid the imaginary axis and using the fact that the integral over the arcs $D_1$ and $D_2$ in Fig.\ref{fig2N}
   vanishes because the
   integrand in (\ref{s4}) vanishes faster than $1/\omega_l$, we obtain, after simple algebra
   \bea
&&\delta \Gamma_c (K_F,P_F) = 4i {\overline{g}}^2
\int\frac{dl}{(2\pi)^2 Z^2_l} \int_0^\infty \frac{d \epsilon_l}{2\pi v_F} \int_0^\infty  \frac{d\omega}{(\omega + \epsilon_l/Z_l)^2} \times \nonumber \\
&&
 (\frac{1}{ \xi^{-2} + l^2 + p^2 - 2 p l \cos \theta + i\gamma \omega} \frac{1}{\xi^{-2} + l^2 + k^2 - 2 k l \cos \theta + i\gamma \omega} - \nonumber\\
&&\frac{1}{ \xi^{-2} + l^2 + p^2 - 2 p l \cos \theta - i\gamma \omega} \frac{1}{\xi^{-2} + l^2 + k^2 - 2 k l \cos \theta - i\gamma \omega})
\label{s5}
\eea
Performing now elementary integrations over $\omega$ and over $\epsilon_l$ (in any order) we obtain
    \beq
\delta \Gamma_c (K_F,P_F)=  \frac{1}{2\pi^2 v_F}
\int\frac{dl}{Z_l} V_{||}(k,l) V_{||}(l,p)
\label{s6}
\eeq
where we defined the static interaction between fermions on the Fermi surface near two conjugated hot spots as
\beq
V_{||}(k,p) = V({\bf k}_F - {\bf p}_F) = {\overline {g}}/(\xi^{-2} + k^2 + p^2 - 2 k p \cos \theta).
\eeq
The notation $V_{||} (k,p)$ is introduced to emphasize that this is a function of momenta along the FS, and each momentum
 is a deviation from the corresponding hot spot.
With this definition,
\beq
V^{eff}(K_F-P_F) \equiv V_{||}(k,p) \left({\vec \sigma}_{\alpha\delta} {\vec \sigma}_{\beta\gamma} + 2{\vec \sigma}_{\alpha\gamma} {\vec \sigma}_{\beta\delta}\right)
\label{ll_1}
\eeq

It is essential for our further discussion, particularly for the analysis of spin and charge susceptibilities in Sec. \ref{sec_7_1},
 that, although the final expression for $\delta \Gamma_c (K_F,P_F)$, Eq. (\ref{s6}), contains the static interaction between the particles on the FS,
  the result does not come from the immediate vicinity of the FS in the sense that typical $\epsilon_k \sim \omega Z_l$ and typical $\gamma \omega \sim \text{max}({k^2,p^2, \xi^{-2}})$, which are at least of order $\xi^{-2}$.  When $\xi = O(1)$, the leading contribution to $\delta \Gamma_c (K_F,P_F)$ comes
   from fermions far from the FS. When $\xi$ is large, the leading contribution comes from states near the FS, but still at some distance from the FS.
    In Sec.\ref{sec_7_1} we will contrast these contributions with the contributions which come from an infinitesimally small region around the FS.

 This $\delta \Gamma_c (K_F,P_F)$ has to be compared with our earlier result $\Gamma_c (K_F,P_F) = (3/2) V(K_F-P_F) = (3/2) V_{||}(k,p)$.
 Taking for simplicity the external fermions to be right at a hot spot (i.e., in our notations setting $k=p=0$ and also setting the bare interaction to be $V_{||}(k,p) = V_{||}(0) = \overline{g}/\xi^{-2}$), we obtain
\begin{equation}
\delta \Gamma_c (K_F,P_F) =\frac{1}{2\pi^2 v_F}  \overline{g}^2\int \frac{dl}{Z_l}\frac{1}{(l^{2}+\xi ^{-2})^2} =
 V_{||}(0) \left(\frac{1}{3\pi}\right)\int_0^\infty dx  \frac{\sqrt{1+x^2 \sin^2\theta}}{(x^2+1)^2}.
   \label{mo8}
\end{equation}%
We see that $\delta \Gamma_c (K_F,P_F)$ is of the same order as
the bare term $V_{||}(0)$. The implication is that, in a CFL, previously forbidden contributions to $\Gamma ^{\omega }(K_{F},P_{F})$ are of the same order
 as $\Gamma_c$ taken from an  ordinary FL.  As we know, we need to increase $\Gamma_c
 (K_{F},P_{F})$ roughly by $\lambda $ to reproduce the result for $Z_{p}$,
Eq. (\ref{ch_5}) using the Ward identities.  We now show that this can be achieved
by summing up a  series of previously forbidden diagrams for $\Gamma ^{\omega }$,
 which contain powers of $\delta \Gamma_c (K_F,P_F)/V(K_F-P_F)$. To select this series, we analyze in the next subsection the
structure of the diagrammatic representation of the Ward identities in a CFL.

Before we proceed, we return momentarily to the evaluation of $\delta \Gamma_c (K_F,P_F)$ and discuss two issues.
 First, in evaluating the integral in  Eq. (\ref{s2}) we assumed that  the dominant contribution to $\delta \Gamma_c$ comes from the interaction between fermions taken right on the FS. From (\ref{s1}), we see that typical transverse momenta in $V(\mathbf{k}-\mathbf{p})$ are of order $(\epsilon/v_F) \sim \omega Z_l$, while typical longitudinal
 momenta along the FS are of order $(\gamma \omega)^{1/2}$.  If $Z_l$ was of order 1, then at large $\xi$ typical transverse momenta would then definitely be smaller than typical longitudinal momenta. In our case the situation is more involved because for $ \xi^{-1} <l< {\overline{g}}/v_F$, $Z_l$ by itself scales as $1/l$, $Z_l \sim {\overline{g}}/v_F l$. A simple analysis shows that then typical transverse and longitudinal momenta are both of order $(\omega {\overline {g}})^{1/2}/v_F$, i.e., the terms we neglected in the evaluation of  the integral in (\ref{s1}) are
  of the same order as the ones we included. However, if we again extend the model to $N >>1$  flavors, the terms that we neglected turn out to be small in $1/N$. In this respect, the approximation we made here by neglecting the transverse term in the interaction is the same as the one we made
  in the calculation of the self-energy, and, like we did there, we use large $N$ as a formal justification of the approximation to neglect the transverse momentum component in the bosonic propagator.

Second, as we already
discussed, the integral (\ref{s1}) involving two dynamical interactions and two
fermionic $G$'s is ultra-violet convergent and can be evaluated by integrating over $\omega_l$ and $\epsilon_l$ in any order
 and within \textit{infinite limits}.
 In the calculation above, we integrated over the frequency first and found that the non-zero result for $\delta \Gamma_c (K_F,P_F)$ comes from the branch cuts in the dynamical interaction.
Integrating over momentum first, we find that
there are two contributions to $\delta\Gamma_c$. One comes from a tiny range where
the poles in the two Green's functions are in different half-planes of the
(complex) momenta and the other comes from the poles in the interactions, viewed
as functions of momenta. For static interaction, the two contributions
cancel each other, just like the two $i\omega $ terms cancel each other in
Eq. (\ref{mo_2}) for the self-energy if one neglects the Landau damping.
However, when $\lambda $ is large and Landau damping cannot be neglected, we found the same result as in our earlier
 calculation of the self-energy, namely that the dominant contribution is the one from
the splitted poles in the Green's function, while the one from the poles in the interaction is smaller by a power of $\lambda$.
 For the contribution from the splitted poles, typical $\omega$ and $\epsilon_l$ are small and we again can treat the
interaction as static. Performing
the integration this way, we obtain
\beq
\delta \Gamma_c (K_F,P_F) = 2 \int \frac{dl}{2\pi} V_{||}(k,l) V_{||}(l,p) \Pi_l,
\eeq
where
\begin{eqnarray}
\Pi _{l} &=&\lim {}_{\omega \rightarrow 0}\frac{1}{v_{F}}\int \frac{%
d\omega _{l}d\epsilon _{l}}{4\pi ^{2}}\frac{1}{i(\omega +\omega
_{l})Z_{l}-\epsilon _{l}}\frac{1}{i\omega _{l}Z_{l}-\epsilon _{l}}  \notag \\
&=&\frac{1}{2\pi v_{F}Z_{l}}  \label{mo_7}
\end{eqnarray}
The final result is the same as Eq. (\ref{s6}), and the computation is easier this way than by integrating over $\omega$ first.
 We caution, however, that integrating over $\epsilon_l$ first and keeping only the contribution from splitted poles gives the false impression that
 $\delta \Gamma_c (K_F,P_F)$ comes from states in an infinitesimally small region around the FS (in the splitted poles contribution, $\omega_l$ and
  $\epsilon_l$ are both of order of external $\omega$, which is vanishingly small).  In reality, one has to include the contribution from the poles
   in the interaction and represent this contribution as the sum of two terms --one from the infinitesimal vicinity of the FS and another from the region a distance $\xi^{-2}$ from the FS. The two terms nearly cancel each other and this is what makes the contribution from the poles in the interaction small. However, taken separately,
    the contribution from the  infinitesimal small vicinity of the FS cancels the one from the splitted poles, and the one which remains comes from states a small but finite distance away from the FS, like we found by integrating over $\omega_l$ first.

\subsection{Ward identities and the equivalence of diagrammatic and FL
expressions for $Z_k$}
\label{sec_4_b}

\begin{figure}[htbp]
\includegraphics[width=0.7\columnwidth]{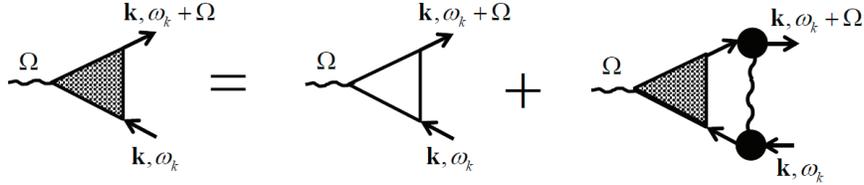}
\caption{Diagrammatic expression of the ladder series for the triple vertex $\Lambda_c (\Omega, K)$.
The shaded triangle represents the full vertex, and the unshaded triangle is the bare vertex which for the same spin projections of incoming and outgoing fermions is just equal to 1.  The interaction (the wave line with big cite circles) is $V^{eff} (K-P)$, which is the sum of the direct RPA spin-mediated interaction and  the two AL terms.}
\label{fig7_1}
\end{figure}

We recall that, in  diagrammatic language, the Ward identity associated with the particle number conservation relates the self-energy $\Sigma
(k,\omega _{k})=i\omega _{k}(Z_{k}-1)$ to a triple charge vertex $\Lambda_c (\Omega,K) \delta_{\alpha\beta}$, where $\Omega $ is set to be infinitesimally small. The relation simply states that
\beq
\Lambda_c (\Omega ,K) = Z_k
\eeq
The issue is which diagrams one should keep  in order to reproduce this relation, given that the
 fermionic $Z_k$ comes from the one-loop
diagram of  Fig. \ref{fig6c_new}. We now demonstrate that, to the same accuracy, the vertex $\Lambda_c (\Omega,K)$ is given by
the ladder series of diagrams shown in Fig. \ref{fig7_1}, with the bare $\Lambda
_{c,0}(\Omega,K)=1$ (for simplicity we set the spin projections of incoiming and outgoing fermions to be equal).
 To see this, we observe that the ladder series gives rise to the integral equation for $\Lambda_c(\Omega,K)$ as follows:
\begin{equation}
\Lambda_c (\Omega,K)=1+
3\int \frac{d^{2}pd\omega _{p}}{(2\pi )^{3}}\Lambda_c (\Omega,P)\left\{ G_{qp}^{2}(P)\right\} _{\Omega } V(P-K)
\label{s8}
\end{equation}%
We assume and then verify that  the dependence on $\Omega$ and $\omega_k$ in $\Lambda_c (\Omega, K)
$  is non-singular and can be safely neglected, i.e., one can  approximate $\Lambda_c (\Omega, K) = \Lambda_c (\Omega, \omega_k, k)$
 by  $\Lambda_c (0,0,k) =\Lambda_c (k)$ and set the external $\omega_k$ to zero in the diagrammatic series in Fig. \ref{fig7}.

\begin{figure}[htbp]
\includegraphics[width=\columnwidth]{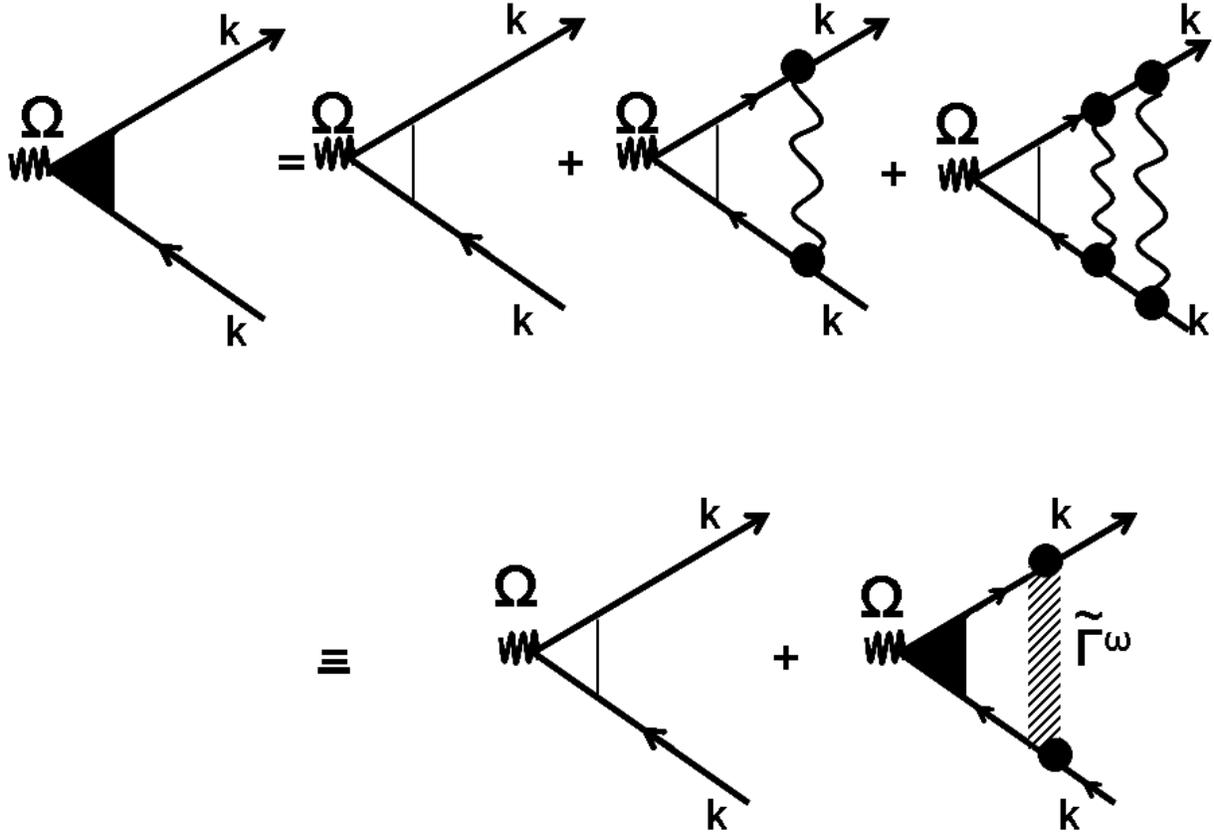}
\caption{The diagrammatic derivation of the relation between the triple
vertex and ${\tilde \Gamma}^\protect\omega$, which we show to be equal to the actual vertex function ${\Gamma}^\protect\omega$. White and black triangles denote bare and full triple vertices. Wavy lines with big circles at the end points are $V^{eff} (K-P)$, and the shaded rectangular vertex with big circles at the end points is the trial vertex function ${\tilde \Gamma}^\omega$. We show in the text that ${\tilde \Gamma}^\omega$ coincides with the actual vertex function $\Gamma^\omega$.}
\label{fig8}
\end{figure}

The integral on the right hand side of (\ref{s8}) is ultra-violet convergent and can be evaluated by integrating over frequency $\omega_p$ and over quasiparticle dispersion
$\epsilon_p$ in any order.
Integrating over frequency first (which, incidentally, is always the safest way to proceed as in any system the frequency integration at $T=0$  extends over infinite limits), we find the same result as in the earlier calculation of the "forbidden" diagram for the vertex function. Namely, the integral contains poles
 coming from the two Green's functions (the $\left\{ G_{qp}^{2}\right\} _{\Omega }$ term) and the branch cut coming from the $|\omega_p|$ term in the interaction.  The poles are in the same half-plane of frequency and can be avoided by closing the integration contour in the half-plane where there are no poles,
  but the branch cuts are in both frequency half-planes and cannot be avoided.  Choosing the integration contour as in Fig.\ref{fig8} and performing the same calculation
   as for the vertex, we obtain
   \beq
   \int \frac{d p_{\perp} d \omega_p}{(2\pi)^3} \left\{ G_{qp}^{2}(P)\right\} _{\Omega } V(P-K+Q) = \frac{{\overline{g}}}{4\pi^2 v_F} \frac{1}{Z_p} \frac{1}{k^{2}+p^{2}-2kp\cos {\theta }+\xi ^{-2}}
 \label{s9}
 \eeq
 The integral comes from internal $\omega_p \sim \epsilon_p/Z_p \sim (k^{2}+p^{2}-2kp\cos {\theta }+\xi ^{-2})/\gamma$, which are small, at least in $\lambda/E_F$.
  This justifies our approximation to neglect the dependence on $\omega_k$ in the triple vertex. The dependence on
   external $\Omega$ appears in $\Omega \gamma/(k^{2}+p^{2}-2kp\cos {\theta }+\xi ^{-2})$ and can also be neglected if $\Omega$ is small enough.
     At the same time, we clearly see that the integral comes from
   fermions located at some finite distance away from the FS rather than from the immediate vicinity of the FS. This is yet another indication that the physics
    associated with the renormalization of the quasiparticle residue is {\it not} confined to the FS. If one would integrate only over an infinitesimally small range around the FS, one would obtain $Z_k =\Lambda_c(k) =1$.   Furthermore, in an ordinary FL the integral in (\ref{s9}) comes from fermions which are generally far away from the FS. Only in a CFL does the integral comes from near the FS: from fermions with $\omega_p \sim \xi^{-2}/\gamma \sim \omega_{sf}$ for external $k$ in a hot region, and from $\omega_p k^{2}/\gamma \leq {\overline{g}} << E_F$ for external $k$ in a lukewarm region $\xi^{-1} < k< {\overline g}/v_F$.

The same result, Eq. (\ref{s9}) can be also obtained by integrating over $\epsilon_p$ first. This way, the computation is easier to carry out as one only has to include the contribution from the poles splitted by $\Omega$, but it requires more efforts to make sure that the integral comes from fermions at some distance away from the FS.

Substituting (\ref{s9}) into (\ref{s8}) and using the definition of $\lambda$ we obtain
\begin{equation}
\Lambda_c (k)=1+\frac{\lambda }{\pi \xi } \int dp \frac{\Lambda_c (p)}{Z_{p}} \frac{%
1}{k^{2}+p^{2}-2kp\cos {\theta }+\xi ^{-2}}  \label{mo_10}
\end{equation}%
We now recall that the one-loop formula for $Z_{k}$ (Eq. (\ref{ch_5})) is
\begin{eqnarray}
Z_{k} &=&1+\frac{\lambda }{\pi \xi }\int dp\frac{1}{k^{2}+p^{2}-2kp\cos {%
\theta }+\xi ^{-2}}  \notag \\
&=&1+\frac{\lambda }{\sqrt{1+(k\xi \sin \theta )^{2}}}  \label{mo_11}
\end{eqnarray}%
One can easily see that the two equations are identical when $Z_{k} = \Lambda_c (k) [= \Lambda_c (\Omega, K)]$,  as it should be, according to (\ref{ch_2_1}).

As the next step, we observe that the same ladder series for the full $%
\Lambda_c (\Omega,K)$ can be re-arranged, as shown in Fig.\ref{fig8}, and
re-expressed in terms of the vertex function (let us call it ${\tilde{\Gamma}}_c (K_{F},P_{F})$), which is the sum of the interaction
 $V^{eff} (K_F-P_F)$ and
the ladder series of diagrams involving the combinations of two G's and the
interaction (see Fig. \ref{fig8}). Taking care of spin
indices, we obtain
\begin{equation}
\Lambda_c (\Omega,K)=1+2\int {\tilde{\Gamma}}_c (K_{F},P)\left\{
G_{qp}^{2}(P)\right\} _{\omega }\frac{d^{3}P}{(2\pi )^{3}}  \label{ch_2_1b}
\end{equation}%
Because $\Lambda_c (\Omega,K) =Z_{k}$,
 this equation takes exactly the same form as the Ward identity, Eq. (\ref{5_3_1}). Then, if we
  identify ${\tilde{\Gamma}}_c (K_{F},P)$  with the actual vertex function $\Gamma_c(K_{F},P)$, the FL formula for $Z_{k}$ is guaranteed to
coincide with the result obtained in the diagrammatic loop expansion. This
simple logics implies that, within our conserving approximation,
the correct $\Gamma_c (K_{F},P)$ is given by the series of ladder diagrams
with the charge component of $V^{eff} (K_F-P_F)$  ($=(3/2) V_{||}(k,p)$) as the leading term.  Below we express the series of ladder diagrams in Fig. \ref{fig8}
through an integral equation, solve it, and obtain $\Gamma_c (K_{F},P_{F})$ for particles on the FS as a function of $k$ and $p$, which, we recall, are the
deviations of the momenta $\mathbf{k}_{F}$ and $\mathbf{p}_{F}$ from the
corresponding hot spots.

\subsection{The structure of the vertex function $\Gamma^\omega_{\alpha\beta,\gamma\delta} %
(K_F,P_F)$.}
\label{sec_4_c}

\begin{figure}[htbp]
\includegraphics[width=\columnwidth]{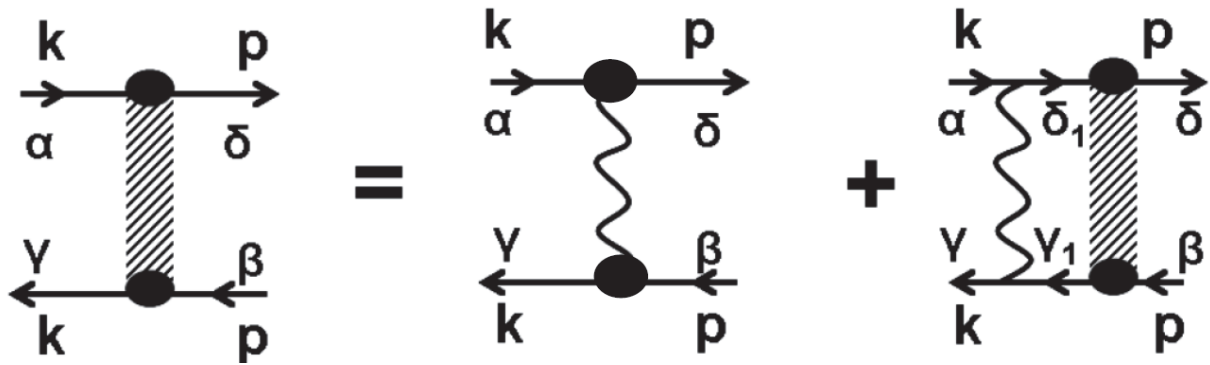}
\caption{Integral equation for $\Gamma^\omega_{K,P}$. The wavy line with big circles at the end points is $V^{eff} (K_F-P_F)$. Its charge and spin components are both equal to   $(3/2) V_{||}(k,p)$.}
\label{fig9}
\end{figure}
 Our reasoning for the selection of a particular ladder series of diagrams for $\Gamma_c$ does not rely on the fact that
 in the previous two subsections we analyzed only the charge component of  the vertex function.
 The same reasoning can be applied to the spin component of the vertex function. One can easily make sure that, by selecting the same set of diagrams for
 $\Gamma_s (K,P)$ as we did for $\Gamma_c (K,P)$, one reproduces the spin Ward identity.
 Each of the two components of the full
 $\Gamma _{\alpha \beta \gamma  \delta }^{\omega}(K_{F},P_{F})$ then can be
 re-expressed, as shown in Fig. \ref{fig9}, and presented as
an integral equation
 \begin{equation}
\Gamma _{a}(K_{F},P_{F})=V_{a}(K_F-P_F)+
 2 \int \frac{V_{a}(K_F-L)}{Z_{l}}\Gamma _{a}(L,P_{F})\left\{
G_{qp}^{2}(L)\right\} _{\omega }\frac{d^{3}L}{(2\pi )^{3}}  \label{sp_3}
\end{equation}%
where $a=c,s$ and $V_a (K,P) = (3/2) V(K-P)$

 Note that in each term in perturbation series the momenta are along the FS
near the corresponding hot spots. In terms
 with odd numbers of the interaction lines, $\mathbf{k}$ and
 $\mathbf{p}$ are close to the hot spots $\mathbf{k}_{F,hs}$ and $\mathbf{k}%
_{F,hs}+\mathbf{q}_\pi$, respectively. In these diagrams, the momentum $k$ is
along the FS around $\mathbf{k}_{F,hs}$ and the momentum $p$ is along the FS
around $\mathbf{k}_{F,hs}+\mathbf{q}_\pi$. In terms
with even numbers of the interaction lines,
 $\mathbf{k}$
and  $\mathbf{p}$ are close to the same hot spot $\mathbf{k}_{hs}$, and
the momenta $k$ and $p$ are along the FS around $\mathbf{k}_{hs}$.

It is convenient to introduce the function $f_{k,p}$ via
\beq
\Gamma _{a}(K_{F},P_{F})=f_{k,p} V_{a}(K_F-P_F) = \frac{3}{2}  V_{||} (k,p)
\label{ya2}
\eeq
The functions $f_{k,p}$ show how much the charge and spin components of the actual
 vertex function in a CFL differ from the corresponding components in an ordinary FL.

For further convenience, we also re-scale the momenta along the FS to ${\bar{k}}=k\xi $, ${\bar{p}}=p\xi$.
There are three
regimes of rescaled momenta near each hot spot. First, there is the true hot
region ${\bar{k}}<1$ (in original units, $|\mathbf{k}_{F}-\mathbf{k}%
_{hs}|\leq \xi ^{-1}$ ). In this regime, $Z_{k}\approx \lambda $ is a
constant. The second is the "lukewarm" region $1<{\bar{k}}<\lambda $ ($\xi
^{-1}<|\mathbf{k}_{F}-\mathbf{k}_{hs}|<\overline{g}/v_{F}$). In this regime $%
Z_{k}\approx \lambda /|{\bar{k}}\sin {\theta }|$ ($=\overline{g}/(v_{F}|%
\mathbf{k}_{F}-\mathbf{k}_{hs}|$) decreases with increasing separation from
the hot spot. Both the first and second regimes correspond to the CFL because
$Z_{k}$ is large compared to one. Finally, the third region is the cold one, $%
{\bar{k}}>\lambda $ ($|\mathbf{k}_{F}-\mathbf{k}_{hs}|>\overline{g}/v_{F}$).
In this region, $Z_{k}\approx 1$, i.e., the system remains in the ordinary
FL regime. The boundary between CFL and the ordinary FL regime is ${\bar{k}}%
\sim \lambda $ ($|\mathbf{k}_{F}-\mathbf{k}_{hs}|\sim {\bar{g}}/v_{F}$).
Note that the corresponding $|\mathbf{k}_{F}-\mathbf{k}_{hs}|$ are smaller
than $k_{F}$, as we assumed from the beginning that $\overline{g}$ is small
compared to $E_{F}$.

Substituting $\Gamma _{a}(K_{F},P_{F})$ in terms of $f_{k,p}$ into (\ref {sp_3})
and integrating over $l-k_{F}$ and $\omega _{l}$ in the same way as
before, we find that the function $f_{\bar{k},\bar{p}}$ satisfies the
integral equation
\begin{equation}
f_{\bar{k},\bar{p}}=1+ \frac{\lambda }{\pi }\int f_{\bar{%
k},\bar{l}}^{a}\frac{K_{\bar{k},\bar{l}}K_{\bar{l}\bar{p}}}{K_{\bar{k},\bar{p%
}}}\frac{d{\bar{l}}}{Z_{\bar{l}}}  \label{ch_n3}
\end{equation}%
where
$Z_{\bar{l}}=1+\lambda /\sqrt{1+{\bar l}^{2}\sin ^{2}{\theta }}$ and
\begin{equation}
K_{\bar{k}\bar{l}}=\frac{V_{||}(k,l)}{\overline{g}\xi ^{2}}=\frac{1}{{\bar{k}}%
^{2}+{\bar{l}}^{2}+1-2\bar{k}\bar{l}\cos \theta }  \label{ch_n5}
\end{equation}%
By construction, $f_{\bar{k},\bar{p}}$\ must satisfy
the FL formula for $Z_{k}$, Eq. (\ref{5_3_1}), which plays the role of a "boundary condition"
for Eq. (\ref{ch_n3})
\begin{equation}
\frac{\lambda }{\pi }\int f_{\bar{k},\bar{p}} K_{\bar{k},\bar{p}}\frac{d%
\bar{p}}{Z_{\bar{p}}}=Z_{\bar{k}}-1 = \frac{\lambda}{\sqrt{1 + {\bar k}^2 \sin^2 \theta}} \label{ch_n4}
\end{equation}

\section{The solution of the integral equation for $f_{\bar{k},\bar{p}}$}
\label{sec_5}

To gain some intuition on how  $f_{\bar{k},\bar{p}}$ looks, consider first
 the perturbation theory in $\lambda $.
At zero order, $f_{\bar{k},\bar{p}}=1$. To first order in $\lambda $, we
have
\begin{equation}
f_{\bar{k},\bar{p}}=1+\lambda \frac{{\bar{k}}^{2}+{\bar{p}}^{2}+1-2\bar{k%
}\bar{p}\cos \theta }{({\bar{k}}-{\bar{p}})^{2}\cos ^{2}{\theta }+(S_{\bar{k}%
}+S_{\bar{p}})^{2}}\left( \frac{S_{\bar{k}}+S_{\bar{p}}}{S_{\bar{k}}S_{\bar{p%
}}}\right)  \label{w_1}
\end{equation}%
where $S_{\bar{k}}=\sqrt{1+{\bar{k}}^{2}\sin ^{2}{\theta }}$. Substituting
this solution into the "boundary condition", we obtain, after
straightforward algebra that the left hand side of Eq. (\ref{ch_n4}) reduces to
\begin{eqnarray}
&&\frac{\lambda }{\pi }\int \frac{d{\bar{p}}}{{\bar{k}}^{2}+{\bar{p}}^{2}+1-2%
\bar{k}\bar{p}\cos \theta }\left( 1-\frac{\lambda }{S_{\bar{p}}}\right)
\notag \\
&&+\frac{\lambda ^{2}}{\pi ^{2}}\int \frac{d{\bar{p}}d{\bar{l}}}{({\bar{l}}%
^{2}+{\bar{p}}^{2}+1-2\bar{l}\bar{p}\cos \theta )({\bar{l}}^{2}+{\bar{k}}%
^{2}+1-2\bar{l}\bar{k}\cos \theta )} +O(\lambda ^{3})  \label{w_2}
\end{eqnarray}%
Performing elementary integrations, we find that the two $O(\lambda ^{2})$
terms cancel out, and (\ref{w_2}) reduces to
\begin{equation}
\frac{\lambda }{\pi }\int \frac{d{\bar{p}}}{{\bar{k}}^{2}+{\bar{p}}^{2}+1-2%
\bar{k}\bar{p}\cos \theta } = \frac{\lambda}{\sqrt{1 + {\bar k}^2 \sin^2\theta}}
\equiv Z_{\bar{k}}-1  \label{w_3}
\end{equation}%
We see that the "boundary condition" is satisfied in the perturbation theory
in $\lambda $, as it indeed should.

The perturbative expansion in $\lambda $ is a nice way to check the internal
self-consistency, but is of little practical use for us as we are interested
in the CFL regime at large $\lambda $. One can try to solve Eq. (\ref{ch_n3})
iteratively at large $\lambda $, using $Z_{\bar{k}}\approx \lambda /\sqrt{1+{%
\bar{k}}^{2}\sin ^{2}{\theta }}$. But then iterations just generate
additional terms $O(1)$ at each subsequent order,
 i.e., this procedure also does not lead to a meaningful result.

We analyzed the integral equation (\ref{ch_n3}) "as it is", i.e., without doing an expansion or iterations,
 and found, after some experimentation, that the trial function
\begin{equation}
f_{\bar{k},\bar{p}}=A \frac{{\bar{k}}^{2}+{\bar{p}}^{2}-2\bar{k}{\bar{p}}%
\cos {\theta }}{|\bar{k}\bar{p}|\sin \theta },  \label{ch_n6}
\end{equation}%
with a constant $A$,
 satisfies (\ref{ch_n3})
  when ${\bar{k}}$ and ${\bar{p}}$ are in the lukewarm regime $1<|\bar{k}|,|\bar{p}|<\lambda $, and one
momentum is much larger than the other, such that $f_{\bar{k},\bar{p}} >>1$.
This can be checked explicitly by noticing that the relevant $\bar{l}$ are
of order $\bar{p}$, and for those $Z_{\bar{l}}=\lambda /|\bar{l}|$. Indeed,
substituting the trial form $f_{\bar{k},\bar{l}}$ into the right hand side of Eq.
(\ref{ch_n3}) we find that the integral gives back the same $f_{\bar{k},\bar{p}}$. Equation
 (\ref{ch_n3}) is then satisfied as long as $f_{\bar{k},\bar{p}}$ is large
compared to 1.

This solution also has to satisfy the "boundary condition" (\ref{ch_n4}), i.e.,
reproduce the large value of $Z_{k} \sim \lambda$ inside the CFL regime. One possibility
would be that the integral over ${\bar{p}}$ in (\ref{ch_n4}) is confined to $
{\bar{p}}\sim {\bar{k}}$, and the overall factor $A$ is large and of order $\lambda $.
This is what happens near a nematic QCP, when the analog of $f_{\bar{k}\bar{p%
}}$ weakly depends on the location of momenta on the FS and is just a constant
of order $\lambda $ (Ref.\cite{cm_nematic}). In our case, however, the
result differs. Substituting our trial solution into the "boundary
condition" (\ref{ch_n4}), we find that for $1 << \bar{k}<<\lambda $ the boundary condition reduces to
\begin{equation}
\frac{2A |\sin {\theta }|}{\pi }\int^{\bar{p}_{max}}d\bar{p}=\frac{2A{\bar{p}}%
_{max}|\sin {\theta }|}{\pi } = \lambda   \label{ch_n7}
\end{equation}%
where ${\bar{p}}_{max} = c\lambda /|\sin {\theta }|$ and $c = O(1)$  is the upper limit of
applicability of the relation $Z_{\bar{p}}=\lambda /(|\bar{p}|\sin \theta )$.
 We see that the overall factor $A = \pi/(2c)$ turns out to be
 $O(1)$ rather
than $O(\lambda )$, and the "boundary condition" on $Z_{k}$ is reproduced
because the width of the region of integration over ${\bar{p}}$ is of order $\lambda $,
 i.e., is large. In
other words, for ${\bar{k}}$ deep inside the CFL regime, the integral over $
\bar{p}$ in (\ref{ch_n7}) is confined not to $O({\bar{k}})$ but rather to
the upper limit of the CFL behavior. The implication of this result is that $
Z_{k}$ for a fermion deep inside the CFL regime is determined by the behavior of
the quasiparticle vertex function when the other momentum $p$ is at the
boundary between CFL and an ordinary FL.

Analyzing the form of Eq. (\ref{ch_n6}) we see that $f_{\bar{k},\bar{p}%
}=O(\lambda )$ only when $\bar{k}=O(1)$ and $\bar{p}=O(\lambda )$. In all other
regions, $f_{\bar{k},\bar{p}}$ is smaller. In particular, when $\bar{k}$ and
$\bar{p}$ are comparable $f_{\bar{k},\bar{p}}=O(1)$, i.e., $\Gamma_c$
 is \textit{not} enhanced compared to the
 interaction $V_{||}(k,p)$.  We also verified explicitly that $f_{\bar{k
},\bar{p}}$ remains $O(1)$ in the hot region when ${\bar k}, {\bar p} <1$.

Assembling the forms of $f_{\bar{k},\bar{p}}$ in\ the various regions, we
obtain
\begin{equation}
f_{\bar{k},\bar{p}}\sim \left\{
\begin{tabular}{ccccc}
$\frac{|\bar{p}|}{|\bar{k}|}$ & when & $\lambda \gg |\bar{p}|\gg |\bar{k}|>1$
&  &  \\
$\frac{|\bar{k}|}{|\bar{p}|}$ & when & $\lambda \gg |\bar{k}|\gg |\bar{p}|>1$
&  &  \\
$O(1)$ & when & $|\bar{p}|\sim |\bar{k}|>1$ & or & $|\bar{p}|,|\bar{k}|<1$
\\
$\frac{\lambda }{|\bar{k}|}$ & when & $|\bar{p}|\gg \lambda ,|\bar{k}|\ll
\lambda $ &  &  \\
$\frac{\lambda }{|\bar{p}|}$ & when & $|\bar{k}|\gg \lambda ,|\bar{p}|\ll
\lambda $ &  &  \\
&  &  &  &
\end{tabular}%
\ \right.   \label{w_4}
\end{equation}%
 Note that this behavior holds independent of the value of $\theta$. The prefactor in each regime, however, depends on $\theta$.

\begin{figure}[htbp]
  $\begin{array}{cc}
\includegraphics[width=0.5\columnwidth]{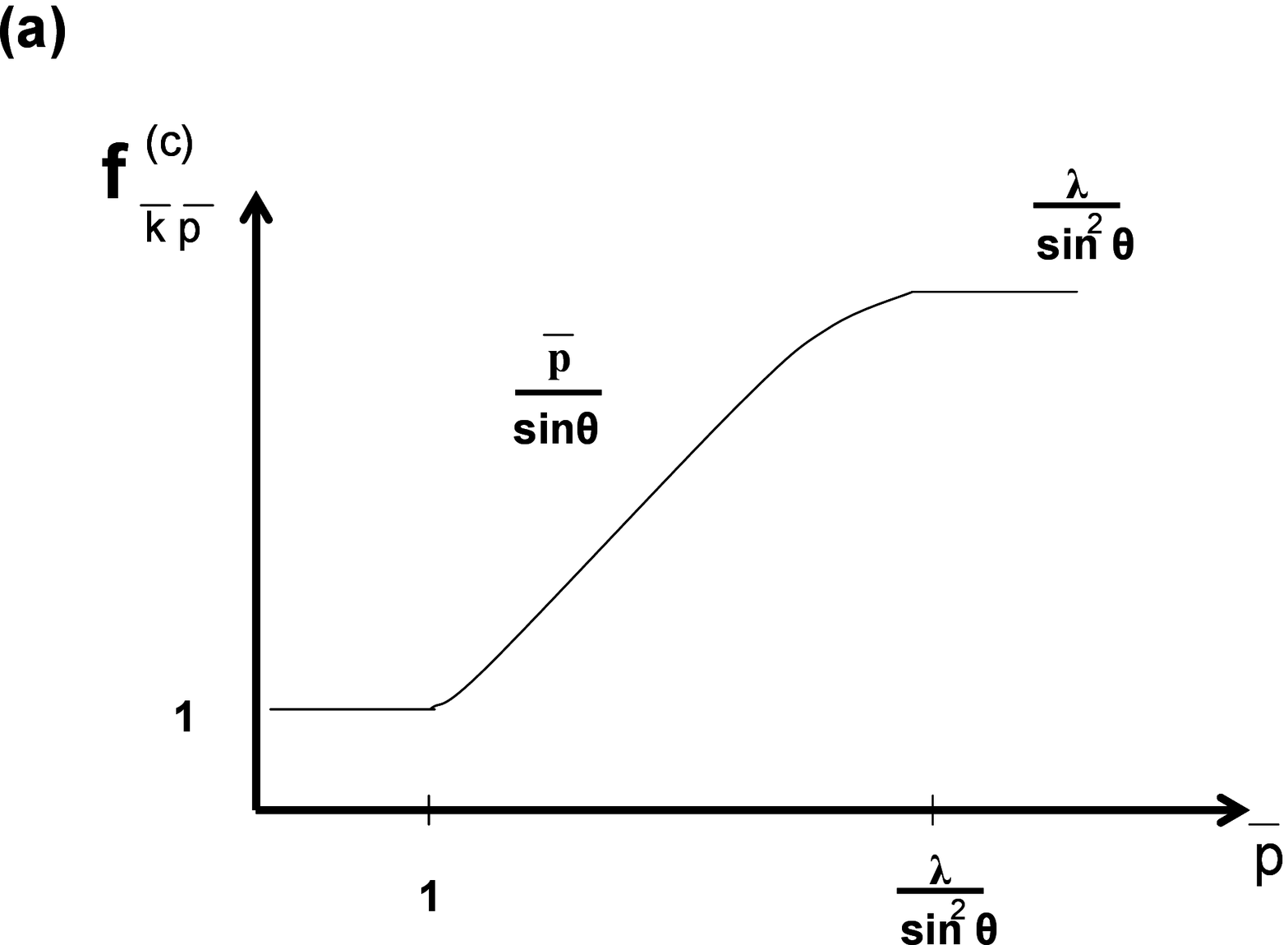}&
\includegraphics[width=0.5\columnwidth]{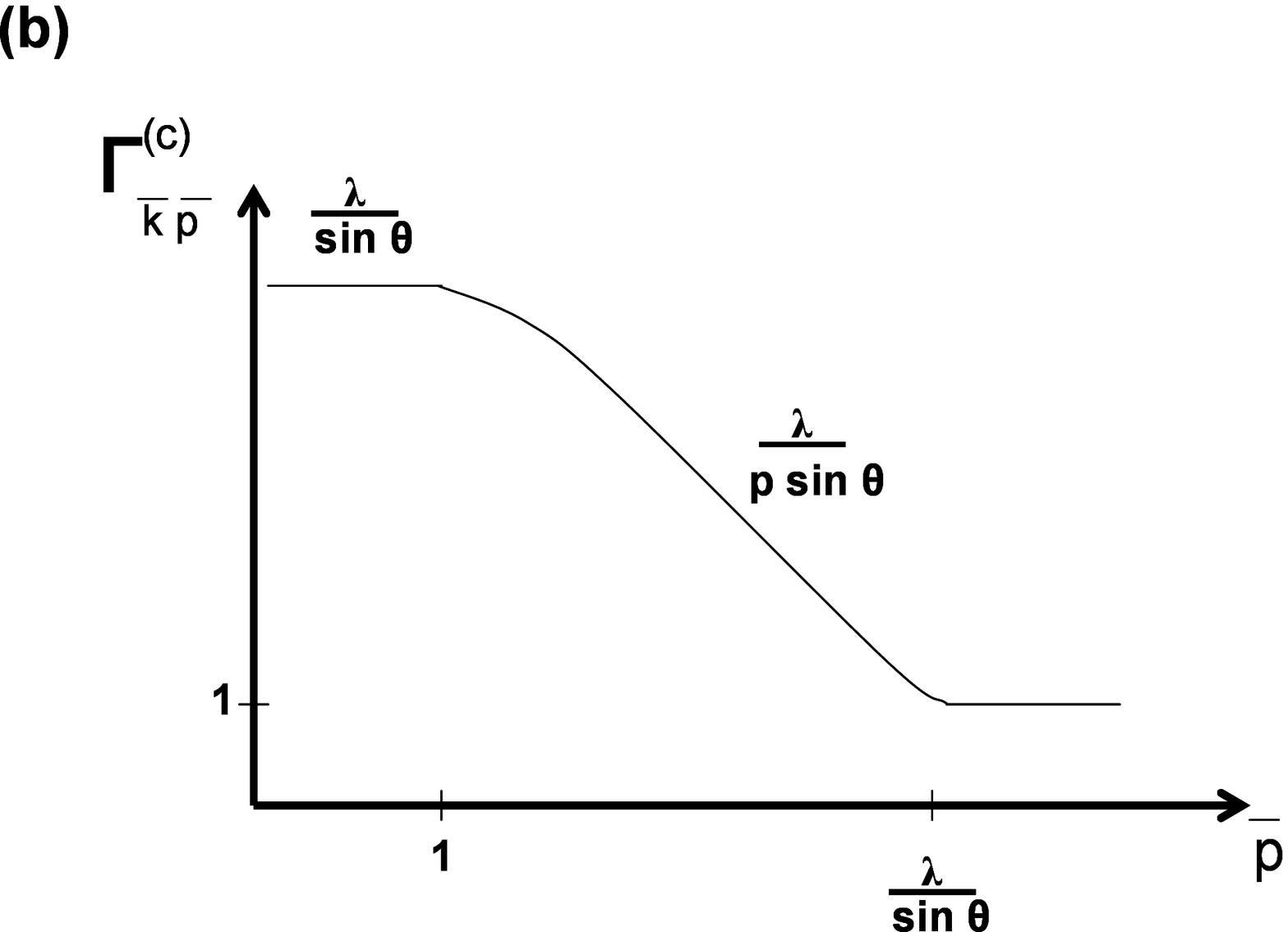}
\end{array}$
\caption{ a) The behavior of $f^{c}_{{\bar k},{\bar p}} = f_{\bar{k},\bar{p}}$ for ${\bar k} = O(1)$, as a function of ${\bar p}$. b) The same for $\Gamma_c (K_F-P_F) \equiv \Gamma^c_{{\bar k}, {\bar p}} = (3/2) V_{||} ({\bar k},{\bar p})
f^{c}_{{\bar k},{\bar p}}$.  The behavior of the spin components $f^{s}_{{\bar k},{\bar p}}$ and $\Gamma^s_{{\bar k}, {\bar p}}$  is the same.
We recall that ${\bar k} = k \xi$, ${\bar p} = p\xi$, and $p$ and $k$ are deviations  from the corresponding hot spots along the FS.}
\label{fig10}
\end{figure}

The full vertex function is $\Gamma^\omega_{\alpha\beta,\gamma\delta}
(K_{F},P_{F})= (3/2) V_{||} ({\bar k},{\bar p}) f_{\bar{k},\bar{p}}
\left(\delta_{\alpha\gamma} \delta_{\beta\delta} +
 {\vec \sigma}_{\alpha\gamma} {\vec \sigma}_{\beta\delta}\right)$.
 Substituting $f_{\bar{k},\bar{p}}$
from (\ref{ch_n6}) and returning to original variables $k$ and $p$ we obtain in the "lukewarm" regime $\lambda >> {\bar k}, {\bar p} >>1$
\begin{equation}
\Gamma^\omega_{\alpha\beta,\gamma\delta}
(K_{F},P_{F}) = \frac{3A}{2}~ \frac{\overline{g}}{|k||p|\sin \theta} ~\left(\delta_{\alpha\gamma} \delta_{\beta\delta} +
 {\vec \sigma}_{\alpha\gamma} {\vec \sigma}_{\beta\delta}\right)
  \label{w_7}
\end{equation}%
We show the behavior of the charge components $f^c_{\bar{k},\bar{p}}=f_{\bar{k},\bar{p}}$ and $\Gamma_c
(K_{F},P_{F})$ for $\bar{k}=O(1)$ as a function of $\bar{p}$ in Fig. \ref{fig10}. The behavior of the spin components is identical.

The selective enhancement of the quasiparticle vertex function is specific
to the SDW QCP and may be essential for the analysis of higher-order
renormalizations of the bosonic propagator~\cite{peter}.

\subsection{Contributions to $\Gamma_{c,s} (K_F, P_F)$ with large and
small $K_F-P_F$.}
\label{sec_5_a}

More information about the physics of the CFL can be obtained if one looks
more carefully into the diagrammatic series for $\Gamma ^{\omega
}(K_{F},P_{F})$ and realizes that the full $\Gamma ^{\omega }(K_{F},P_{F})$
is the sum of the contributions in which momenta $\mathbf{k}_{F}$ and $%
\mathbf{p}_{F}$ are located near hot spots separated by $\mathbf{q}_\pi$
(these are the diagrams with an odd number of spin-fermion
interaction lines), and contributions in which momenta $\mathbf{k}_{F}$ and $%
\mathbf{p}_{F}$ are located near the same hot spot (these are the diagrams
with an even number of spin-fermion interaction lines). We label the
corresponding contributions to $f_{\bar{k},\bar{p}}$ as $f_{\bar{k},\bar{p}%
}^{(\pi )}$ and $f_{\bar{k},\bar{p}}^{(0)}$. The $f_{\bar{k},\bar{p}}$
in (\ref{ch_n6}) is the sum of the two contributions: $f_{\bar{k},\bar{p}%
}=f_{\bar{k},\bar{p}}^{(\pi )}+f_{\bar{k},\bar{%
p}}^{(0)} \equiv f_{\bar{k},\bar{p}}^{(+)}$.

\begin{figure}[htbp]
\includegraphics[width=\columnwidth]{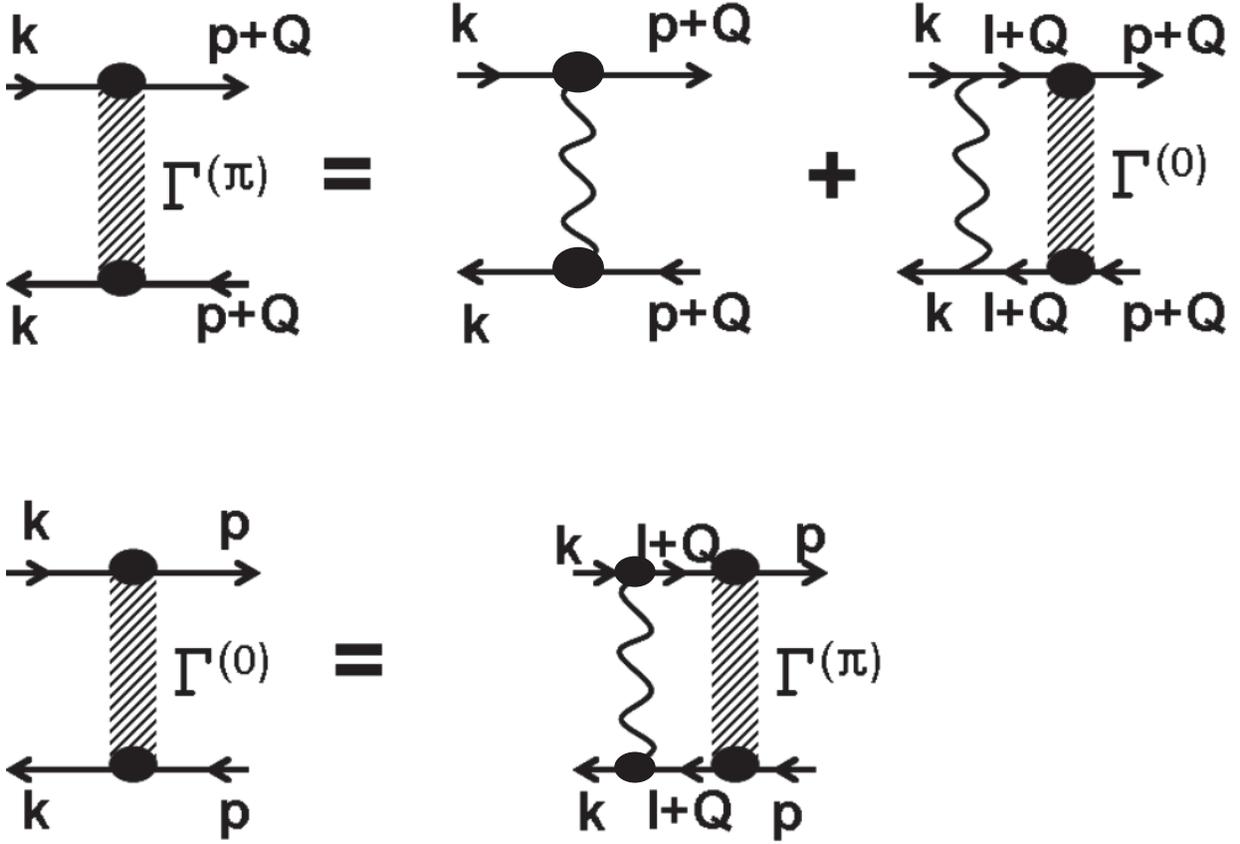}
\caption{ The set of coupled equations for $\Gamma^{(\pi)} (K_F, P_F) = f^{\protect\pi}_{{\bar k},{\bar p}} V_{||} ({\bar k},{\bar p})$ and
$\Gamma^{(0)} (K_F, P_F) = f^{0}_{{\bar k}k,{\bar p}} V_{||} ({\bar k},{\bar p})$.}
\label{fig9a}
\end{figure}

The set of coupled equations for $f_{\bar{k},\bar{p}}^{(\pi )}$ and $f_{\bar{%
k},\bar{p}}^{(0)}$ is readily obtained from Fig. \ref{fig9a}. We have
\begin{eqnarray}
&&f_{\bar{k},\bar{p}}^{(\pi )}=1+\frac{\lambda }{\pi }\int f_{\bar{k},\bar{l}%
}^{(0)}\frac{K_{\bar{k},\bar{l}}K_{\bar{l}\bar{p}}}{K_{\bar{k},\bar{p}}}%
\frac{d{\bar{l}}}{Z_{\bar{l}}}  \notag \\
&&f_{\bar{k},\bar{p}}^{(0)}=\frac{\lambda }{\pi }\int f_{\bar{k},\bar{l}%
}^{(\pi )}\frac{K_{\bar{k},\bar{l}}K_{\bar{l}\bar{p}}}{K_{\bar{k},\bar{p}}}%
\frac{d{\bar{l}}}{Z_{\bar{l}}}  \label{thr9}
\end{eqnarray}%
Summing up the two equations we reproduce Eq. (\ref{ch_n6}). To compare the
relative strength of the two contributions to $f_{\bar{k},\bar{p}}$ it is
also instructive to analyze $f_{\bar{k},\bar{p}}^{(-)}=f_{\bar{k},\bar{p}%
}^{(\pi )}-f_{\bar{k},\bar{p}}^{(0)}$. Subtracting the second equation in (%
\ref{thr9}) from the first one, we obtain that the equation for $f_{\bar{k},%
\bar{p}}^{(-)}$ decouples from that for $f_{\bar{k},\bar{p}}^{(+)}$ and
takes the form
\begin{equation}
f_{\bar{k},\bar{p}}^{(-)}=1-\frac{\lambda }{\pi }\int f_{\bar{k},\bar{l}%
}^{(-)}\frac{K_{\bar{k},\bar{l}}K_{\bar{l}\bar{p}}}{K_{\bar{k},\bar{p}}}%
\frac{d{\bar{l}}}{Z_{\bar{l}}}  \label{ch_n8}
\end{equation}%
The difference with the corresponding expression for $ f_{\bar{k},\bar{p}}^{(+)} \equiv f_{\bar{k},\bar{p}}$,
 Eq. (\ref{ch_n3}), is the negative sign of the integral term in the right hand side of (\ref{ch_n8}).
Just like $f_{\bar{k},\bar{p}}^{(+)}$ is related to the
triple vertex $\Gamma_k = Z_k$ by the "boundary condition", Eq. (\ref{ch_n4}), the function
$f_{\bar{k},\bar{p}}^{(-)}$ is related to the "triple" vertex $\Gamma ^{(-)}_k$,
 which is given by the same set of diagrams as
of $\Gamma_k$ (see Fig. \ref{fig7}), but with different signs of contributions with even and
odd interaction lines.
The "boundary condition" on $f_{\bar{k},\bar{p}}^{(-)}$ is
\begin{equation}
\frac{\lambda }{\pi }\int f_{\bar{k},\bar{p}}^{(-)} K_{\bar{k},\bar{p}}\frac{d%
\bar{p}}{Z_{\bar{p}}}=1 - \Gamma ^{(-)}_{\bar k}
 \label{ch_n4_1}
\end{equation}
and the function $\Gamma ^{(-)}_{\bar k}$ satisfies the integral equation
\begin{equation}
\Gamma ^{(-)}_{\bar k}=1 - \frac{\lambda }{\pi}\int d{\bar p}\frac{\Gamma ^{(-)}_{\bar p}}{Z_{{\bar p}}}\frac{%
1}{{\bar k}^{2}+{\bar p}^{2}-2{\bar k}{\bar p}\cos {\theta }+1}  \label{mo_10_1}
\end{equation}%
which is similar to Eq. (\ref{mo_10}) (after re-scaling), but with different sign of the integral term.

 The function $\Gamma ^{(-)}_k$ has
been analyzed by Hartnoll {\it et al} in the context of vertex corrections for
conductivity~\cite{max_last}. They found that $\Gamma ^{(-)}_k$ has a power-law dependence
$\Gamma ^{(-)}_k = B^{(-)} k^\beta$ for $1 < {\bar k}< \lambda/\sin{\theta}$, with
\begin{eqnarray}
\beta  &=&\frac{\theta }{\pi -\theta },~~0<\theta <\pi /2  \notag \\
&=&\frac{\pi -\theta }{\theta },~~\pi /2<\theta <\pi   \label{ch_n11}
\end{eqnarray}%
 Note that now the exponent does depend on the value of $\theta$.
The prefactor $B^{(-)}$
is determined by the condition $\Gamma ^{(-)}_0 =0$ (or, more accurately, $\Gamma ^{(-)}_0 \ll 1$).
This yields
\beq
1 = \frac{2 B^{(-)} \sin{\theta}}{\pi}\int_{0}^{{\bar p}_{max}} d{\bar p} {\bar p}^{\beta-1}   \label{mo_10_1_1}
\eeq
where, we remind, ${\bar p}_{max} = a \lambda/\sin{\theta}$ and $a = O(1)$.  The integral is confined to the upper limit and
 yields
 \beq
B^{(-)} =  \frac{\pi \beta}{2 \sin{\theta}} \left(\frac{\sin{\theta}}{a \lambda}\right)^{\beta}
\label{ch_n11_1}
\eeq
Using   (\ref{ch_n11_1}) we then find that
\beq
\Gamma^{(-)}_k  =  \frac{\pi \beta}{2 \sin{\theta}} \left(\frac{ {\bar k} \sin{\theta}}{a \lambda}\right)^{\beta} \equiv
\frac{\pi \beta}{2 \sin{\theta}} \left(\frac{ 4\pi v_F k \sin{\theta}}{3 a {\overline{g}}}\right)^{\beta}
\eeq
 At smaller ${\bar k} \leq 1$ the momentum dependence becomes weak and
$\Gamma^{(-)}_k$ saturates at a small value $\Gamma^{(-)}_k \sim (1/\lambda)^{\beta}$. At ${\bar k} > \lambda/(\sin \theta)$ a straightforward analysis shows that
$\Gamma^{(-)}_k$ saturates at  $\Gamma^{(-)}_k \sim 1/(\sin \theta)$.

That $\Gamma^{(-)}_k$ is small at ${\bar k} < 1$ has a direct implication for the form of
 $f_{{\bar k},\bar{p}}^{(-)}$ at ${\bar k} <1$ and $1< {\bar p} < \lambda/\sin{\theta}$.
 Namely,
 the "boundary condition" on $f_{{\bar k},\bar{p}}^{(-)}$, Eq. (\ref{ch_n4_1}), becomes
\beq
\frac{\sin\theta}{\pi }\int \frac{d{\bar p}}{|{\bar p}|} f_{0,\bar{p}}^{(-)} \approx 1
 \label{ch_n4_1_1}
\end{equation}
To satisfy this equation, $f_{{\bar k},\bar{p}}^{(-)}$ must be a decreasing function of ${\bar p}$.  Accordingly, we search for the solution of
Eq. (\ref{ch_n8}) for ${\bar k} \leq 1$ and ${\bar p} \gg 1$  in the form
$f_{{\bar k},\bar{p}}^{(-)} \propto \left({\bar p}\right)^{-\beta^{(-)}}$.  Substituting this form into (\ref{ch_n8}) we obtain after simple algebra the
self-consistency condition on $\beta^{(-)}$ in the form
\begin{equation}
1=\frac{\sin \theta }{\pi }\int_{-\infty }^{\infty }\frac{|x|^{-\beta^{(-)}}{(x-2}%
sgn(x)\cos {\theta })}{x^{2}+1-2x\cos {\theta }}  \label{ch_n9}
\end{equation}%
The integration in (\ref{ch_n9}) can be performed analytically and yields
\begin{equation}
\cot {\frac{\pi \beta^{(-)}}{2}}=\cot \left( \frac{\theta (1+\beta^{(-)})}{2}\right)
~~~{\text{f}or}~0<\theta <\pi /2  \label{thr10}
\end{equation}%
and
\begin{equation}
\cot {\frac{\pi \beta^{(-)}}{2}}=\cot \left( \frac{(\pi -\theta )(1+\beta^{(-)})}{2}%
\right) ~~~{\text{f}or}~\pi /2<\theta <\pi   \label{thr10_1}
\end{equation}%
Solving for $\beta^{(-)}$ we obtain $\beta^{(-)} = \beta$, where
$\beta$ is given by Eq. (\ref{ch_n11}).

When both $\bar k$ and $\bar p$ are between $1$ and $\lambda/\sin{\theta}$, the solution
of (\ref{ch_n8}) is more complex, but we found that it is rather well approximated by
\begin{equation}
f^{(-)}_{\bar k,\bar p} = A^{(-)} \left(\frac{|\bar k \bar p| \sin {\theta}%
}{{\bar k}^2 + {\bar p}^2 -2 \bar k \bar p \cos{\theta}}\right)^{\beta}
\label{ch_n12}
\end{equation}
The prefactor $A^{(-)} = O(1)$ is determined by the "boundary condition",  Eq. (\ref{ch_n4_1}).
For ${\bar k} < \lambda/(\sin \theta)$, $\Gamma^{(-)}_k \sim ({\bar k} \sin \theta/\lambda)^{\beta}$ is still small compared to $1$, and neglecting it, we obtain from (\ref{ch_n4_1})
\begin{eqnarray}
\frac{2^{\beta-1}  A^{(-)} \sin{\theta}}{\pi} \int_{-\infty}^{\infty} \frac{%
|x|^{\beta +1} dx}{(x^2 +1 - 2 x \cos{\theta})^{{\beta} +1}}  =1 - \Gamma^{(-)}_k \approx 1.   \label{w_6}
\end{eqnarray}
where $x = {\bar p}/{\bar k}$. This equation is not an exact one because Eq. (\ref{ch_n12}) is valid only for ${\bar p}, {\bar k} > 1$, but
 it should be good for an estimate of the value of $A^{(-)}$.
 One can easily check that the
prefactor $A^{(-)}$ remains $O(1)$ for all $\theta$ and, according to (\ref{w_6}),  is equal to $A^{(-)} =2$ for $\theta =
\pi/2$, when $\beta =1$.
To reproduce the boundary condition with $\Gamma^{(-)}_k$ included. one has to include subleading terms of order ${\bar k}/\lambda$ and ${\bar p}/\lambda$ into
$f^{(-)}_{\bar k,\bar p}$.

Assembling contributions from various regions, we obtain, for a generic $\sin \theta = O(1)$,
\begin{equation}
f^{(-)}_{\bar k,\bar p} \sim \left\{
\begin{tabular}{ccccc}
$\left(\frac{|\bar{k}|}{|\bar{p}|}\right)^{\beta}$ & when & $\lambda \gg |\bar{p}|\gg |\bar{k}|>1$
&  &  \\
$\left(\frac{|\bar{p}|}{|\bar{k}|}\right)^{\beta}$ & when & $\lambda \gg |\bar{k}|\gg |\bar{p}|>1$
&  &  \\
$O(1)$ & when & $|\bar{p}|\sim |\bar{k}|>1$ & or & $|\bar{p}|,|\bar{k}|<1$
\\
$\left(\frac{|\bar{k}|}{\lambda|}\right)^{\beta}$  & when & $|\bar{p}|\gg \lambda ,|\bar{k}|\ll
\lambda $ &  &  \\
$\left(\frac{|\bar{p}|}{\lambda|}\right)^{\beta}$  & when & $|\bar{k}|\gg \lambda ,|\bar{p}|\ll
\lambda $ &  &  \\
&  &  &  &
\end{tabular}%
\ \right.   \label{w_4_1}
\end{equation}%

\begin{figure}[htbp]
 $\begin{array}{cc}
\includegraphics[width=0.5\columnwidth]{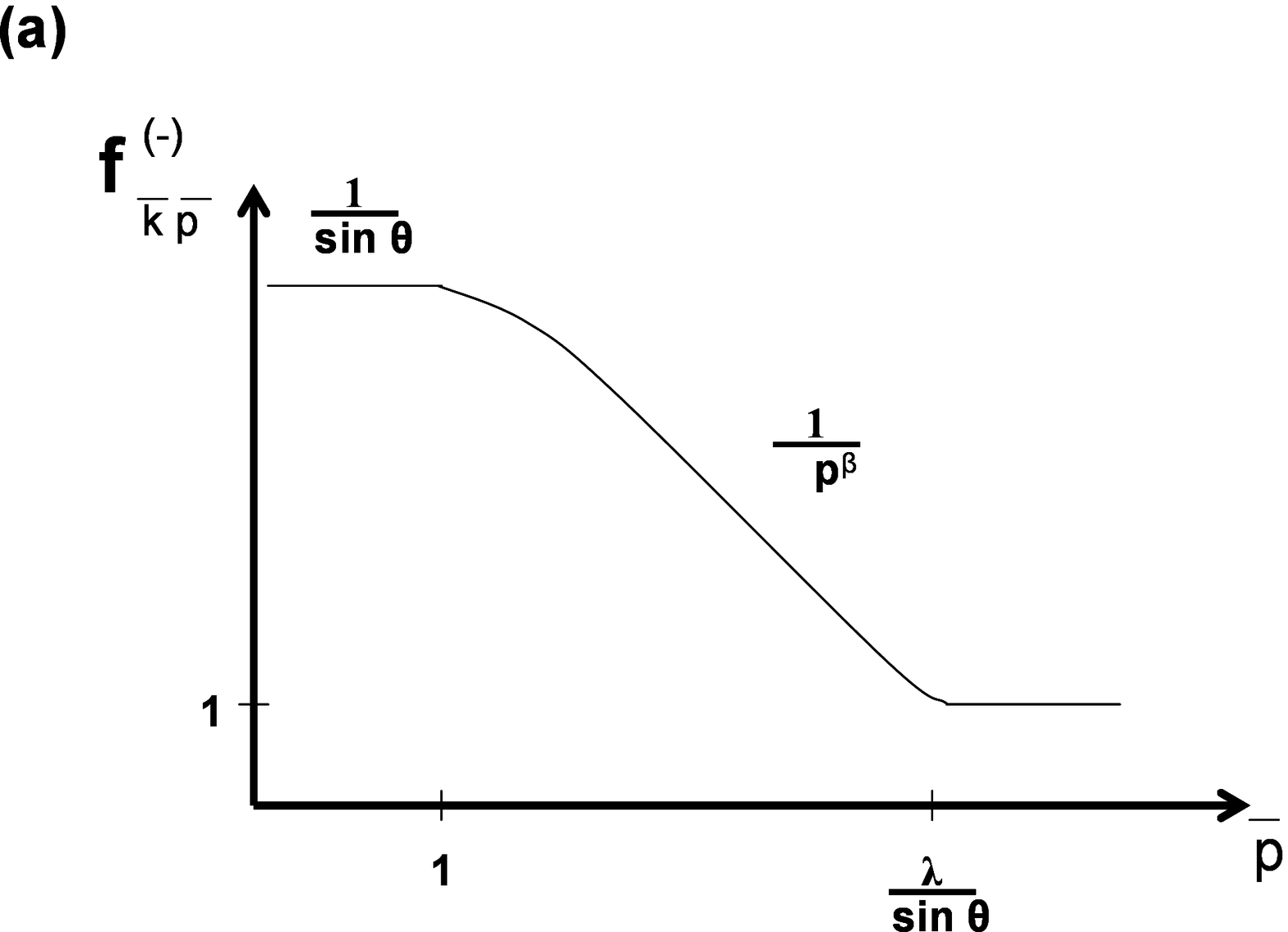}&
\includegraphics[width=0.5\columnwidth]{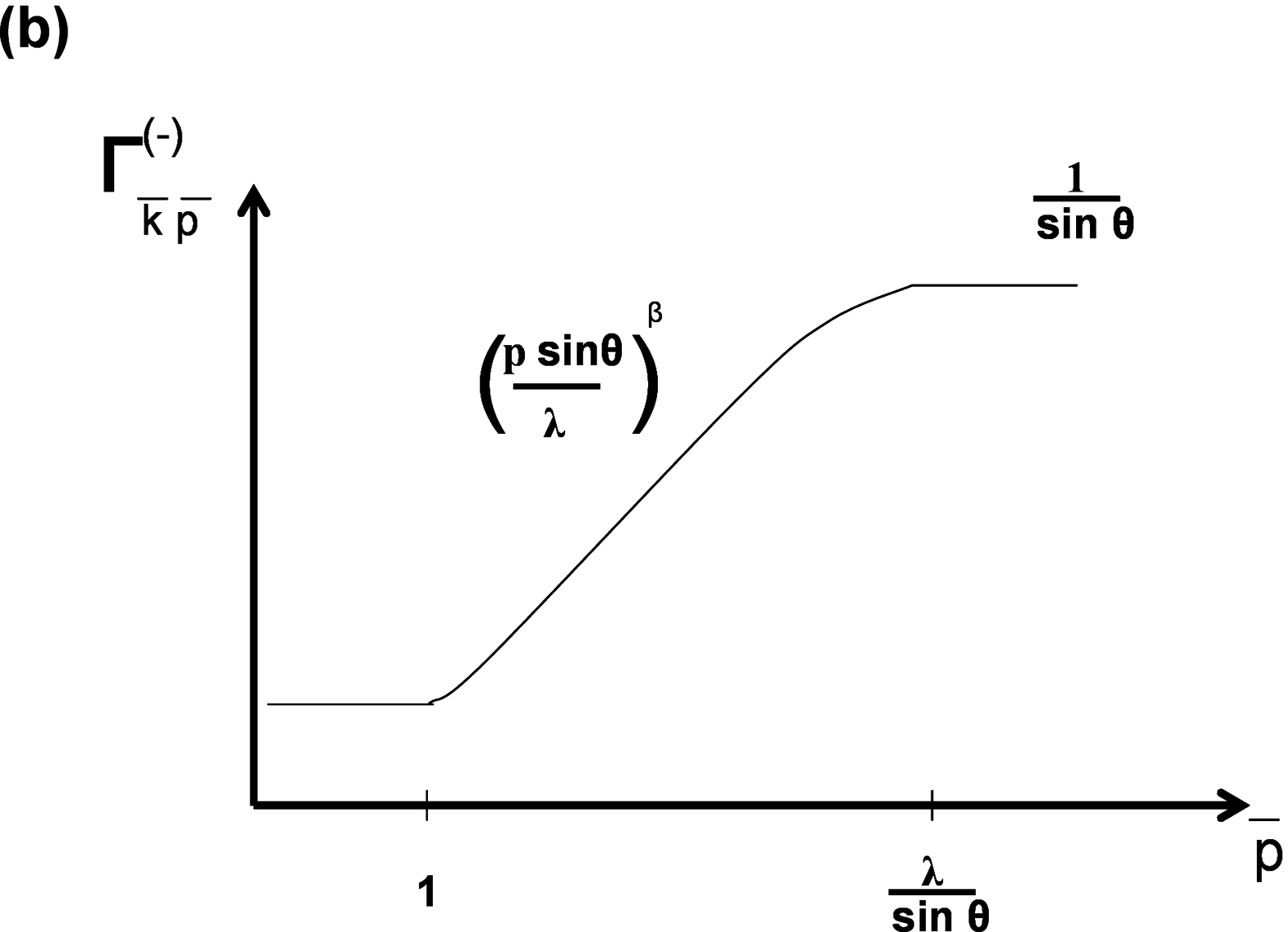}
\end{array}$
\caption{ a) The behavior of $f^{(-)}_{{\bar k},{\bar p}}$ for ${\bar k} =
O(1)$, as a function of ${\bar p}$. b) The same for $\Gamma^{(-)} =
f^{(-)}_{{\bar k},{\bar p}} V_{||}({\bar k}, {\bar p})$}
\label{fig11}
\end{figure}

 We plot $f_{\bar{k},\bar{p}%
}^{(-)}$ for $\bar{k}=O(1)$ as a function of $\bar{p}$ and the corresponding
$\Gamma^{(-)}_c (k,p)= V_c (k-p) f^{(-)}_{k,p}$
in Fig. \ref{fig11}.

The outcome of the analysis of $f^{(-)}_{\bar k,\bar p}$ is that $%
f^{(-)}_{\bar k,\bar p}$
 is of order one when $\bar k$ and $\bar p$ are
comparable, but becomes small when either $\bar k \gg \bar p$ or $\bar p \gg
\bar k$. Comparing this with $f^{(+)}_{\bar k,\bar p}$,
 we see that  both are of order one when $\bar k \sim \bar p$, but when one momentum is
larger than the other, $f^{(+)}_{\bar k,\bar p} >> f^{(-)}_{\bar k,\bar p}$.
In this last case, $f^{(\pi)}_{\bar k,\bar p}$ and $f^{(0)}_{\bar k,\bar p}$
are almost identical and large, i.e., the enhancement of $\Gamma_{c,s}$
compared to the interaction $V_{||}(k,p)$ holds for both components of $%
\Gamma_{c,s}$: the one in which $\mathbf{k}_F$ and $\mathbf{p}_F$ are
located near hot spots at distance $\mathbf{q}_\pi$ from each other, and the one
in which $\mathbf{k}_F$ and $\mathbf{p}_F$ are located near the same hot
spot.

\section{The Landau function in a CFL, the density of states, and the uniform susceptibilities}
\label{sec_7}

\subsection{The Landau function}
\label{sec_7_0}

The full vertex function $\Gamma _{\alpha\beta,\gamma\delta }^{\omega
}(K_{F},P_{F})$ is given by
\beq
 \Gamma _{\alpha\beta,\gamma\delta }^{\omega}(K_{F},P_{F}) =
\frac{3}{2} \frac{\bar g \xi^2} {{\bar k}^2 + {\bar p}^2 + 1 - 2 {\bar k} {\bar p} \cos{\theta}}
f_{\bar{k},\bar{p}} \left(\delta _{\alpha \gamma }\delta _{\beta \delta }+\vec{\sigma }_{\alpha \gamma }\vec{\sigma }_{\beta
\delta }\right)
 \label{thr22_1}
\eeq
In the isotropic case, the  quasiparticle interaction function (the Landau function) $F_{\alpha\beta,\gamma\delta }(K_{F},P_{F})$ is related
to $\Gamma _{\alpha\beta,\gamma\delta }^{\omega
}(K_{F},P_{F})$ by
\beq
F_{\alpha\beta,\gamma\delta }(K_{F},P_{F})= 2 \frac{N_F}{Z^2} \Gamma _{\alpha\beta,\gamma\delta }^{\omega
}(K_{F},P_{F})
\label{s11}
\eeq
 (see, e.g., Ref.\cite{agd}).  The prefactors are essential as in the isotropic case $F_{\alpha\beta,\gamma\delta }(K_{F},P_{F})$ depends
 only on the angle $\gamma$  between ${\bf k}_F$ and ${\bf p}_F$, and may therefore be expanded in the basis of Legendre functions of argument $\cos\gamma$. The corresponding components of $F_{\alpha\beta,\gamma\delta }(\phi)$, labeled as $F^{(l)}_{c}$ and $F^{(l)}_s$  determine
 various observables. For example, the DOS and uniform charge and spin susceptibilities are expressed as
 \beq
 N_F = N_{F,0}(1+F^{(1)}_c), ~\chi_c  = \chi_{c,0} \frac{1+F^{(1)}_c}{1 + F^{(0)}_c}, ~ \chi_s  = \chi_{s,0} \frac{1+F^{(1)}_c}{1 + F^{(0)}_s}
\label{s10}
\eeq
 where $N_{F,0}$, $\chi_{c,0}$, and $\chi_{s,0}$ are the DOS and the susceptibilities of free fermions.

  In our case the situation is a bit more involved because the quasiparticle residue depends on momentum along the FS and also because
 $\Gamma_c (K_F, P_F)$ and $\Gamma_s (K_F,P_F)$ depend separately on $k$ and on $p$ rather than only on $k-p$.
  It is natural to expect that a proper extension of Eq. (\ref{s11}) to our case would be to replace $Z^2$ in (\ref{s11}) by $Z_k Z_p$.
  Keeping $N_F$ intact for the moment and
 combining Eqs. (\ref{y_1}), (\ref{thr1}), (\ref{ya2}), (\ref{w_4}), (\ref{w_4_1}), and  Eq. (\ref{ch_5}) for $Z_{k}$,
  we find after simple algebra that when ${\bar k}$ and ${\bar p}$ satisfy $1 << {\bar k}, {\bar p} << \lambda$, the charge and spin components of the Landau function become momentum-independent,
\begin{equation}
F_{c,s}(K_{F},P_{F}) =
\frac{16 \pi^2 N_{F} A v^2_F \sin\theta }{3\overline{g}}.
\label{y_1_2}
\end{equation}

 However,  how to deal with $N_F$ is a subtle issue because $N_F$ is proportional to the effective mass $m^*$, which in our case  also becomes momentum-dependent. Even more essential, it is a {\it priori} unclear how to properly define $F^{(l)}_c$ and $F^{(l)}_s$ from Eq.
 (\ref{y_1_2})  to reproduce, e.g., Eqs. (\ref{s10}).
 Because of these complications, below we actually derive the expressions for spin and charge susceptibilities and see whether they can be described by
 formulas similar to Eq. (\ref{s10}).

 The susceptibilities in the lattice Hubbard model have been analyzed within the renormalization group scheme by  Halboth and Metzner~\cite{metzner}.
  However, the problem which they studied  differs from ours  -- in their case fermionic self-energy still can be neglected, while in our case it plays the major role.

\subsection{The density of states and uniform susceptibilities}
\label{sec_7_1}

\subsubsection{The density of states}

We first observe that the enhanced effective mass
near the hot spots, $m_{k}^{\ast }/m\propto Z_{k}$\ leads to a logarithmic
enhancement of the  DOS at the Fermi level $N_{F}$ , compared to the
total DOS in an ordinary FL $N_{F,0} = m/(2\pi)$.  We have
\begin{equation}
N_{F} =N_{F,0}\langle Z_{k}\rangle _{k} = 2 N_{F,0}\int_{1/\xi \sin
\theta }^{\lambda /\xi \sin \theta}
\frac{\lambda }{k\xi \sin \theta } \frac{dk}{2 \pi k_F} \\
 =  \frac{3\overline{g} m }{8\pi^3 v_F k_F \sin\theta}
\ln \lambda  \label{7_1}
\end{equation}%
The divergence of the total DOS $N_{F}$ leads to a diverging specific heat coefficient $%
C/T\propto \ln \lambda $.
This agrees with the calculation in Ref.~\cite{acs}.

The divergence of the DOS  raises the question whether the uniform
susceptibilities diverge at a SDW QCP, because in (\ref{s10})  $N_{F}$
appears as an overall factor in both susceptibilities.  If we formally use
 the relation (\ref{y_1_2})  and obtain $F^{(l=0)}_c$ and $F^{(l=0)}_s$
  by averaging over both $k$ and $p$, we find that both $F^{(l=0)}_c$ and $F^{(l=0)}_s$ scale as $N_F$.
   This would imply that
  $N_F$ cancels out between numerator and denominator
   in the expressions for $\chi_{c,s}$ in (\ref{s10}), i.e., $\chi_{c,s}$  tend to  finite values at a SDW QCP.

  To verify this, we perform a diagrammatic order-by-order calculation of the susceptibilities and verify whether the results can be cast into the forms of Eq. (\ref{s10}) with $F_c = F_s$ given by Eq. (\ref{y_1_2}). As a side result of our consideration, we also show how the
  Landau FL formulas for spin and charge susceptibilities are reproduced diagrammatically. To the best of our knowledge, this has not been presented in detail in the textbooks (there is some discussion on this in Ref. \cite{cm_nematic}).

\subsubsection{The uniform susceptibilities}

  Because spin and charge components of our vertex function are the same (to leading order in $\xi$) the static charge and spin susceptibilities are the same up
  to the extra factor $\mu^2_B$ in the spin magnetic susceptibility.

\begin{figure}[htbp]
  $\begin{array}{cc}
\includegraphics[width=0.65\columnwidth]{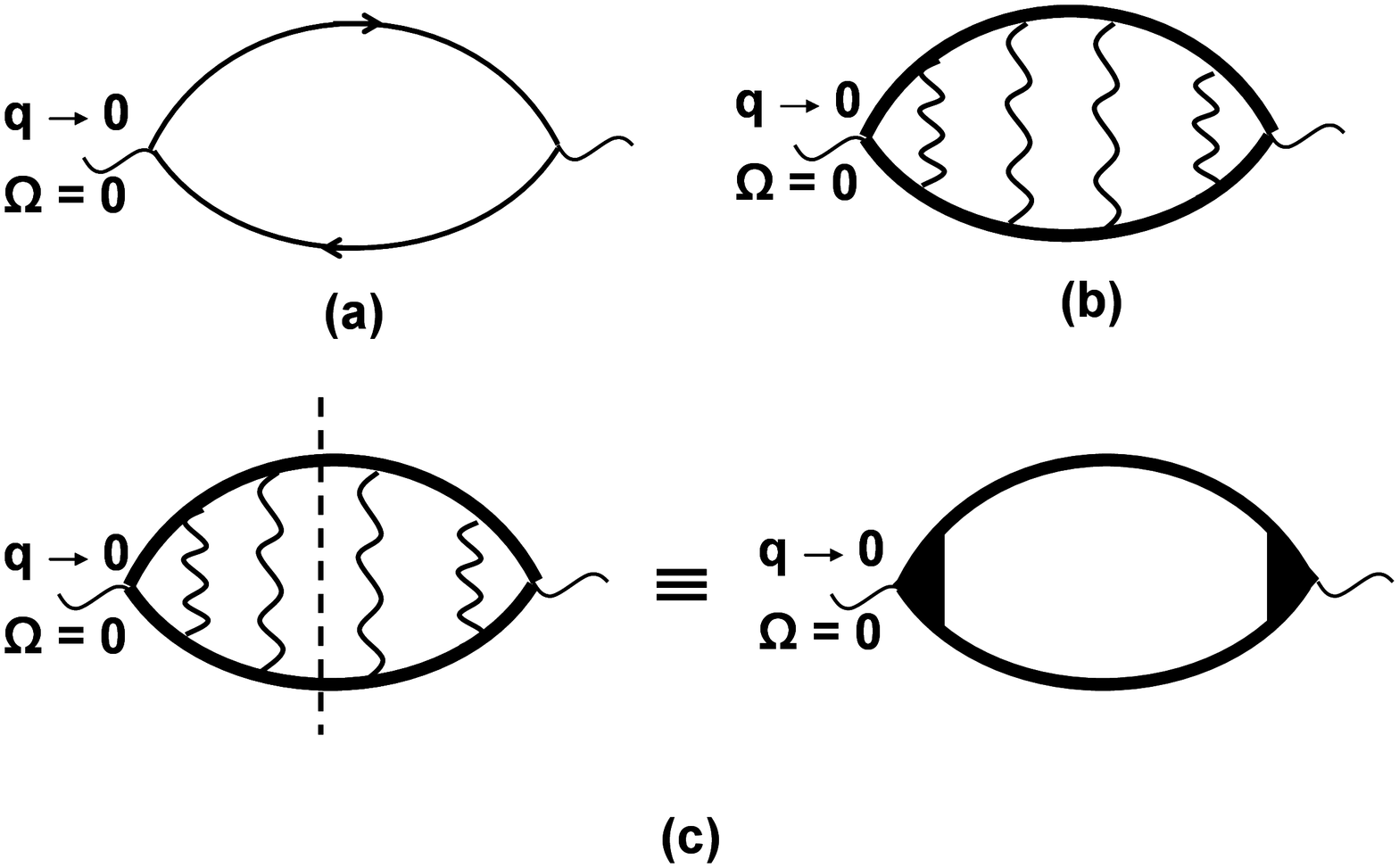}\\
\includegraphics[width=0.65\columnwidth]{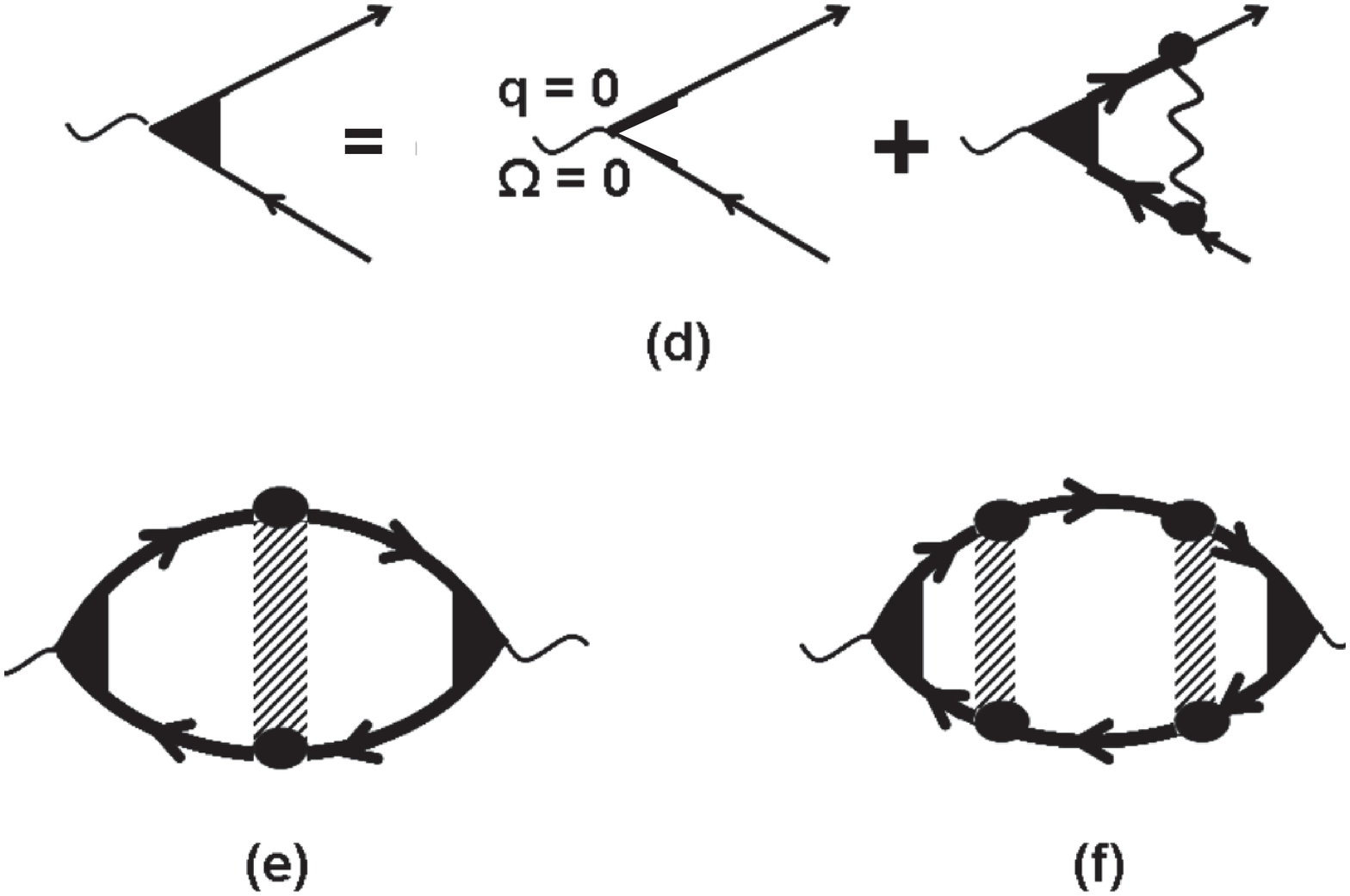}
\end{array}$
\caption{Diagrams for the particle-hole polarization bubble with either spin $\delta-$functions or spin $\sigma-$matrices in side vertices (for charge or spin bubbles, respectively). (a) - particle-hole bubble for free fermions.  At $\Omega =0$ and $q \to 0$, the result comes from fermions right on the FS.
(b) particle-hole bubble with self-energy and vertex correction insertions. The factor $Z_k$ is included into the fermionic propagator. There are two contributions from
 each cross-section: One comes from fermions right on the FS and  another comes from fermions away from the FS. For the latter one can safely set $q=0$.
 (c) The same skeleton diagram as in b), but with one selected cross-section (marked by the dashed line),
   in which the contribution comes from the FS. In all other cross-sections the contributions come from fermions away from the FS.
    The shaded vertices here and in (d) represent the infinite sum of such contributions. (e) and (f) -- Diagrams with two and three cross-sections in which the contribution comes from fermions at the FS.              }
\label{fig3N}
\end{figure}

  The charge susceptibility $\chi_c = dn/d\mu$ (equal to charge compressibility, up to a sign) describes the change of the number density $n$ of particles under a change of the chemical potential.
 For free fermions, $\chi_{c,0}$ is given diagrammatically by a particle-hole bubble with zero incoming frequency, in the limit of vanishingly small momentum (Fig. \ref{fig3N}a):
 \beq
 \chi_{c,0} = -2 ~{\text lim}_{q \to 0} \int \frac{d^2p d \omega_p}{2\pi^3} \frac{1}{(i\omega_p - \epsilon_p) (i\omega_p - \epsilon_{p+q})}
\label{s12}
\eeq
where the overall factor $2$ comes from spin summation and a factor $-1$ from the fermionic loop.
Using $d^2p = d p (d \epsilon_p/v_F)$ and $\epsilon_{p+q} \approx \epsilon_p + v_F q_{\perp}$ and integrating over $\omega_p$ first (which, we note, is always the safest way to proceed), we find that the integral over $\omega_p$ is confined to a tiny range of $|\omega_p|, |\epsilon_p| < v_F |q_{\perp}|$ where the poles in the two Green's functions are in two different half-planes of frequency.  Integrating over $\omega_p$ and then over $\epsilon_p$, we obtain $\chi_c$ as the integral over the FS
  \beq
 \chi_{c,0} = 2 \int \frac{dp}{4\pi^2 v_F}
\label{s14}
\eeq
For a circular FS, $v_F$ is a constant along the FS, the integral over $p$ gives the length of the FS $2\pi k_F$, and we obtain the well-known result
$\chi_{c,0} = 2 N{F,0} = m/\pi$, where $m = k_F/v_F$.   The key point here is that the integration
 in (\ref{s12})  is truly confined to an infinitesimally small range
 near the FS, with the energy width of order $v_F q_{\perp}$.  This should be contrasted with the integrals for $Z_k$ and $\Gamma^\omega$,
   which  come from small but finite distances from the FS~\cite{comm_sept}

 Now let us perform the same calculation, but use the renormalized Green's function instead and also add vertex corrections.
  Self-energy corrections obviously replace $i\omega_p$ by $i\omega_p Z_p$.  Vertex corrections are more tricky. At first glance, they form a ladder series, shown in Fig. \ref{fig3N}b,  and one has to dress up only one out of two side vertices  to avoid double counting.
   The ladder series
    yield the dressed vertex
  $\Lambda_c (p) = Z_p$. Evaluating the integral with the product of two dressed Green's functions and one $\Lambda_c (p)$ we find that the factors $Z_p$ cancel out and the result remains the same as Eq. (\ref{s14}).  This does not give us the expected proportionality to $N_F$.

  A more careful analysis, however,reveals that both side vertices have to be renormalized.  The argument is the following:  Once we set the external $q$ to be small but finite, we have two
  contributions from each cross-section in the diagram in Fig. \ref{fig3N}b. One comes from the infinitesimally small range near the FS and another comes
   from a small but finite distance from the FS.  In this situation, to get the result proportional to the bare susceptibility,  we have to choose one cross-section in which we
    take the contribution coming from the FS and sum up all cross-sections
    on both sides of
     the selected one, each time taking only the contribution from a finite distance from the FS [see Figs. \ref{fig3N}(c) and \ref{fig3N}(d)].  In this computational procedure, there is no double counting, and the result is
\bea
 \chi_{c,1} &=& -2~ {\text lim}_{q \to 0} \int \frac{d^2p d \omega_p}{(2\pi)^3} \left(\Lambda_c (k) \right)^2 \frac{1}{(i\omega_p Z_p - \epsilon_p) (i\omega_p Z_p- \epsilon_{p+q})} = \nonumber \\
 && \int \frac{d^2p d \omega_p}{(2\pi)^3}  \frac{Z^2_p}{(i\omega_p Z_p - \epsilon_p) (i\omega_p Z_p- \epsilon_{p+q})} =\nonumber\\
 && 2 \int \frac{dp}{4\pi^2 v_F} Z_p   = \frac{1}{\pi^2 v_F \sin{\theta}} \frac{\lambda}{\xi} \ln{\lambda} = 2 N_{F}
\label{s15}
\eea
precisely as expected.  The same result can be also obtained on physics grounds, once we use  the fact that $\Gamma^c_k = Z_k$ is the renormalization factor in the coupling between fermionic density operator $c^\dagger_k c_k$ and the change of the chemical potential.

 We see therefore that at this stage we reproduced the numerator of (\ref{s10}) by  choosing one cross-section in which the integration is confined to the FS.
 We now show that the denominator is reproduced by selecting terms with two, three, and so on cross-sections of this kind, and each time summing up an infinite series of contributions between such cross-sections and keeping only the terms coming from a finite distance from the FS.  These intermediate contributions sum up exactly into $\Gamma^\omega$, and this is how the Landau function enters into the diagrammatic loop expansion.
 The term with two cross-sections in which  contributions are confined to the FS is shown in Fig. \ref{fig3N}e.  The vertex between the two cross-sections is $\Gamma^c (K_F,P_F)$.  Using the form of $\Gamma^c (K_F,P_F)$, Eq. (\ref{w_7}),  assembling the spin factors, and evaluating the integrals, we obtain
\beq
 \chi_{c,2} = -6 \frac{ A {\overline{g}}}{(4\pi^2 v_F)^2 \sin{\theta}} \int_{1/\xi \sin \theta }^{\lambda /\xi \sin \theta} \frac{dk}{k} \int_{1/\xi \sin \theta }^{\lambda /\xi \sin \theta} \frac{dp}{p} =  -\frac{3 A {\overline{g}}}{2\pi^4 v^2_F \sin{\theta}} (\ln{\lambda})^2
\label{s17}
\eeq
We emphasize that the quadratic dependence on $\ln{\lambda}$ is a direct consequence of the fact that $\Gamma_c (K_F,P_F)$ scales as $1/|k p|$. If $\Gamma_c$ would be the same as $V(K_F-P_F) = V_{||} (k,p)$,
  we would only obtain one power of the logarithm.

The diagram with three cross-sections in which  contributions are confined to the FS is shown in Fig. \ref{fig3N}f. Evaluating the integrals, we obtain
\beq
 \chi_{c,3} = \frac{9 A^2 {\overline{g}}^2}{4\pi^6 v^3_F \sin {\theta}} \frac{\xi}{\lambda} (\ln{\lambda})^3
\label{s17_1}
\eeq
and so on.
We now notice that
\beq
 \chi_{c,2} =  -\chi_{c,1} S,  ~~\chi_{c,3} =  \chi_{c,1} S^2,
\eeq
where
\beq
S = \frac{3 A {\overline{g}}}{(2\pi^2 v_F)} \frac{\xi}{\lambda}~\ln{\lambda}
\label{s18}
\eeq
i.e., the first two terms in the series form a geometric progression. One can easily make sure that this continues to higher loops, and the full $\chi_c$ is
 given by
 \beq
 \chi_c = \frac{\chi_{c,0}}{1 + S} \propto \frac{N_F}{1 + S}
 \label{s19}
 \eeq

 The term $S(\theta)$ is proportional to $F_c (K_F,P_F)$. One can easily make sure that for a rotationally-invariant system it is exactly $F^{(l=0)}_c$.
  From this perspective, our analysis is the diagrammatic derivation of the FL formula for the charge susceptibility.
  A very similar reasoning was used by A. Finkelstein in his analysis of the charge susceptibility in a  disordered electron liquid with interactions~\cite{sasha}.

   For our case, using $\lambda/\xi = 3{\overline{g}}/(4\pi v_F)$, we obtain
   \beq
   S = \frac{2A}{\pi}~\ln{\lambda}.
   \label{s20}
   \eeq
   Substituting the expressions for $N_F$ and $S$ into (\ref{s19}) we find that $\chi_c$ remains finite at the SDW QCP and is
   \beq
   \chi_c = 2N_{F,0} \frac{3 \overline{g}}{8 \pi A v_F k_F \sin \theta} \sim N_{F,0} \frac{\overline{g}}{E_F}
   \label{s22}
   \eeq
 This expression is parametrically smaller than for free fermions, i.e.,  although $\chi_c$ remains finite at the SDW QCP, it gets
  reduced by renormalizations.

     It is instructive to compare $S(\theta)$ with $F_c (K_F,P_F)$, averaged over $k$ and $p$. The result depends over what interval we average.
     If we average only over the interval where  $F_c (K_F,P_F)$ is given by  Eq. (\ref{y_1_2}) and is independent on momenta, we obtain
     \bea
     <F_c (K_F,P_F)> &=& N_F \frac{16 \pi^2 A v^2_F \sin \theta}{3{\overline {g}}} \nonumber \\
    && = \frac{16 \pi^2 Av^2_F \sin \theta}{3{\overline {g}}}~ \frac{3\overline{g} m }{8\pi^3 v_F k_F \sin\theta} \ln{\lambda} = \frac{2A}{\pi} \ln{\lambda},
   \label{s21}
   \eea
   which is exactly the same as (\ref{s20}). Then $ <F_c (K_F,P_F)>$  is the same as  $F^{(l=0)}_c$ in the FL theory.
   If, however, we average over the whole FS, we get a smaller $ <F_c (K_F,P_F)>$. This uncertainty implies that in lattice systems
    there is no universal formula like  Eq. (\ref{s10}), and the only way to  obtain $\chi_c$ is to
     explicitly
     sum up the series of terms with one, two, three, and so on cross-sections, in which the integral comes from the FS.

The same result holds for the spin susceptibility $\chi_s$, which, as we said, differs from a charge susceptibility only by a factor $\mu^2_B$.
 \beq
   \chi_s = 2 \mu^2_B N_{F,0}~  \frac{3 \overline{g}}{8 \pi A v_F k_F \sin \theta} \sim \mu^2_B N_{F,0} \frac{\overline{g}}{E_F}
   \label{s22_1}
   \eeq
The reduction of $\chi_s$ compared to free-fermion value $\mu^{2}_B N_{F,0}$ agrees with the analysis in~\cite{peter3}.

We emphasize that the  equivalence between the two is a direct consequence of the fact that the fully renormalized vertex function has the same functional form in the charge and spin channels, in which case Ward identities impose equivalence between
spin and charge components of $\Gamma^\omega$ (and of the Landau function).

\section{ Conclusions.}
\label{sec_8}

In this paper we obtained the quasiparticle vertex function $\Gamma ^{\omega
}(K_{F},P_{F})$
 for a system of interacting fermions in 2D  near a spin-density wave instability at wave
vector $\mathbf{q}_\pi=(\pi ,\pi )$ ($K_{F}$ is a 3D vector in
momentum/frequency space with components $(\mathbf{k}_{F},0)$). Near the SDW instability,
 the interaction between low-energy fermions, $V(K-P)$, is mediated
by overdamped collective excitations in the spin channel. We identified two regimes
 near a SDW QCP: (i) an ordinary FL regime, in which renormalizations induced by the spin-mediated interaction,
  are weak, and (ii) a CFL regime, in which they are strong. We argued that at the boundary between the two regimes
 the self-energy changes from  predominantly  static (in an ordinary FL) to predominantly dynamic (in a CFL).
 In the latter case, the
 quasiparticle $Z$
factor (the inverse residue) and the effective mass ratio $m^{\ast }/m$
coincide and are large in hot and lukewarm regions near a hot spot on the FS.
 We argued that $\Gamma ^{\omega }_{\alpha\beta,\gamma\delta}(K_{F},P_{F})$ differs from the
effective spin-mediated interaction $V(K_F-P_F) {\vec \sigma}_{\alpha\delta} {\vec \sigma}_{\beta\gamma}$ already
 in the ordinary FL regime due to additional contributions from AL-type diagrams.
The latter, although nominally of higher power in the (small) interaction constant, turn out to be of similar
structure as the first order spin-fluctuation exchange term
 due to
 a singularity, which leads to the
 cancelation of one power of the coupling. The dominant effect of the AL-terms is to change the sign of the quasiparticle interaction function
 in the spin channel.  We demonstrated that the full $\Gamma^\omega$ in an ordinary FL (the sum of the direct spin-mediated interaction and the AL terms) has equal spin and charge components and satisfies the constraint on these two components, imposed by the Ward identities related to the conservation of the total number of particles and the total spin.

We further considered the vertex function in a CFL.  We argued that the equivalence between charge and spin components of $\Gamma^\omega$ still holds, but each component gets strongly renormalized due to contributions which vanish for a
static interaction and are negligible in an ordinary FL, but become essential in a CFL where the
Landau damping term plays a central role.
We show that  $\Gamma _{c}(K_{F},P_{F}) = \Gamma_s (K_{F},P_{F})$ are enhanced
compared to $V(K_F-P_F)$, but the enhancement is
 present only when one of the fermionic momenta, either $\mathbf{k}_{F}$ or $%
\mathbf{p}_{F}$, is at much larger distance from the corresponding hot spot
than the other. When the deviations are comparable, $\Gamma _{c,s}$ and $V(K_F-P_F)$ are of the same order.
We used this renormalized $\Gamma_{c,s}$ to obtain the quasiparticle interaction function (the Landau function)
 $F(K_{F},P_{F})$ and showed that near a CFL it is
essentially independent of momenta $\mathbf{k}_{F}$ and $\mathbf{p}_{F}$.

We further showed that the residue $Z_{k}$ of a fermion with momentum $\mathbf{%
k}_{F}$ deep in the CFL region near a hot spot is determined by contributions
from $\Gamma_{c,s}(K_{F},P_{F})$ for which the other fermion with momentum $%
\mathbf{p}_{F}$ is located at much larger deviations from a hot spot, namely
near the boundary between lukewarm and cold regions of the FS.
 We also
demonstrated that in the momentum range where $\Gamma _{c,s}$ is enhanced
compared to $V(K_{F}-P_F)$, the enhancement holds for the part of
 $\Gamma _{c,s}$  in which $\mathbf{k}_{F}$ and $%
\mathbf{p}_{F}$ are located near different hot spots at a distance
$q_\pi$  from each other, and the part in which $\mathbf{k}_{F}$ and $\mathbf{p}%
_{F} $ are located near the same hot spot.

As an immediate application of this result we considered  the density of states and the uniform charge
and spin susceptibilities. We showed that the DOS  diverges as $\log \xi $ upon
approaching the SDW QCP. We introduced the Landau function $F (K_F,P_F)$ by straightforward extension of the formula for the isotropic
 case but cautioned that $F(K_F,P_F)$ depends on both $\mathbf{k}_F$ and $\mathbf{p}_F$ along the FS and not only  on their difference. In this situation,
  one cannot use the standard FL formulas and has to obtain charge and spin susceptibilities by explicitly summing up bubble diagrams with self-energy
   and vertex corrections. We demonstrated how to do this and paid special attention to the difference between contributions coming from the infinitesimal vicinity of the FS and from states at a small but still finite distance from the FS. We showed that the charge and spin susceptibilities
    tend to exactly the same value at the SDW QCP (modulo the additional factor $\mu^2_B$ in the spin susceptibility).
    We demonstrated that  higher-loop terms for $\chi_{c,s}$ form a geometrical series,
    like in an isotropic FL. We argued that, in this situation, one can effectively describe $\chi_{c,s}$ by a FL-like formula in which the
    $<F_{c,s} (K_F,P_F)>$ play the role of the $l=0$  Landau parameters.

We thank E. Abrahams, L.P. Gorkov, A. Finkelstein, P. Hirschfeld, S. Maiti, D. Maslov, J. Schmalian, and A. Varlamov for fruitful discussions. We are thankful to S. Maiti for the help with the figures.  A.V.C. was
supported by the DOE Grant No. DE-FG02-ER46900. PW is grateful for the
hospitality extended to him as a visiting professor at the University of
Wisconsin, Madison, and acknowledges support through an ICAM Senior
Scientist Fellowship.


\begin{thebibliography}{99}
\bibitem{landau} L.D. Landau, Sov. Phys. JETP {\bf 5}, 101 (1957); {\bf 8}, 70 (1959).

\bibitem{agd} A.\ A.\ Abrikosov, L.\ P.\ Gorkov, and I.\ E.\ Dzyaloshinski,
\emph{Methods of Quantum Field Theory in Statistical Physics}, (Dover, New York, 1963); E. M. Lifshitz and L. P. Pitaevski, \emph{%
Statistical Physics}, (Pergamon Press, 1980).

\bibitem{baym} D. Pines and P. Nozieres, \emph{The Theory of Quantum Liquids}%
, Addison-Wesley, Menlo Park, 1966; P.W. Anderson  \emph{Basic Notions of
Condensed Matter Physics}, Benjamin-Cummings, Melno Park, 1984; G. Baym and
C. Pethick, \emph{Landau Fermi Liquid Theory}, Wiley, New York, 1991.

\bibitem{rg} R.
Shankar, Rev. Mod. Phys. 66, 129 (1994); W.
Metzner, C. Castellani, and C. Di Castro, Adv. Phys. 47, 317 (1998).

\bibitem{pitaevskii}  L.P. Pitaevskii, Sov. Phys. JETP {\bf 10}, 1267 (1960).

\bibitem{kondratenko} P.S. Kondratenko, Sov. Phys. JETP {\bf 19}, 972 (1964).
See also O. Betbeder-Matibet and P. Nozieres, Annals of Physics, {\bf 37}, 17 (1966);
I. E. Dzyaloshinskii and P. S. Kondratenko, Sov. Phys. JETP, {\bf 43}, 1036 (1976).

\bibitem{nozieres} P. Nozieres "Theory of interacting
Fermi systems'' Frontiers of Physics, New York, W.A. Benjamin, 1964.

\bibitem{acs} A. Abanov, A. V. Chubukov, and J. Schmalian, Adv. Phys.,
\textbf{52}, 119 (2003).

\bibitem{ms} M.A. Metlitski and S. Sachdev, \prb {\bf 82}, 075128 (2010)

\bibitem{efetov} K. B. Efetov, H. Meier, and C. P\'{}epin, Nature Physics {\bf 9} 442, (2013).

\bibitem{wang} Y. Wang and A. V. Chubukov Phys. Rev. Lett. {\bf 110}, 127001
(2013).

\bibitem{peter} E. Abrahams and P. W\"{o}lfle, Proc. Nat. Acad. Sc. {\bf 109},
3238 (2012); E. Abrahams, J. Schmalian, and P. W\"{o}lfle, arXiv:1303.3926.

\bibitem{cm_nematic} D. L. Maslov and A. V. Chubukov, \prb {\bf 81},
045110 (2010);  see also M. Zacharias, P. W{\"o}lfle, and M. Garst, \prb {\bf 80}, 165116 (2009).

\bibitem{scalapino} see, e.g., D. J. Scalapino, Rev. Mod. Phys. {\bf 84}, 1383 (2012) and references therein.

\bibitem{max_last} S. A. Hartnoll, D. M. Hofman, M. A. Metlitski, and S.
Sachdev, \prb {\bf 84}, 125115 (2011).

\bibitem{ms1} M.A. Metlitski and S. Sachdev, \prb {\bf 82}, 075127 (2010).

\bibitem{senthil} D. F. Mross, J. McGreevy, H. Liu, and T. Senthil, \prb {\bf 82}, 045121 (2010).

\bibitem{baym_1} G. Baym and L.P. Kadanoff, Phys. Rev. {\bf 124}, 287 (1961).

\bibitem{cm_fm} A. V. Chubukov and D. L. Maslov Phys. Rev. Lett. 103, 216401
(2009).

\bibitem{sasha} A. M. Finkelstein, in "50 Years of Anderson Localization", World Scientific Review, {\bf 9}, 385 (2010); Int. J. of Mod. Phys. B
{\bf 24},  1855  (2010).

\bibitem{abrikosov} A.A. Abrikosov "Fundamentals of the Theory of Metals", North-Holland, 1988.

\bibitem{maslov_last} A. V. Chubukov and D. L. Maslov,
\prb {\bf 81}, 245102 (2010).

\bibitem{al}  L.G.Aslamazov and  A.I.Larkin, Soviet Phys.: Solid State {\bf 10}, 875 (1968); Phys. Lett.
{\bf 26A}, 238 (1968).

\bibitem{varlamov}
A. I Larkin and A.A. Varlamov, "Fluctuation Phenomena in Superconductors" in
 "Handbook on Superconductivity: Conventional and Unconventional Superconductors" edited by K.-H.Bennemann and J.B. Ketterson,
  Springer, 2002.


\bibitem{martin} P. C. Martin and J. Schwinger, Phys. Rev. {\bf 115}, 1342 (1959);
L.P. Kadanoff and P. C. Martin, Phys. Rev. {\bf 124}, 670 (1961).


\bibitem{gorkov} M. Dzero and L. P. Gor'kov Phys. Rev. B {\bf 69}, 092501 (2004).

\bibitem{comm_1} A. V. Chubukov, \prb {\bf 72}, 085113 (2005); \prb {\bf 71}, 245123 (2005). 
 Note that  the identification of the ordinary FL to CFL
 crossover with the crossover from $\Sigma (k, \omega) \approx \Sigma (k)$ to $\Sigma(k, \omega) \approx \Sigma(\omega)$ is specific to the SDW
transition. For a nematic transition one can also identify regimes of predominantly $k$ and $\omega$ dependence of the self-energy~\cite{cm_nematic}, however, the crossover
between the two regimes occurs already within an ordinary FL, at $%
\overline{g}/(v_{F}\xi ^{-1})\sim (a/\xi )\ll 1$.

\bibitem{ss_lee} S.-S. Lee, Phys. Rev. B {\bf 80}, 165102 (2009);

\bibitem{sri_last} A. L. Fitzpatrick, S. Kachru, J. Kaplan, and S. Raghu, Phys. Rev. B 88, 125116 (2013).

\bibitem{metzner}  C. J. Halboth and W. Metzner
Phys. Rev. B {\bf 61}, 7364 (2000).

\bibitem{comm_sept}
For free fermions,  $\chi_0$ can also obtained by integrating first over the fermionic dispersion $\epsilon_k$
 in finite limits $-\Lambda < \epsilon_k <\Lambda$ and then integrating over frequency.  The result of such integration does not depend on
 $\Lambda$ and is the same as Eq. (\ref{s14}). This reflects the fact that for free fermions, the 2D integral over $\omega$ and quasiprticle $\epsilon_k$
  is logarithmically divergent in both infra-red and ultra-violet limits, and can computed using either infra-red regularization (integrating over frequency first) ot unlta-violet regularization (integrating over quasiparticle dispersion first) [A. V. Chubukov, D. L. Maslov, and R. Saha
Phys. Rev. B 77, 085109 (2008)].  For higher-order diagrams, the integral over $\epsilon_k$ converges already at $|\epsilon_k| \ll \Lambda$
 and the integration can be done in any order and in infinite limits. For our purposes, it is more convenient to evaluate higher-order diagrams for the susceptibility by integrating first over frequency.

\bibitem{peter3} P. W\"{o}lfle and E. Abrahams, Phys. Rev. B {\bf 80}, 235112
(2009).

\end{thebibliography}
\end{document}